\makeatletter
\def\input@path{{latex/}}
\makeatother
\documentclass[cernpreprint,atlasdraft=false,texlive=2016,UKenglish,texmf,texmf]{atlasdoc}
 
\usepackage[biblatex=true,subcaption]{atlaspackage}
\usepackage{atlasbiblatex}
 
\usepackage{atlascontribute}
 
\usepackage{atlasphysics}
 
\usepackage{atlasphysics}
\usepackage{multicol}
\usepackage{multirow}
\usepackage{physics}
 
\usepackage{cleveref}

\graphicspath{{logos/}{figures/}}
 
\usepackage{ANA-IDTR-2019-05-PAPER-defs}
\usepackage{mwe}

\definecolor{orange}{rgb}{1,0.5,0}
\definecolor{green1}{rgb}{0.39,0.66,0.35}
\usepackage{tikz}
\usetikzlibrary{arrows.meta}
 
\AtlasTitle{Alignment of the ATLAS Inner Detector in Run-2 }
 
\AtlasAbstract{
The performance of the ATLAS Inner Detector alignment has been studied
using $pp$ collision data at $\sqrt{s} = \SI{13}{TeV}$ collected by
the ATLAS experiment during Run 2 (2015 to 2018) of the Large Hadron Collider (LHC).
The goal of the detector alignment is to determine the detector geometry as accurately as possible and correct for time-dependent movements.
The Inner Detector alignment is based on the minimization of track-hit residuals
in a sequence of hierarchical levels, from global mechanical assembly structures to local sensors.
Subsequent levels have increasing numbers of degrees of freedom; in total there are almost \num{750000}.
The alignment determines detector geometry on both short and long timescales, where short timescales
describe movements within an LHC fill.
The performance and possible track parameter biases originating from systematic detector deformations are evaluated.
Momentum biases are studied using resonances decaying to muons or to electrons. The residual sagitta bias and momentum scale bias after alignment are reduced to less than $\sim$\SI{0.1}{TeV^{-1}} and 0.9$\times10^{-3}$, respectively. Impact parameter biases are also evaluated using tracks within jets.
}
 
\author{The ATLAS Collaboration}
 
\AtlasRefCode{IDTR-2019-05}

\AtlasJournal{EPJC}
 
\PreprintIdNumber{CERN-EP-2020-108}
 
\AtlasJournalRef{Eur. Phys. J. C 80 (2020) 1194}
 
\AtlasDOI{10.1140/epjc/s10052-020-08700-6}

\hypersetup{pdftitle={ATLAS document},pdfauthor={The ATLAS Collaboration}}
 
\begin{document}
 
\maketitle
 
\tableofcontents

\section{Introduction}
\label{sec:intro}
% The next lines are included from the .//chapters/Introduction.tex input file
The precise reconstruction of the trajectories of charged particles created in proton--proton ($pp$) and heavy-ion collisions at CERN's Large Hadron Collider (LHC) is a key ingredient
in many of the physics processes studied by the ATLAS Collaboration.
Almost every measurement performed using the ATLAS detector~\cite{ATLASTechnicalPaper}, from Standard Model processes to searches for new physics phenomena,
relies on the accurate reconstruction of charged particles.

In order to reconstruct the trajectories of charged particles,
ATLAS uses the Inner Detector (ID) tracking system to provide efficient, robust and precise position measurements of charged particles as they traverse the detector.
The energy deposits from charged particles (hits) recorded in individual detector elements of the ID are
used to reconstruct their trajectories (tracks) and estimate the associated track parameters.
The precision achieved for the track parameters is determined by several factors: the intrinsic resolution of sensitive devices; the knowledge of the magnetic field;
the distribution of material in and before the ID and the knowledge of it; and the knowledge of the geometry, i.e. the location and orientation, of the detector elements.
The purpose of the detector alignment is to determine, as precisely as possible, the actual geometry of the active detector elements of the tracking system, and to follow changes in the geometry with time.
 
Poor knowledge of the actual geometry of the active detector elements results in
a deterioration of the resolution of reconstructed track parameters.
The criteria for the minimum precision required were defined in order to limit the degradation of the resolution of the track parameters for high-momentum tracks to less than 20\% in comparison to a perfectly aligned detector~\cite{ATLAS-TDR-4}.
In addition, correlated geometrical distortions can lead to systematic biases in the reconstructed track parameters.
Correlated systematic biases can be introduced either by real detector deformations to which the alignment procedure has little sensitivity
or by the procedure used to determine the alignment parameters.
These correlated biases are referred to as `weak modes' of the alignment.
 
In this document, the ATLAS ID alignment procedure and its performance during Run 2 of the LHC is presented.
A new layer of pixel sensors was included in the detector for Run 2, which posed additional challenges for the alignment of the detector compared to those faced during Run 1~\cite{ATLAS-CONF-2012-141,ATLAS-COM-CONF-2014-046}.
The greatest new challenge was the short-timescale movement of parts of the detector during data taking.
 
This paper is organised as follows: a brief description of the ATLAS detector is given in \Cref{sec:detector}. \Cref{sec:theFormalism} presents the formalism of the ATLAS track-based ID alignment.
\Cref{sec:IDalign} introduces the different alignment levels and
\Cref{sec:TimeAlign} discusses the detector stability and describes the time-dependent alignment.
The performance of the ATLAS Run 2 alignment is presented in terms of track parameter biases in \Cref{sec:MomentumBias,sec:IPbias}.
Concluding remarks are made in \Cref{sec:conclusion}.

% End of text imported from the .//chapters/Introduction.tex input file

\section{The ATLAS detector}
\label{sec:detector}
% The next lines are included from the .//chapters/Detector.tex input file
The ATLAS detector~\cite{ATLASTechnicalPaper} at the LHC is a multipurpose particle detector with a
forward--backward symmetric cylindrical geometry that covers nearly the entire solid angle around the collision point.
The global ATLAS reference frame is a right-handed Cartesian coordinate system, where the origin is at the
nominal $pp$ interaction point, corresponding to the centre of the detector. The positive $x$-axis points
to the centre of the LHC ring, the positive $y$-axis points upwards and the $z$-axis points
along the beam direction.
Polar coordinates $(r,\phi)$ are used in the transverse plane, $\phi$ being the azimuthal angle around the beam pipe.
The pseudorapidity is defined in terms of the polar angle
$\theta$ as $\eta=-\ln\tan(\theta/2)$. Angular distance is  measured in units of $\Delta R \equiv  \sqrt{(\Delta\eta)^2 + (\Delta\phi)^2} $.
 
ATLAS consists of the ID (described in \Cref{subsec:ID}),
electromagnetic and hadronic calorimeters, a muon spectrometer and a magnet system.
Lead/liquid-argon sampling calorimeters provide electromagnetic
energy measurements with high granularity and a steel/scintillator-tile hadronic
calorimeter covers the central pseudorapidity range of $\abs{\eta} < 1.7$.
The endcap and forward regions are instrumented with liquid-argon calorimeters for measurements of
both electromagnetic and hadronic showers up to $\abs{\eta} = 4.9$.
The outer part of the detector consists of a muon spectrometer with high-precision tracking chambers for
coverage up to $\abs{\eta} = 2.7$, fast detectors for triggering over $\abs{\eta} < 2.4$, and three large superconducting toroid magnets with eight coils each.
The ATLAS detector has a two-level trigger system to
select events for offline analysis~\cite{ATL-DAQ-PUB-2016-001}.

\subsection{Inner Detector structure}\label{subsec:ID}
\begin{figure}[ht!]
\centering
\includegraphics[width=0.6\textwidth]{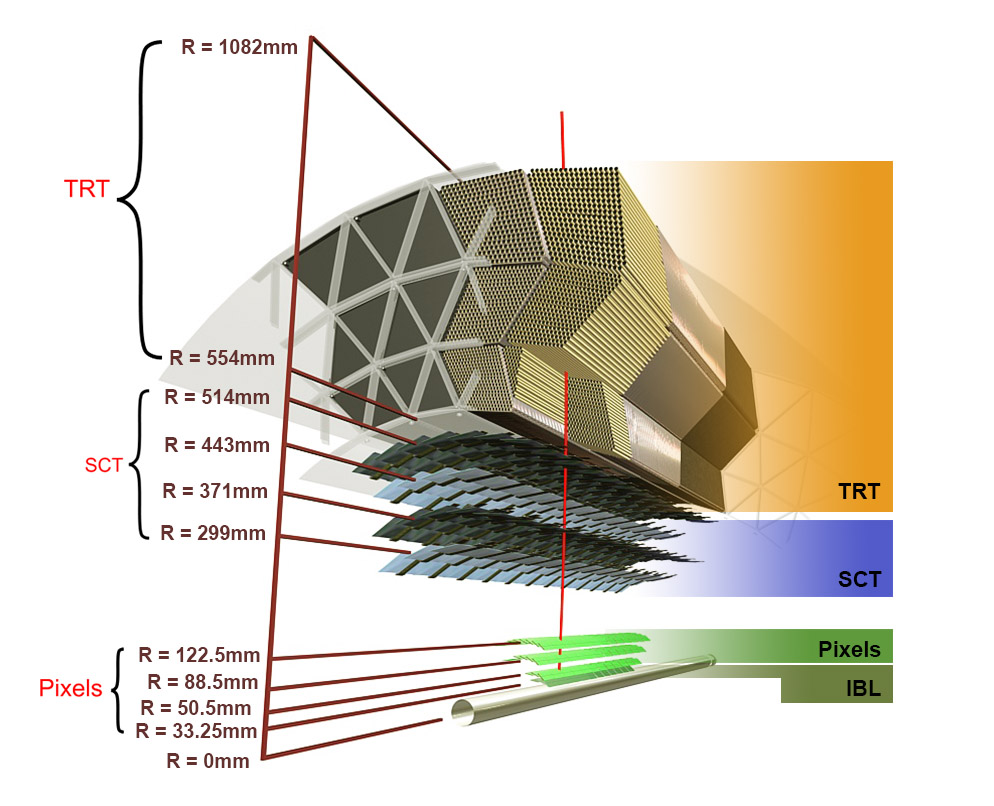}
\caption{A 3D visualisation of the structure of the barrel of the ID. The beam pipe, the IBL, the Pixel layers, the four cylindrical layers of the SCT and the three layers of TRT barrel modules consisting of 72 straw layers are shown.}
\label{fig:3Dview}
\end{figure}

The ATLAS ID~\cite{ATLAS-TDR-4,ATLAS-TDR-5} consists of three subdetectors utilising three technologies:  silicon pixel detectors,
silicon strip detectors and straw drift tubes,
all surrounded by a thin superconducting solenoid providing a \SI{2}{T} axial magnetic field~\cite{ATLAS_solenoid_paper}.
The ID is designed to reconstruct charged particles within a pseudorapidity range of $\abs{\eta} < 2.5$ (see \Cref{fig:3Dview} for a schematic view of the ID barrel region and \Cref{Tab:IDcharacteristics} for a list of the main detector characteristics).
The material distribution inside the ID has been studied in data through use of hadronic interactions and photon conversion vertices
\cite{PERF-2015-06,PERF-2015-07}.
During the second LHC data-taking run (2015--2018) with $pp$ collisions at a centre-of-mass energy $\sqrt{s}=\SI{13}{\TeV}$,
the ID collected data with an efficiency greater than \SI{99}{\percent} \cite{ATLASRun2DataQuality}. 
 
The innermost part of the inner detector consists of a high-granularity silicon pixel detector and includes the
insertable B-layer (IBL)~\cite{IBLref,PIX-2018-001}, a new tracking layer added for Run 2 which is closest to the beam line and designed to improve the precision and robustness of track reconstruction.
The IBL consists of 280 silicon pixel modules arranged on 14 azimuthal carbon fibre staves surrounding the beam pipe at a radius of \SI{33.25}{mm}.
Each stave is instrumented with 12 two-chip planar modules, covering the region of $\abs{\eta}<2.7$, and 8 single-chip modules with 3D sensors \cite{IBL-3d-sensors,IBL-3d-perf},
four at each end of the stave ($2.7<\abs{\eta}<3$).
The remainder of the Pixel detector ~\cite{ATLAS-TDR-4,ATLAS-TDR-5,pixels-det-sensors} consists of 1744  silicon pixel modules arranged in three barrel layers and two endcaps with three disks each.
Each pixel module comprises 16 front-end chips bump-bonded to the sensor substrate.
The barrel modules were assembled on staves of 13 modules each,
whilst the endcap modules were assembled directly on the disks.
In order to simplify the notation throughout the rest of the paper,
the term Pixel is used to refer only to the detector already in place during Run 1 and the new layer is referred to explicitly as the IBL.
 
The Semiconductor Tracker (SCT) \cite{sct-barrel-modules,sct-ec-modules, IDET-2013-01} consists of \num{4088} silicon strip modules.
They are arranged in four barrel layers and two endcaps with nine disks each.
Each module consists of two pairs of single-sided strip sensors glued back-to-back with a \SI{40}{mrad} angle between them.
Each module comprises 12 128-channel chips.
Due to the stringent build tolerances each SCT module is considered a solid object for the purposes of alignment.
The barrel modules are mounted directly on the cylindrical support structures of each layer in 12 rings,
whilst the endcap modules are assembled in 3 rings on the disks.
The barrel SCT sensors have a uniform pitch strip of \SI{80}{\micro\metre},
while the endcap sensor strips run radially with a \SI{161.5}{\micro\radian} angular pitch.
 
The Transition Radiation Tracker (TRT) \cite{trt-barrel-det, trt-ec-det, IDET-2015-01} is the outermost subdetector and extends track reconstruction radially outwards to a radius of 1082~mm.
It is made of \num{350848} gas-filled straw tubes of 4~mm diameter.
The tubes are arranged in 96 barrel modules in 3 layers (32 modules per layer) and 40 disks in each endcap.
The expected hit resolutions for each subdetector are summarised in \Cref{Tab:IDcharacteristics}.

\begin{table}[ht!]
\centering
\caption{Summary of the main characteristics of the ID subdetectors.
The intrinsic resolution of the IBL and the Pixel sensors are reported along $r$--$\phi$ and $z$, while for SCT and TRT only the resolution along $r$--$\phi$ is given \cite{IBLref,ATLASTechnicalPaper}. For SCT and TRT the element size refers to the spacing of the read-out strips and the diameter of the straw tube, respectively.}
\scriptsize
\begin{tabular}{lccll}
\toprule
Subdetector&Element size&Intrinsic resolution [\si{\micro\metre}]&Barrel layer radii [\si{mm}] &Disk layer |z| [\si{mm}]\\
\midrule
IBL& $\SI{50}{\micro\metre} \times \SI{250}{\micro\metre}$ &10 $\times$ 60 &33.25 &\\
Pixel& $\SI{50}{\micro\metre} \times \SI{400}{\micro\metre}$ &10 $\times$ 115&50.5,\,\,88.5,\,\,122.5 & 495, 580, 650\\
SCT& \SI{80}{\micro\metre} &17&299,\,\,\,371,\,\,\,443,\,\,\,514& from 839 to 2735 \\
TRT& \SI{4}{mm} &130&from 554 to 1082& from 848 to 2710\\
\bottomrule
\end{tabular}
\label{Tab:IDcharacteristics}
\end{table}
 
\subsection{Local coordinate system}
\begin{figure}[ht!]
\centering
\includegraphics[width=0.96\textwidth]{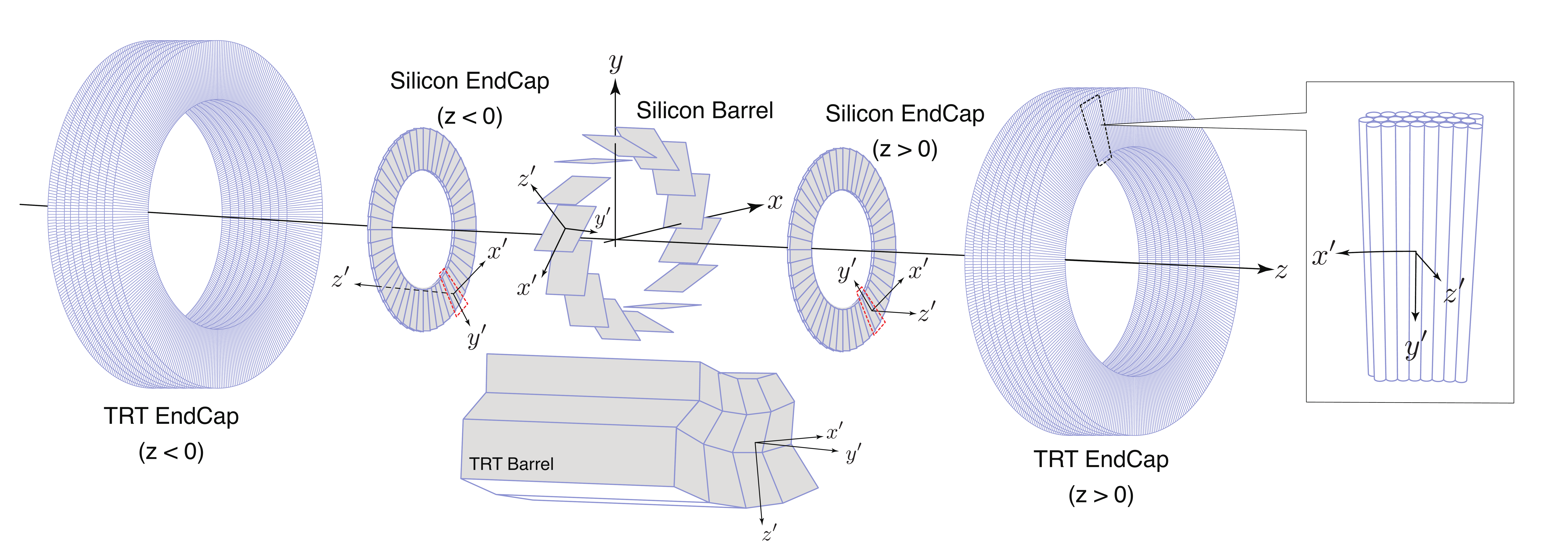}
\caption{Schematic representation of the ATLAS global reference frame $(x,y,z)$
and the local reference frame of each component of the ID.
The Pixel, IBL, and SCT modules are grouped in the `Silicon' category. For each component, the local-$x$ axis points along the most sensitive direction; the local-$z$ axis points away of the ATLAS centre; and the local-$y$ direction  is chosen according to the right-handed frame. For TRT tubes, the local reference frame is determined by the orientation of the module they are mounted on. For visualisation purposes only, the local reference frame is referred to as $(x',y',z')$ in the drawing.}
\label{fig:coordinates}
\end{figure}

The local coordinate system of an individual sensor of the detector is a
right-handed system frame with the origin placed in the geometrical centre of the sensor.
The local coordinate system for each subsystem component is illustrated in \Cref{fig:coordinates}.
The convention used is the following:
the local-$x$ axis points along the most sensitive direction of the sensor.
This corresponds to the shorter pitch side for Pixel and IBL modules,
and perpendicular to the strip-orientation for the SCT.
In the silicon detectors, the local-$y$ axis is oriented along the long side of the sensor
(i.e.\ longer pitch direction for the Pixels and IBL and the strip direction in the SCT),
while the local-$z$ direction is orthogonal to the local $x$--$y$ plane.
In the case of the TRT,
the local-$y$ axis points along the wire:
either in the same direction as the global $z$-axis (barrel) or radially outwards (endcaps).
In the barrel, the local-$z$ axis points radially outwards (from the origin of the global frame to the straw centre).
In the endcaps, the local-$z$ axis points outwards (parallel to the beam line).
The local-$x$ axis is perpendicular to both the TRT wire and the radial direction.
 
Hits are reconstructed in the local reference frame.
The TRT measures the radial distance of the primary ionisation from the wire as $\sqrt{x^2+z^2}$, taking both $x$ and $z$ in the local frame.
 
% End of text imported from the .//chapters/Detector.tex input file

\section{Alignment principles and formalism}
\label{sec:theFormalism}
% The next lines are included from the .//chapters/Formalism.tex input file
This section reviews the formalism for in situ alignment of the ATLAS ID using reconstructed tracks.
The concept of Global $\chi^2$ alignment is introduced, followed by a discussion of `weak modes' (\Cref{sec:constraints,sec:MomentumBias}) and how they can be avoided by adding constraints on track parameters.
The section closes with a detailed description of the alignment procedure and its implementation within the ATLAS software.
 
The approach used is based on the Newton--Raphson method and determines both the trajectory parameters and a set of alignment parameters, $\ali $.
In this context, $\ali $ are chosen as the six degrees of freedom (DoF) of each alignable structure that uniquely define its position and orientation in space.
These correspond to three translations $(T_x,T_y,T_z)$ and three rotations $(R_x,R_y,R_z)$. Translations are relative to the origin of the reference frame of each alignable structure and rotations are around the Cartesian axes.
 
\subsection{Track fitting with the Newton--Raphson method}
\label{sec:trackfitting}
The Newton--Raphson method uses an iterative approach to find the best fit to a set
of measurements of a track left in the detector by a charged particle traversing active detector elements.
The quality of the fit is characterised by a track $\chi^2$,
determined from the distances between the hits in the detector, which constitutes the track measurements, and the fitted track (residuals).
The trajectory of a track in a magnetic field is parameterised by a set of five parameters.  The chosen parameterisation  in ATLAS is:
$\per=(d_0,~z_0,~\phi_0,~\theta_0,~q/p)$,
where $d_0$ and $z_0$ are the transverse and longitudinal impact parameters and
$\phi_0$ and $\theta_0$ the azimuthal and polar angles of the track, all defined at the point of closest approach to the $z$-axis of the reference frame~\cite{TrackingEDM}.
The ratio $q/p$ is the inverse of the particle momentum ($p$) multiplied by its charge ($q$)
(see Ref.~\cite{ATLAS-COM-CONF-2014-046} for more details).

The track $\chi^2$ is calculated
from all measured track-hit residuals, $r_i=e_i(\per)-m_i$.
where $m_i$ is the position of the
\ith\ measurement,
and $e_i$ is the position of the intersection of the fitted track with
the surface on which the \ith\ measurement is made.
The determination of the intersection position ($e_i$) includes the measurement in question,
which causes $r_i$ to be a biased residual.
The track $\chi^2$ is defined, using vector notation, as
 
\begin{linenomath}
\begin{equation}
\chi^2 = \rer^\top \Omega^{-1} \rer,
\label{eq:basicTrackChi2}
\end{equation}
\end{linenomath}
 
where $\rer$ is the vector of track residuals and $\Omega$ is the covariance matrix
of the corresponding measurements.\footnote{The local position and uncertainty of each measurement are provided by the corresponding subsystem after applying its own clustering and hit reconstruction techniques. The Pixel detector uses an artificial neural network, trained on simulation, to determine the position of a cluster and its uncertainty~\cite{PERF-2012-05}. The SCT parameterises the position of the cluster and its uncertainty, using simulation, as a function of the number of strips in the cluster and the incident angle of the particle. For the TRT the drift radius and its uncertainty is calibrated using an iterative procedure in data and simulation~\cite{IDET-2015-01}.}
The parameters of a track's trajectory, $\per$, are those that minimise this $\chi^2$.
The minimisation is done using the first and second derivatives of the $\chi^2$
with respect to $\per$.  Defining the derivative $G=\mbox{d}\rer/\mbox{d}\per$, the
condition for the minimisation of the $\chi^2$ is
 
\begin{linenomath}
\begin{equation}
\left(\frac{\mbox{d}\chi^2}{\mbox{d}\per}\right)^\top = 2~G^\top \Omega^{-1}\rer = 0\,\,.
\label{ref:chi2fd}
\end{equation}
\end{linenomath}
 
\begin{figure}[ht!]
\centering
\setlength{\unitlength}{0.1\textwidth}
\includegraphics[width=0.30\textwidth]{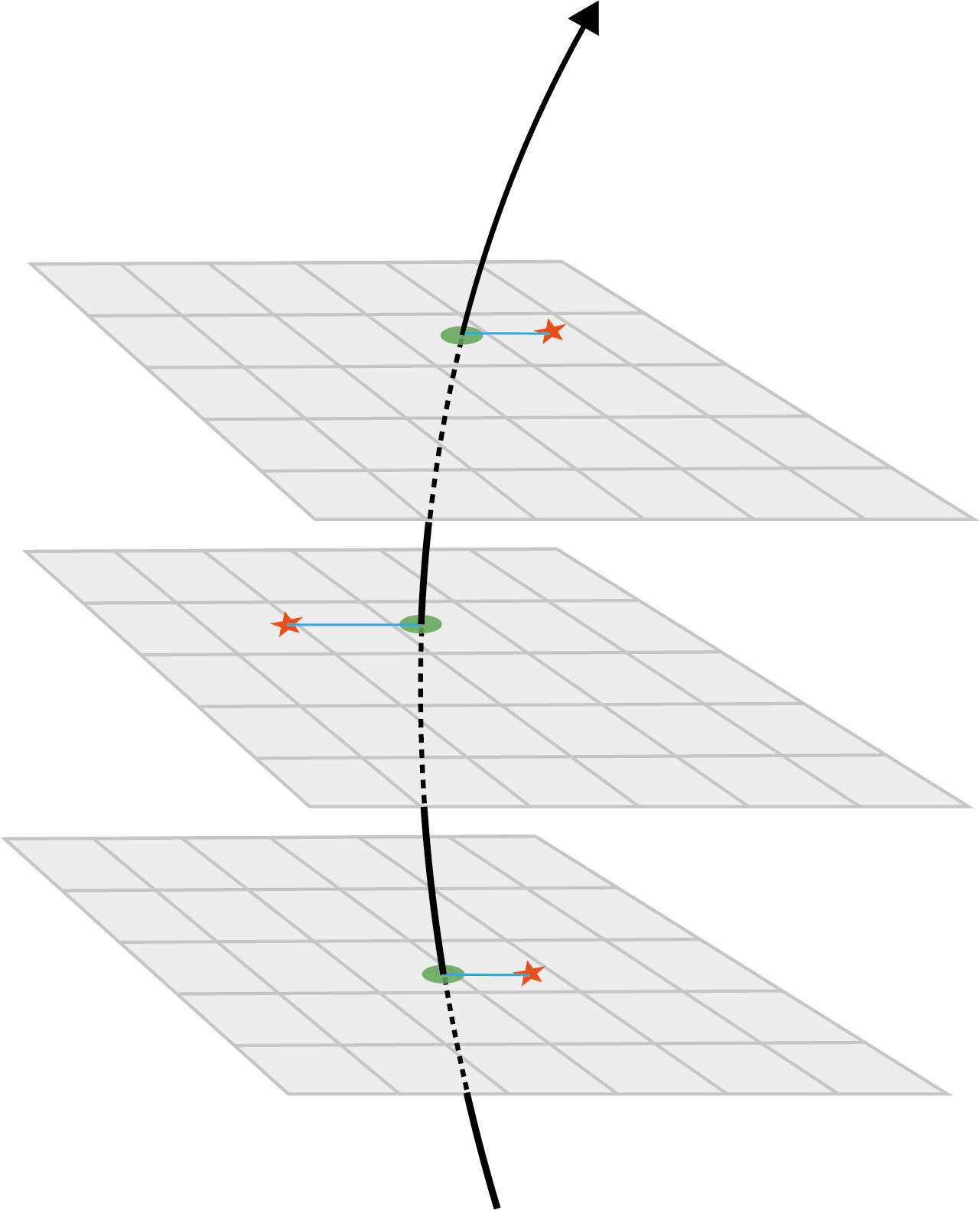}
\put(-1.37, 0.08){$\per$}
\put(-1.20, 0.67){\color{red}{$m_{i-2}$}}
\put(-1.51, 1.75){{\color{green1}{$e_{i-1}($}}$\per${\color{green1}{$)$}}}
\put(-1.45, 2.45){{\color{cyan}{$r_i$}} $=$ {\color{green1}{$e_i($}}$\per${\color{green1}{$)$}} $-$ {\color{red}{$m_i$}}}
\caption{Schematic representation of a charged particle crossing detector planes. The measurement, $m_i$, on each the \ith\ layer is indicated by a red star. Also shown are the fitted track trajectory for a given set of track parameters, $\per$ (black line), the position of the intersection of the fitted track with the surface on which the \ith\ measurement is made, $e_i(\per)$ (green ellipse), and the residuals, $r_i$ (blue line).}
\label{fig:cartoon_track_3planes}
\end{figure}
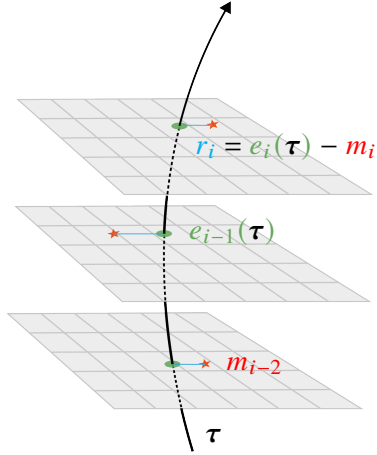
 
In practical terms, the values of $\per$ satisfying \cref{ref:chi2fd} are found using an iterative procedure by evaluating the first and second derivatives of $\chi^2$ with respect
to the track parameters of the current iteration, $\per_0$.
If the derivative $G$ were constant, then the problem would be linear and the solution would be exact.
In general, the derivative $G$ depends on the track parameters themselves.
Therefore, the procedure is repeated until a convergence criterion is met.
 
The track fit is further improved by taking into account the impact of material interactions on the trajectory of the particle. Energy loss is treated as point-like at the center of material layers and for hadrons and muons, which are used during the alignment, deterministic as the variance of the energy loss processes is small. Additional parameters, $\tet$, are added to account for the effects of multiple Coulomb scattering (MCS) of the particle with the detector components,
as detailed in \Cref{sec:trackfittingMCS}.
Consequently, the residuals now also depend on $\tet$ and the variance of the scattering angles, $\Theta$:
\begin{linenomath}
\begin{equation}
\XTrk = \rer^\top \Omega^{-1} \rer + \tet^\top \Theta^{-1} \tet.
\label{eq:TrackChi2Scatter}
\end{equation}
\end{linenomath}
Thus, \XTrk ~has to be minimised for $\per$ and $\tet$ simultaneously.
The derivatives of residuals with respect to track and scattering parameters are defined as $G \equiv \partial \rer / \partial \per$ and $S \equiv \partial \rer / \partial \tet$, respectively.
In the following, the \GX\ method for alignment including MCS is described and the following simplified notation is adopted:
 
\begin{linenomath}
\begin{equation}
\res \equiv \left(\begin{array}{c} \rer \\ \tet \end{array} \right), \hspace{5 pt}
V \equiv \left(
\begin{array}{cc}
\Omega & 0 \\
0  &  \Theta
\end{array}
\right),   \hspace{5 pt}
\trk \equiv \left(\begin{array}{c} \per \\ \tet \end{array} \right),  \hspace{5 pt} \mathrm{and} \hspace{5 pt}
H \equiv \left(
\begin{array}{cc}
G & S \\
\frac{\partial \tet}{\partial \per}=0 & \frac{\partial \tet}{\partial \tet}=\textbf{I}
\end{array}
\right).
\label{ref:auxdef}
\end{equation}
\end{linenomath}
 
It should be noted that refinements are made to the track fit by performing multiple iterations of the fit during which: material effects are recalculated; measurements are recalibrated based on the particles incident angle; and outliers are removed.

\subsection{The Global $\chi^2$ method for alignment}
\label{sec:globalchi2}
 
\subsubsection{General definition of the Global $\chi^2$ method for alignment}
The Global $\chi^2$ is a track-based alignment method which uses a $\chi^2$ built from a large sample of reconstructed tracks and their associated hits in the detector elements being aligned.
The alignment parameters are determined by minimising the Global $\chi^2$ with respect to the alignment parameters:
\begin{linenomath}
\begin{equation}
\XGlo = \sum_i \XTrkI~,
\label{eq:global_chi2}
\end{equation}
\end{linenomath}
where \XTrkI ~is the $\chi^2$ of the \ith\ track as given by \cref{eq:TrackChi2Scatter}.
The residuals used in \cref{eq:global_chi2} depend on the alignment parameters ($\ali$) as both, the measurements and the track extrapolations depend on $\ali$, the former directly and the latter through the fitted track parameters.
Therefore, the minimisation of \XGlo ~with respect to $\ali$ uses the total derivative operator with respect to $\ali$, which can be expressed as:
\begin{linenomath}
\begin{equation}
\frac{\full }{\full \alpha_i } = \frac{\partial }{\partial \alpha_i } +  \sum_{j}\frac{\full \pi_j }{\full \alpha_i }\frac{\partial }{\partial \pi_j }~.
\label{eq:full_deriv}
\end{equation}
\end{linenomath}
The  $\full \trk / \full \ali$ term is determined from the condition that,
once \XGlo ~is at a minimum, \XTrk ~is also at a minimum with respect to the track parameters:
\begin{linenomath}
\begin{equation}
\frac{\full }{\full \ali } \frac{\partial \XTrk}{\partial \trk } = 0~.
\label{eq:trackatminimum}
\end{equation}
\end{linenomath}
 
Using \cref{eq:trackatminimum} in \cref{eq:full_deriv}, this results in:
\begin{linenomath}
\begin{equation*}
\frac{\full \trk}{\full \ali } = - \left(\frac{\partial^2 \XTrk}{\partial \trk ^2}\right)^{-1} \frac{\partial^2   \XTrk}{\partial \ali  \partial \trk }~,
\end{equation*}
\end{linenomath}
which allows the nested dependence of the \trk ~on \ali ~to be resolved,
thereby removing the need to determine both (the track parameters and alignment parameters) simultaneously.

Ignoring second-order derivatives in the residuals, using the covariance matrix of the track parameters, $C$, expressed as
\begin{linenomath}
\begin{equation}
C = 2 \left( \frac{\partial ^2\XTrk}{\partial \trk ^2}\right)^{-1} = \left( H^\top V^{-1} H \right)^{-1} ,
\label{eq:trackpars_covmatrix}
\end{equation}
\end{linenomath}
and defining $A$ as the derivative of the residuals with respect to the alignment parameters:
\begin{equation}
A \equiv \frac{\partial \res}{\partial \ali }~,
\label{eq:drhodalpha}
\end{equation}
the total derivative operator with respect to $\ali$ can be written as:
\begin{linenomath}
\begin{equation*}
\left(\frac{\full }{\full \ali }\right)^\top = \left(\frac{\partial }{\partial \ali } \right)^\top - A^\top V^{-1} H C \left( \frac{\partial }{\partial \trk } \right)^{\top}~.
\end{equation*}
\end{linenomath}
The first- and second-order derivatives of $\XGlo$ with respect to $\ali $ are thus:
\begin{linenomath}
\begin{equation}
\boldsymbol{Y}  \equiv  \left( \frac{\full \XGlo}{\full \ali } \right)^\top = 2 \sum_{\mathrm{tracks}} A^\top V^{-1} (V - HCH^\top ) V^{-1} \res~,
\label{eq:alignfd1st}
\end{equation}
\begin{equation}
\mathcal{M}  \equiv \frac{\full ^2\XGlo}{\full \ali ^2} = 2 \sum_{\mathrm{tracks}} A^\top V^{-1} (V - HCH^\top ) V^{-1} A~.
\label{eq:alignfd2nd}
\end{equation}
\end{linenomath}
Here, the term $HCH^\top $ represents the covariance of the track parameters in the measurement space, whereas the covariance of the residuals of the track fit is given by
\begin{linenomath}
\begin{equation*}
R = V - HCH^\top ~.
\end{equation*}
\end{linenomath}

\subsubsection{Newton--Raphson method for Global $\chi^2$ alignment}
\label{sec:N-Rglobalchi2}
In analogy to the general method for track fitting (\Cref{sec:trackfitting}) an iterative approach is used to solve for the alignment parameters.
The first- and second-order derivatives are obtained using \cref{eq:alignfd1st,eq:alignfd2nd} and evaluated for an initial set of alignment parameters, $\ali _0$.
Such an initial geometry description is available from design drawings, survey measurements,
or previous alignment results.
The alignment corrections, to the initial geometry, are given by
\begin{linenomath}
\begin{equation}
\boldsymbol{X} \equiv \Delta \ali = - \left( \left.\frac{\full ^2\XGlo}{\full \ali ^2} \right|_{\ali _0}\right)^{-1} \left.\left(\frac{\full \XGlo}{\full \ali }\right)^{\top}\right|_{\ali _0} \equiv - \mathcal{M}^{-1} \boldsymbol{Y}.
\label{eq:align_solution}
\end{equation}
\end{linenomath}
The above step is repeated for successive iterations until a convergence criterion is met and $\Delta \ali $ is negligible\footnote{This typically refers to a correction threshold of  $<\SI{0.1}{\micro\metre}$. However, it should be noted that the convergence criteria can vary between specific alignment level and degrees of freedom chosen.}.
This requires re-fitting the tracks using the updated geometry (initial alignment constants $\ali _0$ plus their corrections $\Delta \ali$),
to obtain new residuals and new derivatives,
and solving again to compute the next set of corrections to the alignment constants.

\subsubsection{Locality ansatz}
\label{sec:localityansatz}
If the initial track parameters, $\trk _0$, minimise $\XGlo$ for a given $\ali _0$,
\cref{eq:alignfd1st} simplifies to
\begin{linenomath}
\begin{equation}
\left.\left( \frac{\full \XGlo}{\full \ali }\right)^{\top}\right|_{\trk _o, \ali _0}
= 2 \sum_{\mathrm{tracks}} A^\top V^{-1} \res \, ,
\label{eq:locality_V}
\end{equation}
\end{linenomath}
as the term $H^\top V^{-1}\res $ is zero.
Consequently, if the measurements are independent and $V$ is diagonal, the derivative with respect to a particular parameter $\ali_i$ only receives contributions from residuals for which the related entries in the derivative matrix $A$ are non-zero.
In other words,
if $\ali_i$ is an alignment parameter of a given detector module,
only the measurements in this module contribute to the first-order derivative of $\XGlo$ with respect to $\ali_i$.
Therefore, contributions to the $\XGlo$ from measurements in other subdetectors and MCS effects can be ignored.
This useful property is labelled as the so-called \emph{locality ansatz}~\cite{TRTAlignment} and provides an important simplification for the software implementation.
 
\subsubsection{Adding constraints on track parameters}
\label{sec:constraints}
 
It is of particular importance to assure that the determination of the track parameters is free from systematic biases that can occur due to poorly determined `weak modes' of the alignment.
These modes are geometry distortions that leave the $\chi^2$ of the fitted tracks nearly unchanged and
typically lead to an incorrect solution of the alignment.
They can be controlled by imposing constraints on track parameters~\cite{Siena_proc}.
Examples of such constraints, discussed in detail in \Cref{sec:MomentumBias,sec:IPbias}, are the beam-spot constraint, track parameter constraints from external detector systems (e.g.\ calorimeters), and constraints determined using  reconstructed physics events (e.g.\ mass constraints from narrow resonances).
These constraints are included in the Global $\chi^2$ method by adding extra terms to the expression for
the $\chi^2$ in \cref{eq:global_chi2}. For one track the modified contribution to $\chi^2$ is
\begin{linenomath}
\begin{equation}
\chi^2_{\mathrm{cons}} = \res^\top V^{-1}\res + (\trk  - \ctr )^\top T^{-1} (\trk  - \ctr )~,
\label{eq:constrained_chi2}
\end{equation}
\end{linenomath}
where \ctr\ is a vector defining the constraint on $\trk $ and $T$ is its covariance matrix.
 
In the ATLAS implementation, this constraint is implemented by adding a pseudo-measurement on a track~\cite{TrackingEDM}.
The solution for the alignment parameters is given by \cref{eq:align_solution}, where for each constrained
track the covariance matrix is now defined as
\begin{linenomath}
\begin{equation*}
C = 2 \left(   \frac{\full ^2\chi^2}{\full \trk ^2}_{\mathrm{cons}} \right)^{-1} =  \left( H^\top V^{-1} H + T^{-1} \right)^{-1} \, .
\end{equation*}
\end{linenomath}
In this context, the first-order derivative of the Global $\chi^2$ is given by
\begin{linenomath}
\begin{equation}
\left. \left( \frac{\full \chi^2}{\full \ali } \right)^{\top} \right|_{\ali _0}
= 2 \sum_{\mathrm{tracks}}  A^\top V^{-1} (V - HCH^\top ) V^{-1} \res (\ali _0)
- A^\top V^{-1} H C T^{-1}(\trk (\ali _0) - \ctr)~.
\label{eq:constrained_V}
\end{equation}
\end{linenomath}
 
If the tracks have been re-fitted with the imposed constraint, the locality ansatz drastically
simplifies \cref{eq:constrained_V}, reducing it to \cref{eq:locality_V}. This property is used in the ATLAS implementation.

\subsubsection{Constraints on alignment parameters}
\label{sec:alignconst}
 
Often one has some prior knowledge of the geometry from either survey measurements or mechanical constraints.
These constraints can be included by adding terms to the $\chi^2$ in \cref{eq:global_chi2}.
In the general case, one can write
\begin{linenomath}
\begin{equation}
\chi^2_{\mathrm{cons}} =  \sum_{\mathrm{tracks}} \res^\top V^{-1}\res + (\ali  - \ali_{\mathrm{cons}} )^\top W^{-1} (\ali  - \ali_{\mathrm{cons}} )~,
\label{eq:alignconst_chi2}
\end{equation}
\end{linenomath}
where $\ali_{\mathrm{cons}}$ is a vector defining the constraint on $\ali $ and $W$ is its covariance matrix.
The added term leads to extended expressions for the first and second derivatives of $\chi^2$ with respect
to $\ali$ (\cref{eq:alignfd1st,eq:alignfd2nd}):
\begin{linenomath}
\begin{equation}
\boldsymbol{Y}  \longrightarrow \boldsymbol{Y} + 2~W^{-1} (\ali  - \cali )~,
\label{ref:alignfd1stalicons}
\end{equation}
\begin{equation*}
\mathcal{M}  \longrightarrow \mathcal{M} + 2~W^{-1},\label{ref:alignfd2ndalicons}
\end{equation*}
\end{linenomath}
while the solution is computed using $\boldsymbol{Y}$ and $\mathcal{M}$ in \cref{eq:align_solution}.
 
The special case when $\cali \equiv \ali_0$ and $W$ is diagonal, i.e.\ when the alignment parameters are
constrained to their initial values, is further discussed in \Cref{sec:SMC}.

\subsubsection{The \LX\ method}
\label{sec:localchi2}
The main advantage of the \GX\ method arises from its rigorous treatment of correlations between
alignable objects through the tracks connecting them.
However, this approach becomes technically challenging when the number
of alignment parameters is very large,
such as in the case of the alignment of individual TRT straws
($\approx \num{700000}$ parameters).
In order to overcome this challenge, a simplified version of the $\chi^2$ approach (the \LX method) is used.
It is based on the minimisation of the same $\chi^2$, \cref{eq:global_chi2}, but the implicit
dependence on the fitted track parameters is dropped, reducing \cref{eq:full_deriv} to a simpler form:
\begin{linenomath}
\begin{equation*}
\frac{\full }{\full \ali } = \frac{\partial}{\partial \ali }~.
\end{equation*}
\end{linenomath}
Consequently, \cref{eq:alignfd1st,eq:alignfd2nd} are reduced to:
\begin{linenomath}
\begin{equation*}
\begin{split}
\left( \frac{\full \XLoc}{\full \ali }  \right)^\top
& = 2 \sum_{\mathrm{tracks}} A^\top \Omega^{-1} \rer \,\\
\frac{\full ^2 \XLoc}{\full \ali ^2}
& = 2 \sum_{\mathrm{tracks}} A^\top \Omega^{-1} A~.
\end{split}
\end{equation*}
\end{linenomath}
In addition, the problem is reduced to separate systems of equations describing individual alignable modules.
The \LX\ method eliminates the numerical challenges of the \GX
since only systems of equations with up to six parameters (albeit many of them) need to be solved.
However, due to the loss of the correlations between alignable objects,
the \LX method needs a much larger number of iterations to converge.

\subsection{Solving the linear system of alignment equations}
 
In general, the properties of the matrix representing a system of linear equations determine the most suitable solution technique.
The matrix $\mathcal{M}$ in \cref{eq:alignfd2nd} as defined in the~\GX ansatz is found to be symmetric and singular and to have a poor matrix condition number if no constraints are applied. The addition of appropriate constraints generally renders the matrix positive definite.
The singular nature of the matrix is the result of detector movements that leave a track's~$\chi^2$ unchanged.
The simplest examples are global transformations of the detector (either translations or rotations),
which are generally singular
modes.\footnote{Rotations within a magnetic field or translations in an inhomogeneous magnetic field may not be
singular modes but for practical purposes may essentially be so. They are typically extremely poorly constrained because track trajectories are not significantly modified by small changes in the magnetic field.}
A trivial way to remove these global degrees of freedom is to fix a detector element, making it the reference point for all other detector elements.
This method has the unwanted drawback of arbitrarily selecting one detector module as the reference frame.
In the following section, two methods used to obtain a solution to the alignment system of linear equations are discussed along with how  `weak modes' are removed or mitigated.
 
\subsubsection{Diagonalisation of the matrix}
The symmetric matrix $\mathcal{M}$ is decomposed into its diagonal basis:
$P\mathcal{D}P^{\top}$ where $\mathcal{D}$ is a diagonal matrix containing the eigenvalues of  $\mathcal{M}$,
and $P$ is a matrix containing the eigenvectors of  $\mathcal{M}$.
Of course, in the diagonal basis all parameters (directions) are linearly independent,
and the solutions plus their associated uncertainties are given by the eigenvalues ($\lambda_i$) with:
\begin{linenomath}
\begin{equation}
X_{\mathcal{D}}^i=-\frac{1}{\lambda_i}Y_{\mathcal{D}}^i~~~~~{\mathrm{and}}~~~~~\sigma(X_{\mathcal{D}}^i)=\frac{1}{\sqrt{\lambda_i}}\,,
\label{eq:5}
\end{equation}
\end{linenomath}
where $X_{\mathcal{D}}^i$ and $Y_{\mathcal{D}}^i$ are the \ith\ component of vectors $\pmb{X}_{\mathcal{D}}$ and
$\pmb{Y}_{\mathcal{D}}$ in the diagonal basis, with $\pmb{Y}_{\mathcal{D}} = P^{\top}~\pmb{Y}$.
 
Singular and weak modes must be excluded as their eigenvalues are zero or have an arbitrarily large associated uncertainty, respectively.
Although this can be achieved in many ways, the primary method employed is to set ${\mathcal{D}}^{-1}_{i,i} = 0$ for the modes that need to be removed, thereby creating a new diagonal matrix ${\mathcal{D'}}^{-1}$ which provides the solution:
 
\begin{linenomath}
\begin{equation*}
\pmb{X} =  - P\mathcal{D'}^{-1}P^{\top}\pmb{Y} .\\
\end{equation*}
\end{linenomath}

The DSPEV function in the LAPACK~\cite{LAPACK} software package is used as a baseline in the  ATLAS implementation to diagonalise large matrices.
Alternative implementations using ROOT~\cite{root}, EIGEN~\cite{eigenweb}  and CLHEP~\cite{CLHEP} linear algebra classes are also available.
In general, the computation time for matrix diagonalisation scales as $\mathcal{O}(\mathrm{DoF}^3)$ and
solutions for very large systems become untenable on a single machine.
If the initial matrix is poorly conditioned,
the accuracy of the numerical solution can be limited by the precision of 64-bit
floating-point computations for problems exceeding $\mathcal{O}$(\num{10000}) DoFs.
 
\subsubsection{Direct solving}
Even for very large problems, direct solvers offer an accurate and CPU-efficient method for solving sparse linear equations. In addition, less memory is required as no matrix is inverted or diagonalised in the process.
The LDLT Cholesky factorisation method provided within EIGEN~\cite{eigenweb} is used within the ATLAS ID alignment and
takes less than 10 minutes to solve an alignment problem with \num{35000} parameters (the approximate number of parameters needed to align all modules in the ID simultaneously) on a modern CPU.
Direct solving is used when aligning thousands of degrees of freedom (usually when aligning at individual module level).
Obtaining a direct solution does not offer the possibility of eliminating specific eigenmodes. Thus, other preconditioning techniques are used in order to extract a meaningful solution (e.g.\ \Cref{sec:SMC}).
It is noteworthy that, although not extensively utilised within ATLAS, it is possible to iteratively find the eigenvalues and associated eigenvectors of large systems by solving $\mathcal{M}\mathbf{x}=\lambda \mathbf{x}$ for $\mathbf{x}$ and $\lambda$ \cite{FEAST1}, which can be useful in understanding the weak modes of very large systems
and identifying the underconstrained degrees of freedom.
 
\subsubsection{Constraining alignment parameters in the solution (the Soft Mode Cut) \label{sec:SMC}}
 
As introduced in \Cref{sec:alignconst}, setting $\cali \equiv \ali_0$ and having a diagonal $W$
constrains the alignment parameters to their initial values.
Here, $W$ denotes a diagonal matrix with diagonal elements: $\sigma(\alpha_i)^2$, providing the tolerances to the corrections of the alignment parameters.
 
For this special case,
the top row of \cref{ref:alignfd1stalicons} simplifies to \cref{eq:alignfd1st} and the diagonal of the matrix $\mathcal{M}$ in \cref{eq:alignfd2nd} is incremented by the reciprocal of assumed variances of alignment corrections:
\begin{linenomath}
\begin{equation*}
\left(\mathcal{M} + 2~W^{-1} \right) \pmb{X} = -\pmb{Y} ~.
\end{equation*}
\end{linenomath}
 
The above equation can be rearranged as
\begin{linenomath}
\begin{equation}
\sum_{j} \left(\sigma(\alpha_i)\sigma(\alpha_j)~\mathcal{M}_{ij} + 2~I_{ij}\right) ~ \frac{X_j}{\sigma(\alpha_j)} = -\sigma(\alpha_i)~Y_i \\
\label{eq:solve:12}
\end{equation}
\end{linenomath}
yielding an equation in which the corrections to the alignment parameters are normalised to their assumed uncertainties  $\Delta \alpha_i \longrightarrow  \Delta  \alpha_i/{\sigma(\alpha_i)}$.
Apart from the extra identity matrix $I$, \cref{eq:solve:12} is exactly equivalent to \cref{eq:align_solution}.

To illustrate the effect of such a constraint,
consider the case that all $\sigma(\alpha_i)$ are equal ($\sigma(\alpha_i)=\sigma_c$).
The extra identity matrix does not affect the eigenmodes of $\mathcal{M}$,
but adds an offset to its spectrum of eigenvalues:
\begin{linenomath}
\begin{equation*}
\mathcal{M'}=\mathcal{M}+2~I/\sigma_c^2, ~~~~ \mathcal{D'}=\mathcal{D}+ I /\sigma_c^2, ~~~~ \lambda'_i=\lambda_i + 1/\sigma_c^2.
\end{equation*}
\end{linenomath}
The solution in the diagonal basis, \cref{eq:5}, takes the form:
\begin{linenomath}
\begin{equation}
X_{\mathcal{D}}^i=\frac{1}{\lambda_i+1/\sigma_c^2}~Y_{\mathcal{D}}^i~~~~~{\mathrm{and}}~~~~~\sigma(X_{\mathcal{D}}^i)=\frac{1}{\sqrt{\lambda_i+1/    \sigma_c^2}}\,.
\label{solve:14}
\end{equation}
\end{linenomath}
Hence, one obtains a solution explicitly free from ill-defined (weak) modes. This operation does not require an explicit diagonalisation and can be used as preconditioning prior to
fast solving, providing powerful control over solutions for an arbitrarily large number of DoFs. Due to the
typically exponential nature of the eigenspectrum, \cref{solve:14} represents a solution with a clear cut-off in the diagonal basis for $\lambda_i \ll 1/\sigma_c^2$.
This technique is extensively used in the ATLAS implementation.

% End of text imported from the .//chapters/Formalism.tex input file

\section{Inner Detector alignment}
\label{sec:IDalign}
% The next lines are included from the .//chapters/IDAlignment.tex input file
The ID is composed of a large number of active detector components (see \Cref{subsec:ID} for details).
Each component or grouped collection of modules (e.g.\ a subdetector) can be treated as an alignable structure.
The alignment is performed at different hierarchical levels following the assembly structure of the ID.
Starting with the largest physical structures at \emph{level} 1, the detector subsystems are aligned separated into endcaps and barrel regions in order to correct for collective movements.
\emph{Level}  2 treats individual barrel layers and endcap disks as physical structures (barrel modules and endcap wheels in the case of the TRT).
\emph{Level} 3 corresponds to a silicon module or TRT wire alignment. In this context, the SCT modules are considered as a single element in the alignment procedure due to their high construction precision~\cite{sct-barrel-modules,sct-ec-modules}.
The levels are addressed sequentially during the alignment procedure, see \Cref{Tab:AlignLevels}.
 
In Run 2, the alignment levels were updated to accommodate the IBL. These changes are straightforward for levels 2 and 3, as the IBL represents merely an additional Pixel layer or additional silicon modules, respectively.
The support structure of the IBL is mechanically independent from the previously installed Pixel subdetector, so IBL movement is not expected to be correlated with collective Pixel movements.  Consequently, the IBL is treated as a separate physical structure at level 1.
 
\subsection{Time-dependent alignment}
\label{TimeDependentGeneral}
 
Time-dependent alignment is performed for each LHC fill prior to data reconstruction to determine if the detector, or individual subsystems, have moved significantly compared to a reference alignment.
Such detector movements occur on different timescales, which are classified as short, medium, or long.
 
Short timescales describe movements during a single LHC fill that are a result of variations of the thermal load of the ID.
These movements are caused by fluctuations in the power consumption of the front-end electronics, due to variations in the trigger rate, that additionally affect the temperature of the cooling system.
On medium timescales, in the range of days to a month, changes to the environmental conditions of the detector, such as 
ramping the magnetic field or cycling the power or cooling systems,
often lead to significant movements of the detector. Slow gradual movements of the subsystems over several months (long timescales) were also observed and attributed to mechanical relaxations after sudden changes.
 
An automated time-dependent level 1 alignment is performed within the ATLAS prompt calibration
loop~\cite{ATLASRun2DataQuality} to address all known time-dependent movements, as detailed in \Cref{sec:TimeAlign}.
These results are monitored and new alignment corrections are automatically obtained during the calibration period.
They serve as input for the bulk reconstruction of the corresponding dataset.

\subsection{Baseline alignment constants}
\label{BaselineConstants}
The baseline alignment constants are a set of reference constants that serve as initial estimates for the
time-dependent refinements of the alignment.
In order to achieve an accurate detector alignment and a minimisation of track parameter biases over a data-taking period,
a large quantity of data are used (typically $\sim \SI{2}{fb^{-1}}$).
The levels of alignment performed are summarised in \Cref{Tab:AlignLevels}.
The alignment using the global $\chi^{2}$ method typically converges within two to four iterations for levels 1 and 2,
while at least four iterations are required at level 3 (silicon).
The TRT level 3 (straw level) uses the local $\chi^{2}$ method and requires up to 30 iterations to converge,
owing to the large number of DoFs.
 
Depending on the alignment level, some DoFs may be fixed during the alignment procedure if poor sensitivity is expected. Alignment levels targeting the silicon subdetectors use all tracks, whereas alignment levels including the TRT require tracks based on silicon and TRT hits.
In order to remove weak modes from the alignment solution, appropriate constraints are added to the global
$\chi^{2}$ method (see \Cref{sec:constraints}).
Different constraints are considered depending on the
expected misalignment and DoF for each alignable structure, listed in \Cref{Tab:AlignLevels}.
Additionally, each subsystem can be aligned at any required level independently from the others.
Further subdivision of alignment levels 
into smaller physical detector components, e.g.\ the division of individual barrel layers into staves, is also supported and used.
At level 1, the SCT barrel is kept fixed due to its good stability and to serve as reference for the rest of the structures.
 
\begin{table}[ht!]
\begin{center}
\caption{Typical alignment configurations used throughout Run 2 data taking to derive baseline alignment
constants. Translational degrees of freedom (DoF) are denoted by a $T$, rotational ones by an $R$. As shown in \Cref{fig:coordinates}, TRT barrel straws run parallel to the beam line. That corresponds to  $T_z$ at \emph{level} 1 and $T_y$ at  \emph{level} 2.}
\small
\begin{tabular}{llcll}
\toprule
Level & Description & Structures & DoF & Additional constraints\\
\midrule
1	&   IBL   							& ~~1   &  All 		 		& \\
&   Pixel detector   					& ~~1   &  All 				&  \\
&   SCT endcaps (SCT barrel fixed)		& ~~2   &  All except $T_z$ 		& \\
&   TRT split into barrel and 2 endcaps	& ~~3   &  All except $T_z$ 		& \\
\midrule
Si 2	&   Pixel and IBL barrel split into layers   	& ~~4   &  All 				& Beam spot,\\
&   Pixel endcaps split into disks   		& ~~6   &  All 				& momentum bias, and \\
&   SCT barrel split into layers			& ~~4   &  All 				& impact parameter bias \\
&   SCT endcaps split into disks		& 18 &  All 				&  \\
\midrule
Si 3	&   Pixel and IBL barrel modules		& 1736~~~~   &  All 				& Beam spot,\\
&   Pixel endcaps modules   		 	& 288~~   &  $T_x,T_y,R_z$ 		& momentum bias, \\
&   SCT barrel modules				& 2112~~~~   &  All 				& impact parameter bias, and \\
&   SCT endcaps modules				& 1976~~~~ &  $T_x,T_y,R_z$		& module placement accuracy \\
\midrule
TRT 2&   TRT barrel split into barrel modules   	& 96   &  All except $T_y$ 	& Momentum bias and\\
&   TRT endcaps split into wheels   		& 80   &  $T_x,T_y,R_z$  		& impact parameter bias \\
&  Pixel and SCT detectors fixed  		&	 &					& \\
\midrule
TRT 3&  TRT straws						& 351k & $T_x,R_z$			& \\
&  Pixel and SCT detectors fixed  		&	 &					& \\
\bottomrule
\end{tabular}
\label{Tab:AlignLevels}
\end{center}
\end{table}

\subsection{Residuals}
\begin{figure}[ht!]
\centering
\includegraphics[width=0.49\textwidth]{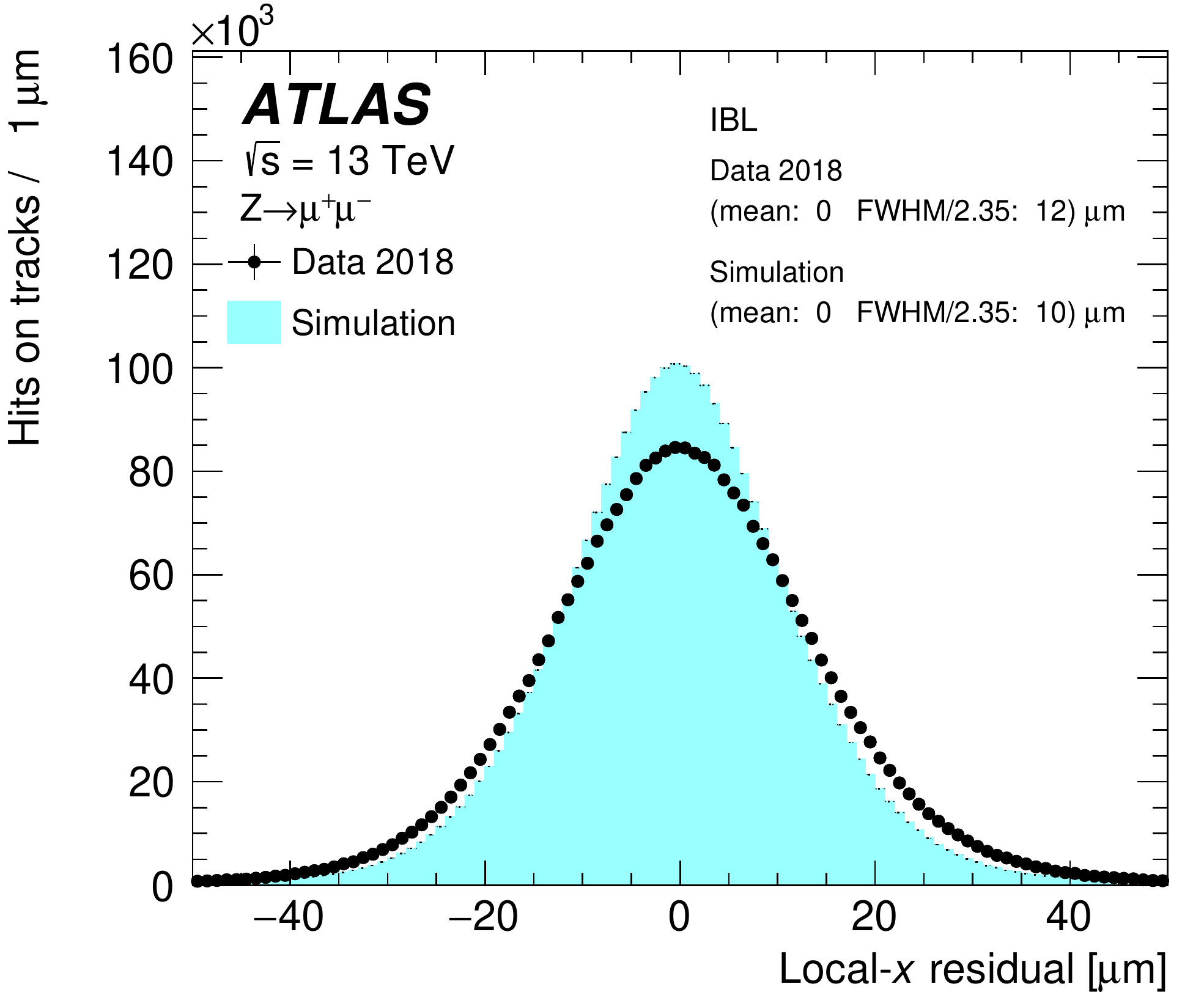}\hfill
\includegraphics[width=0.49\textwidth]{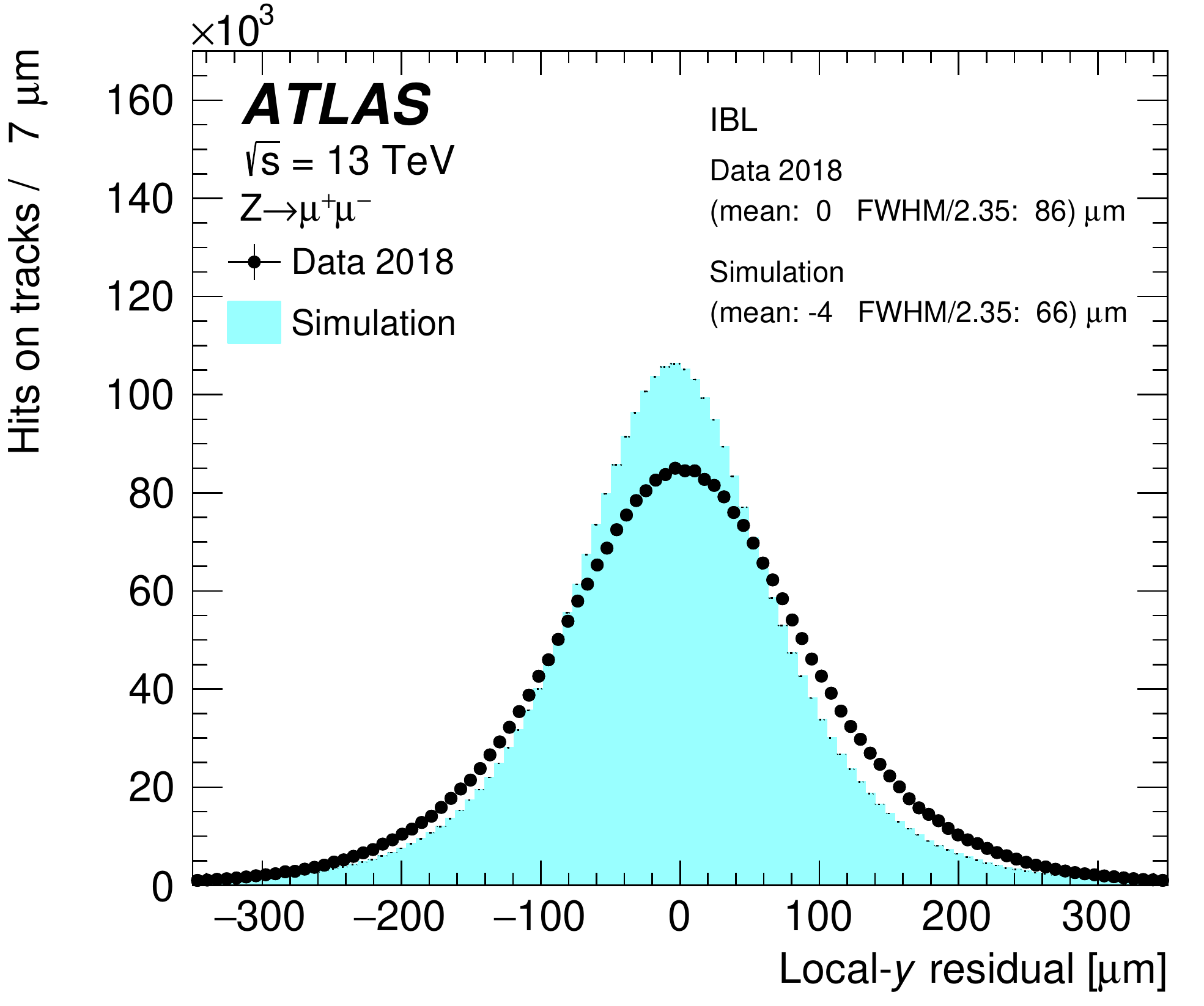}\hfill
\caption{The IBL local-$x$ (left) and local-$y$ (right) residual distributions for the $Z\rightarrow \mu^+\mu^-$ data sample compared with simulated data. The distributions are integrated over all hits on tracks in barrel modules.}
\label{fig:Residuals_local_xy_IBL}
\end{figure}

\begin{figure}[ht!]
\centering
\includegraphics[width=0.49\textwidth]{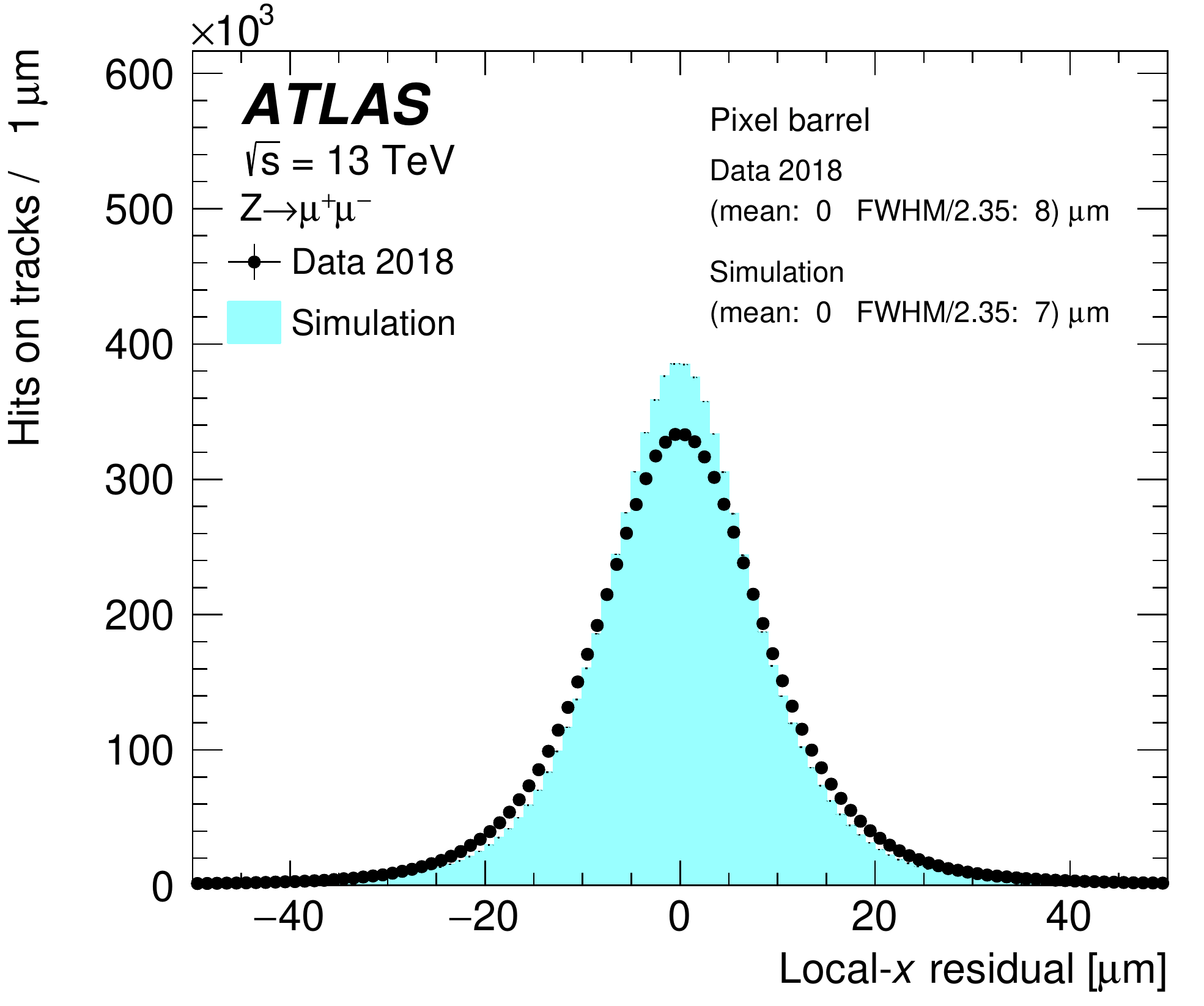}\hfill
\includegraphics[width=0.49\textwidth]{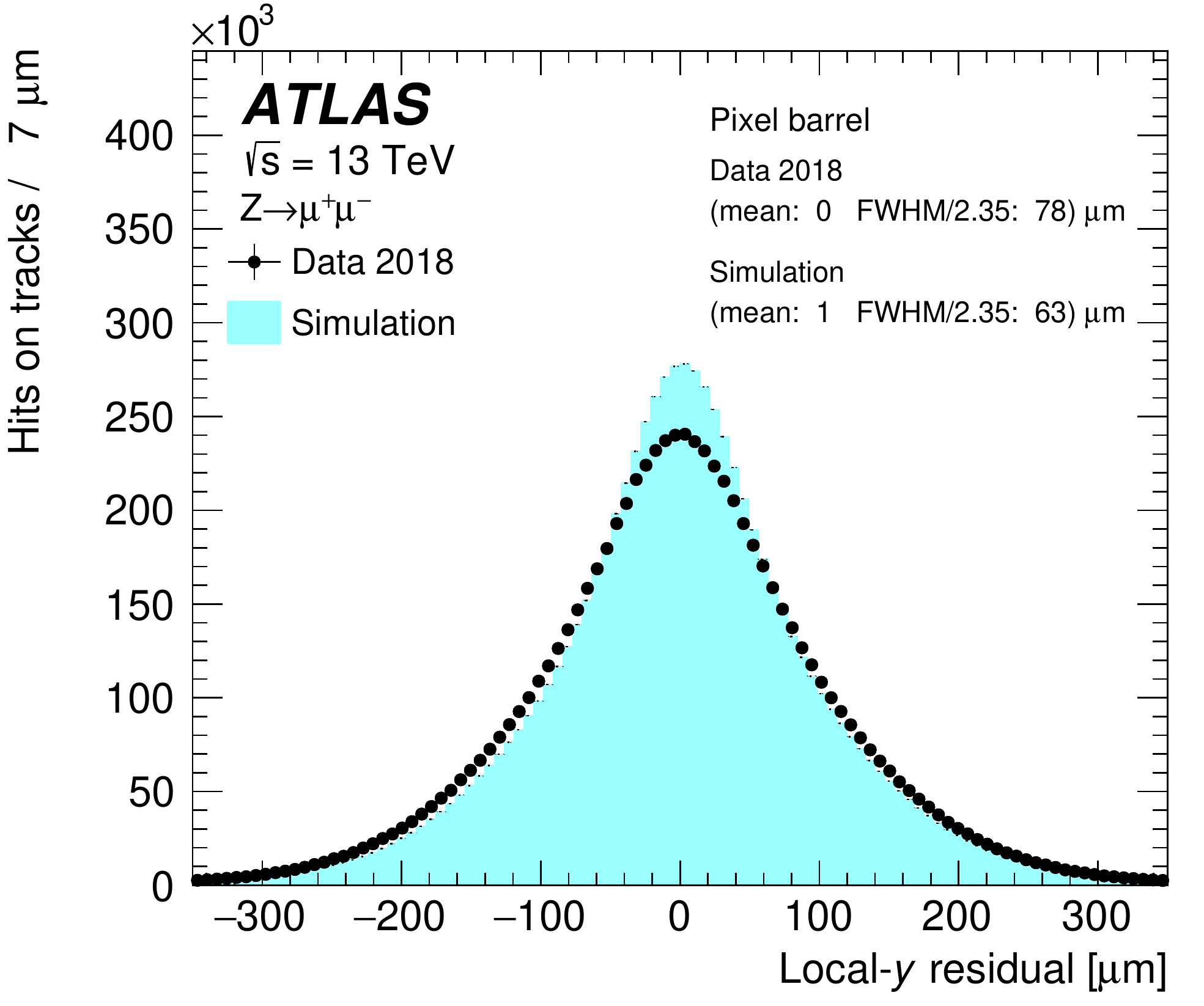}
\caption{The Pixel local-$x$ (left) and local-$y$ (right) residual distributions for the $Z\rightarrow \mu^+\mu^-$ data sample compared with simulated data. The distributions are integrated over all hits on tracks in barrel modules.}
\label{fig:Residuals_local_xy_Pixels}
\end{figure}

\begin{figure}[ht!]
\centering
\includegraphics[width=0.49\textwidth]{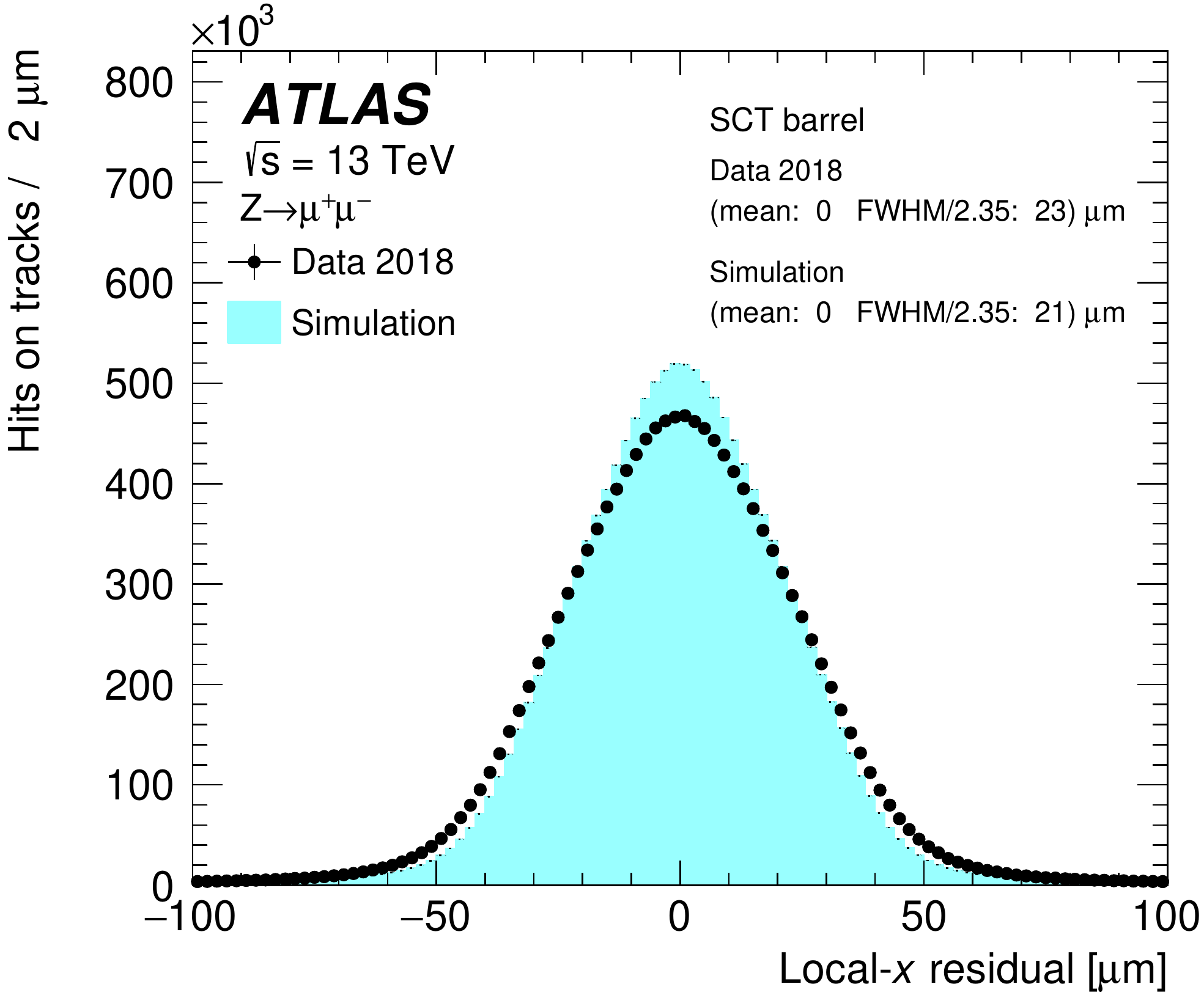}\hfill
\includegraphics[width=0.49\textwidth]{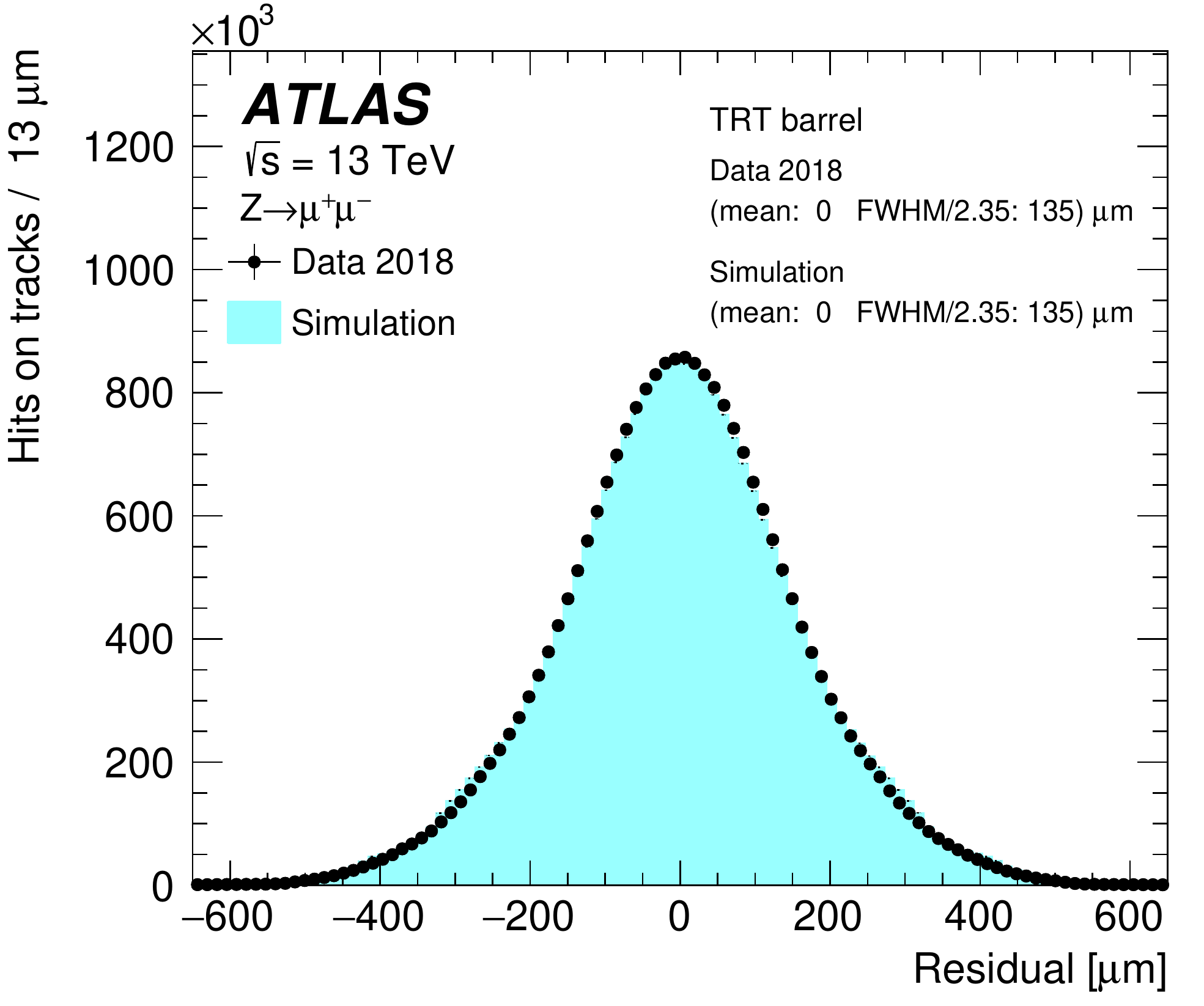}
\caption{The local-$x$ residual distributions in the SCT (left) and TRT (right) for the $Z\rightarrow \mu^+\mu^-$ data sample compared with simulated data. The distributions are integrated over all hits on tracks in barrel modules.}
\label{fig:localX_SCT_TRT}
\end{figure}

As described in \Cref{sec:globalchi2}, the solution of the Global $\chi^2$ is the one that minimises the unbiased\footnote{The unbiased residual does not include the measurement in question when determining the intersection position ($e_i$) of the fitted track with the surface.}  track-hit residuals.
\Cref{fig:Residuals_local_xy_IBL,fig:Residuals_local_xy_Pixels,fig:localX_SCT_TRT} show track-hit residual distributions for data and simulation in different ID subdetectors.
 
Data and simulation correspond to a set of muons selected in $Z\rightarrow \mu^+\mu^-$ candidate events triggered by the lowest-threshold unprescaled single and dimuon triggers. The simulation sample was generated with the \POWHEGBOXV{v1} Monte Carlo event generator~\cite{Nason:2004rx, Frixione:2007vw, Alioli:2010xd} at next-to-leading order (NLO) in $\alpha_\mathrm{S}$ interfaced to \PYTHIAV{8.186}~\cite{Sjostrand:2014zea} for the modelling of the parton shower, hadronisation, and underlying event, with parameter values set according to the AZNLO tune~\cite{STDM-2012-23}. The CT10 (NLO) set of parton distribution functions (PDF)~\cite{Lai:1999wy} was used for the hard-scattering processes, whereas the CTEQ6L1 PDF set~\cite{Pumplin:2002vw} was used for the parton shower.
Events are required to contain two muons (satisfying `medium' quality criteria as defined in Ref.~\cite{atlas_MCPpaper}) with opposite charge and $\pt>\SI{20}{\GeV}$. In addition, requirements on the opening angle between the two muons, $\gamma(\mu^{+}, \mu^{-})> 45^\circ$, and their invariant mass, $70~\GeV < m_{\mu^{+} \mu^{-}}< 110~\GeV$, are imposed.
In \Cref{fig:Residuals_local_xy_IBL,fig:Residuals_local_xy_Pixels,fig:localX_SCT_TRT},
both data and simulation correspond to \SI{2}{fb^{-1}} of data collected during 2018.
Statistical uncertainties in data and simulation are included in all the figures, although barely visible as they are negligible.
 
Adequate agreement is seen between data and simulation in the residual distributions, where differences are quantified in terms of the `full width at half maximum' (FWHM) figure of merit.
A similar level of agreement is observed for the data collected during the other years of Run 2.
In the case of the IBL, Pixel and SCT barrel, larger residual widths are observed in data.
As shown in \Cref{DynamicResults}, the Run\,2 alignment accuracy and stability in the central pseudorapidity range for both the Pixel and SCT barrel modules is controlled to a precision better than \SI{0.5}{\micro\metre} and \SI{2}{\micro\metre} in local-$x$ and local-$y$, respectively.
Consequently, several other possible causes of the observed discrepancy between data and simulation are considered, such as imperfect modelling of the interactions of muons with detector material in the simulation, the material description, delta ray production modelling, mis-modelling of the detector response (and resolution) in simulation, and residual biases not uniform across individual modules in data.
The latter particularly impacts the local-$y$ track-hit residuals in \Cref{fig:Residuals_local_xy_IBL}.
The poorest agreement is seen for the IBL residuals, which have not yet been corrected for sensor distortions, in contrast to the Pixel layers. The sensor distortion can result in track-hit residual biases of up to \SI{10}{\micro\metre} within a given module, thus causing a broadening of the overall distribution~\cite{Wollrath:2019rrw}. The shape of the IBL modules was recently parameterised with Bernstein--B\'ezier functions and will be corrected in the track fitting procedure for Run 3 data taking.
The cause of the small bias of \SI{4}{\micro\metre} in the IBL local-$y$ track-hit residuals in simulation in \Cref{fig:Residuals_local_xy_IBL} is currently unidentified.
Simulated samples use a perfectly aligned detector with no track-based alignment correction, hence this bias originates from the track or cluster reconstruction. On data, this small reconstruction bias is removed by the alignment without a significant effect on alignment precision.

% End of text imported from the .//chapters/IDAlignment.tex input file

\section{Detector stability and time-dependent alignment}
\label{sec:TimeAlign}
% The next lines are included from the .//chapters/TimeDependence.tex input file
This section discusses the main sources of time variation in ID geometry
and the methods implemented to mitigate these effects within the ATLAS prompt calibration loop~\cite{ATLASRun2DataQuality}.
In addition, the stability of the ID in Run 2 is summarised,
final time-dependent corrections for all subsystems are presented, and the precision of the alignment is determined.
All results use $pp$ collision data at $\sqrt{s} = \SI{13}{\TeV}$. The alignment precision for heavy-ion data in Run 2 is at least as good as the final precision of $pp$ collision data, as the instantaneous luminosity, and therefore the thermal load variations in the ID, is typically lower.
 
\subsection{Short-timescale movements}
\label{ShortTimeMovements}
 
Detector movements on short timescales are particularly challenging,
since the ID track-based alignment calculates an average position
correction for the time interval under study.\footnote{This time interval varies from a few minutes to several hours depending on the configuration of the alignment task.}
 
The procedure used to correct for rapid movements must balance two competing effects: the alignment corrections
must be determined in time intervals that are short enough to capture the motion of the particular deformation,
but long enough to include sufficient data to obtain precise corrections.
 
\subsubsection{Temperature-dependent IBL distortions}
\label{IBLdistortion}
 
\begin{figure}[ht!]
\centering
\includegraphics[width=0.7\textwidth]{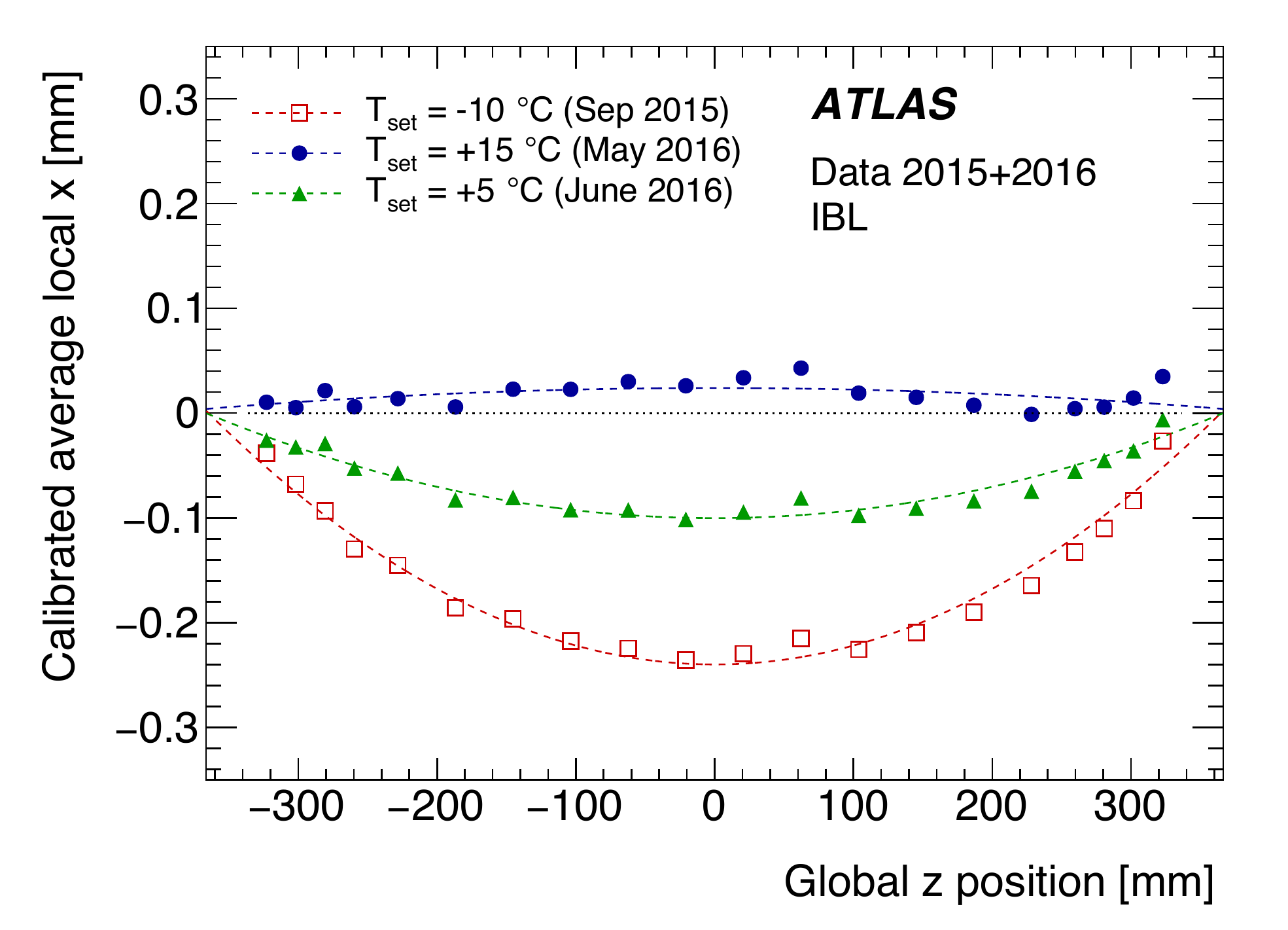}
\caption{IBL local-$x$ position in the transverse plane averaged over all 14 IBL staves for 2015 data
using $T_{\mathrm{set}}= \SI{-10}{\celsius}$ (red open squares), and for 2016 data using different
$T_{\mathrm{set}}$ ($+\SI{15}{\celsius}$, solid blue circles; $+\SI{5}{\celsius}$, solid green triangles). No error
bars associated with data are shown. The IBL distortion was constant during all three LHC
fills. Here, only the correction due to the IBL distortion is shown. The baseline, which describes the
overall translation of the whole stave, is subtracted using \cref{eq:IBLbow}. The fit represents only a first-order correction. Additional corrections are computed as part of the detailed alignment corrections at a later stage.}
\label{fig:IBLTempDependent}
\end{figure}

During the Run 2 commissioning of the IBL,
it was already noticed that the IBL staves can be distorted by hundreds of micrometers relative to the design geometry.
It was soon observed that these distortions depend on the operating temperature and correspond to
module displacement in the azimuthal direction of the staves, equivalent to their local-$x$ direction.
The distortion was understood to be caused by tight, asymmetric mechanical coupling of materials with different
coefficients of thermal expansion  (CTE).
The correlation between temperature and the size of the IBL distortion
was studied using cosmic-ray data in March 2015 with a controlled variation of the IBL temperature,
$T_{\mathrm{set}}$, in the range $\SI{-20}{\celsius}$ to $\SI{+15}{\celsius}$~\cite{IBLDistTaskForce}.
The size of the distortion was measured in situ using the track-based alignment and a fit to a model determined from a three-dimensional finite-element analysis.
This model parameterises the IBL distortion in local-$x$, $\delta x(z)$, using a parabolic function,
 
\begin{linenomath}
\begin{equation}
\delta x(z) = B - \frac{M}{z_{0}^{2}} \left( z^2 - z_{0}^{2} \right) \, ,
\label{eq:IBLbow}
\end{equation}
\end{linenomath}
 
where $z$ is the global-$z$ coordinate of the module, $z_0 \equiv \SI{366.5}{mm}$ is the coordinate of the
stave mount at both ends,
$B$ is the baseline describing the overall translation of the stave in
local-$x$, and $M$ is the magnitude of the distortion at the stave centre.
The temperature gradient of
$M$ with respect to $T_{\mathrm{set}}$ is found to be $\mathrm{d}M/\mathrm{d}T_{\mathrm{set}} =
\SI{-10.6(7)}{\micro\metre / K}$. The local-$y$ position shows no temperature-dependent effect within
\SI{20}{\micro\metre}
uncertainty, whereas the local-$z$ (bending out of the plane of the stave) was not included in this study.
The IBL
distortion is shown in \Cref{fig:IBLTempDependent} for different $T_{\mathrm{set}}$ values using 2015 and 2016 $pp$ collision data.
 
From the initial data taking in Run 2 through September 2015,
the IBL power consumption per module was found to be stable,
and fluctuations in $T_{\mathrm{set}}$ were within $\sim \SI{0.2}{K}$ resulting in a stable detector $(\delta x(z) < \SI{3}{\micro\metre})$~\cite{IBLDistTaskForce}.
This situation changed with the rapid increase in integrated luminosity per LHC fill after September,
which induced an increase in the low-voltage (LV) currents in the IBL module front-end electronics.
This increase was traced back to radiation-induced leakage current in transistors~\cite{IBLradiation}.
The change in LV currents depends on the total ionisation dose.
Studies show that the increase reaches a peak value for radiation doses between 10 and 30 kGy and decreases for higher doses to a value close to the pre-irradiation case.
 
These variations in the LV currents caused an increase in IBL module temperatures that resulted in changes in IBL distortions on short timescales.
In this context, values of $\delta x(z)$ of up to
\SI{30}{\micro\metre} were observed between LHC fills and up to \SI{10}{\micro\metre} within a single fill, corresponding to a
variation of \SI{0.5}{\micro\metre\per\hour}.

\subsubsection{Vertical movements of the Pixel detector}
\label{VerticalPixel}
\begin{figure}[ht!]
\centering
\includegraphics[width=0.75\textwidth]{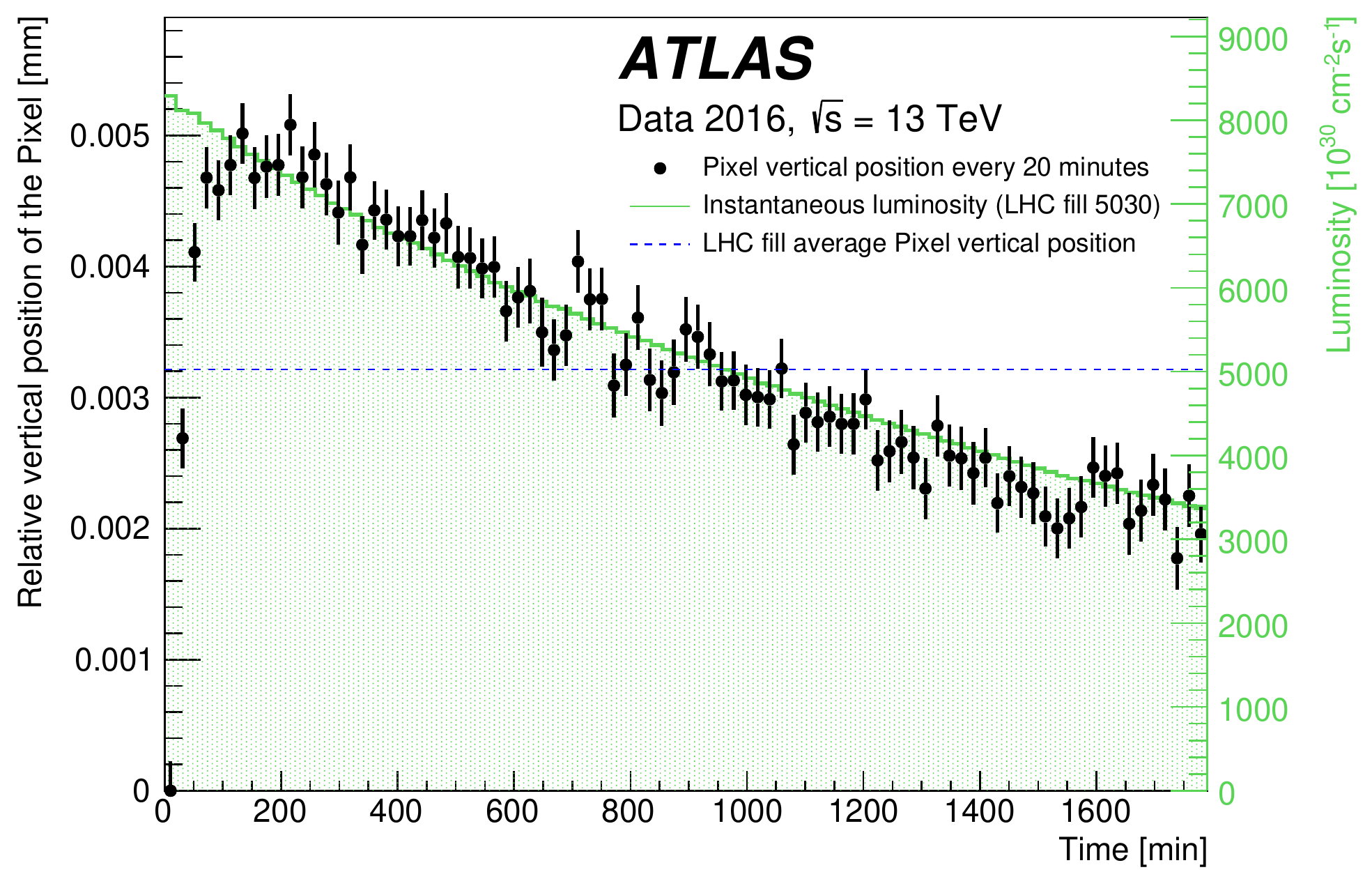}
\caption{The Pixel detector vertical ($T_{y}$) movement as a function of the time since the start of an LHC fill.
The average Pixel $T_{y}$ for the entire run (dashed blue line) is compared with its time evolution and with the instantaneous LHC luminosity. The error bars represent the statistical uncertainty.}
\label{fig:pixelsTy_run}
\end{figure}

Another systematic deformation on short timescales is a change in the vertical position (global-$y$
direction) of the Pixel detector by up to \SI{8}{\micro\metre} at the start of an LHC fill.
\Cref{fig:pixelsTy_run} shows the Pixel detector vertical movement from the start of an LHC fill.
The position is computed every 20 minutes, which is the shortest time interval used in the ATLAS prompt calibration loop.
As is evident from \Cref{fig:pixelsTy_run}, the average position across an LHC fill does not accurately describe the position of the Pixel detector.
 
\begin{figure}[ht!]
\centering
\includegraphics[width=0.7\textwidth]{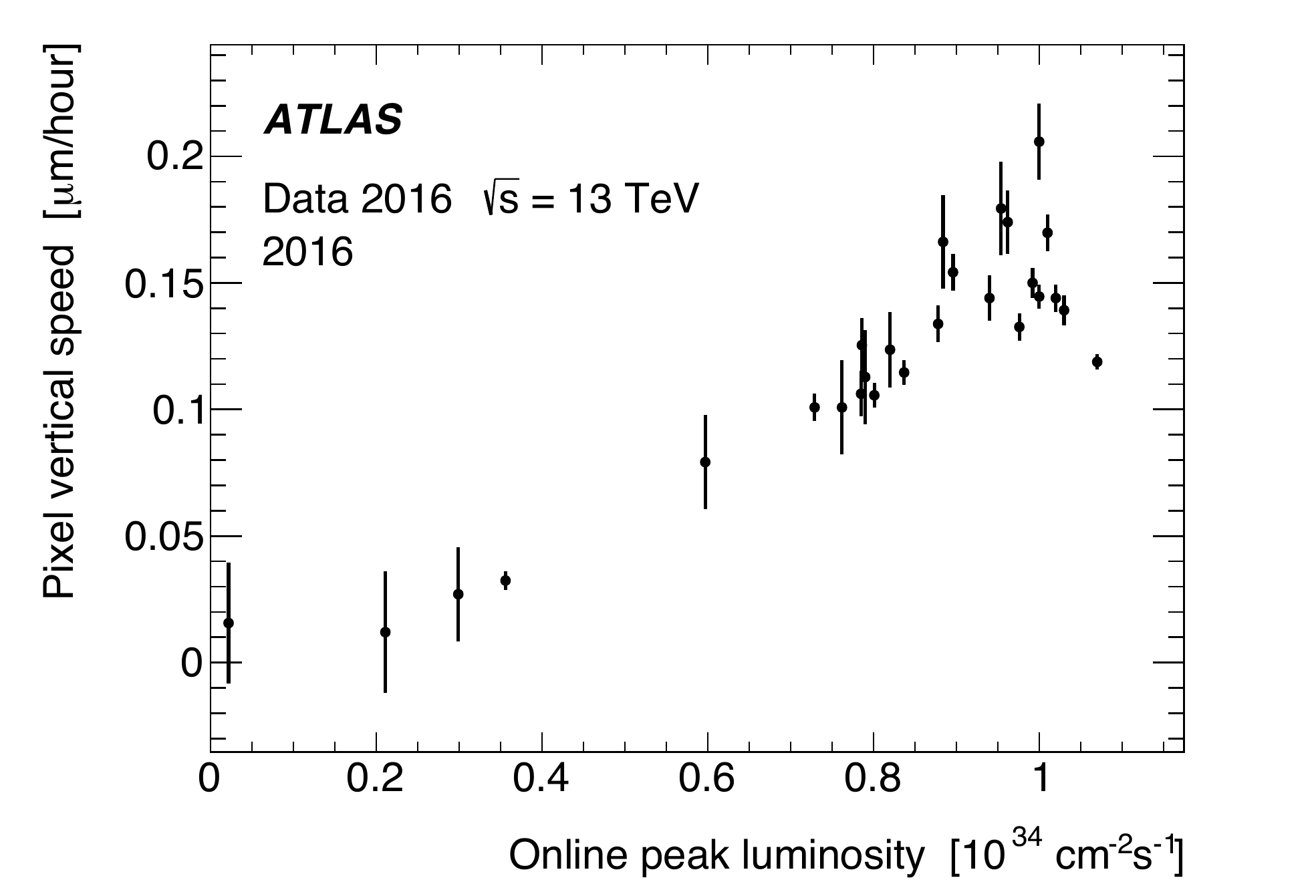}
\caption{Vertical speed of the Pixel detector as a function of the peak luminosity of an LHC fill, extracted from alignment corrections. Only fills longer than 8 hours are considered.}
\label{fig:PixelMovement}
\end{figure}

The cause of this movement is understood to be the following.
When the Pixel detector is switched on at the start of a fill, modules reach their new temperature
almost immediately as a result of the strong thermal coupling between the modules and the evaporative cooling
system~\cite{ATLASTechnicalPaper,JimenezPenaPhd}.
The LV current in the read-out electronics also increases immediately,
while the temperature in the Pixel detector volume rises gradually during the first 60 minutes.
The smaller mass load due to the change in density of the bi-phase cooling liquid causes the Pixel detector to rise.
After this initial rise,
as the instantaneous luminosity and thus the occupancy decreases over the course of the fill, LV digital currents, module temperature and Pixel volume temperature gradually decrease as well.
This in turn causes an additional slow drift in the direction opposite to the initial movement.
The speed of this slow drift depends on the peak luminosity per LHC fill.
This speed increased during 2016 to reach values of \SI{0.2}{\micro\metre\per\hour}, as shown in \Cref{fig:PixelMovement}.
The vertical speed is determined as the average speed of the Pixel detector excluding the first hour after the start of data taking.
This vertical drift was monitored and corrected for throughout Run 2.
 
\subsection{Dynamic alignment on short timescales}
\label{DynamicAlignment}

In an effort to mitigate the effects of systematic short-timescale distortions and ensure adequate data
quality for all analyses relying on tracking, conceptual improvements within the
alignment framework and strategy were made. A key improvement was the introduction of a new alignment
DoF, $B_x$, to parameterise the IBL distortion deformation magnitude $M$.  The $B_x$ DoF correlates the
local-$x$ coordinate of each module along the IBL stave using the parabolic function defined in
\cref{eq:IBLbow}. Minimising the global $\chi^{2}$ with respect to $B_x$ provides corrections for varying degrees of
IBL stave distortion using a single DoF, which can be done with small amounts of data.
In contrast, a full level 3 alignment, which relies on a large amount of data, had been required previously,
which did not allow short-timescale movements to be determined.
 
The automated alignment scheme that is performed within the ATLAS prompt calibration loop in Run 2 data
taking determines level 1 and IBL $B_x$ (per stave) dynamic alignment constants every 20 minutes
at the start of a fill and every 100 minutes for the rest of the fill.
This level of granularity in time is adequate to mitigate the effects of short-timescale
vertical movements on track parameter resolution.
The alignment is
performed in two iterations of the level 1 calibration loop (level 1 CL) followed by two dedicated
iterations to correct for IBL distortions.
The $B_x$ correction in the level 1 CL corresponds to a collective, uniform correction for all IBL staves.
The dedicated IBL bowing iterations determine $B_x$ individually for each stave,
as summarised in \Cref{Tab:AlignLevelsCL}.
The SCT barrel is used as the reference in the dynamic alignment.
 
\begin{table}[t!]
\begin{center}
\caption{Typical alignment configurations used throughout Run 2 data taking to derive dynamic alignment corrections. }
\small
\begin{tabular}{llcl}
\toprule
Level & Description & Structures & DoF \\
\midrule
1 CL	&   IBL   							& 1   &  All DoF incl. $B_x$, except $R_z$	 		 \\
&   Pixel detector   					& 1   &  All 				  \\
&   SCT endcaps (SCT barrel fixed)		& 2   &  All except $T_z$ 		 \\
&   TRT split into barrel and 2 endcaps	& 3   &  All except $T_z$ 		 \\
\midrule
IBL bowing	&   IBL staves   		& 14~~   &  $B_x$				 \\
&   Pixel, SCT, and TRT detectors fixed   		&     &  				  \\
\bottomrule
\end{tabular}
\label{Tab:AlignLevelsCL}
\end{center}
\end{table}

\subsection{Inner Detector stability during Run 2 data taking}
\label{DynamicResults}
 
\subsubsection{Time-dependent corrections for all subsystems}
\label{ResultsCorrections}
 
\begin{figure}[ht!]
\centering
\includegraphics[width=0.8\textwidth]{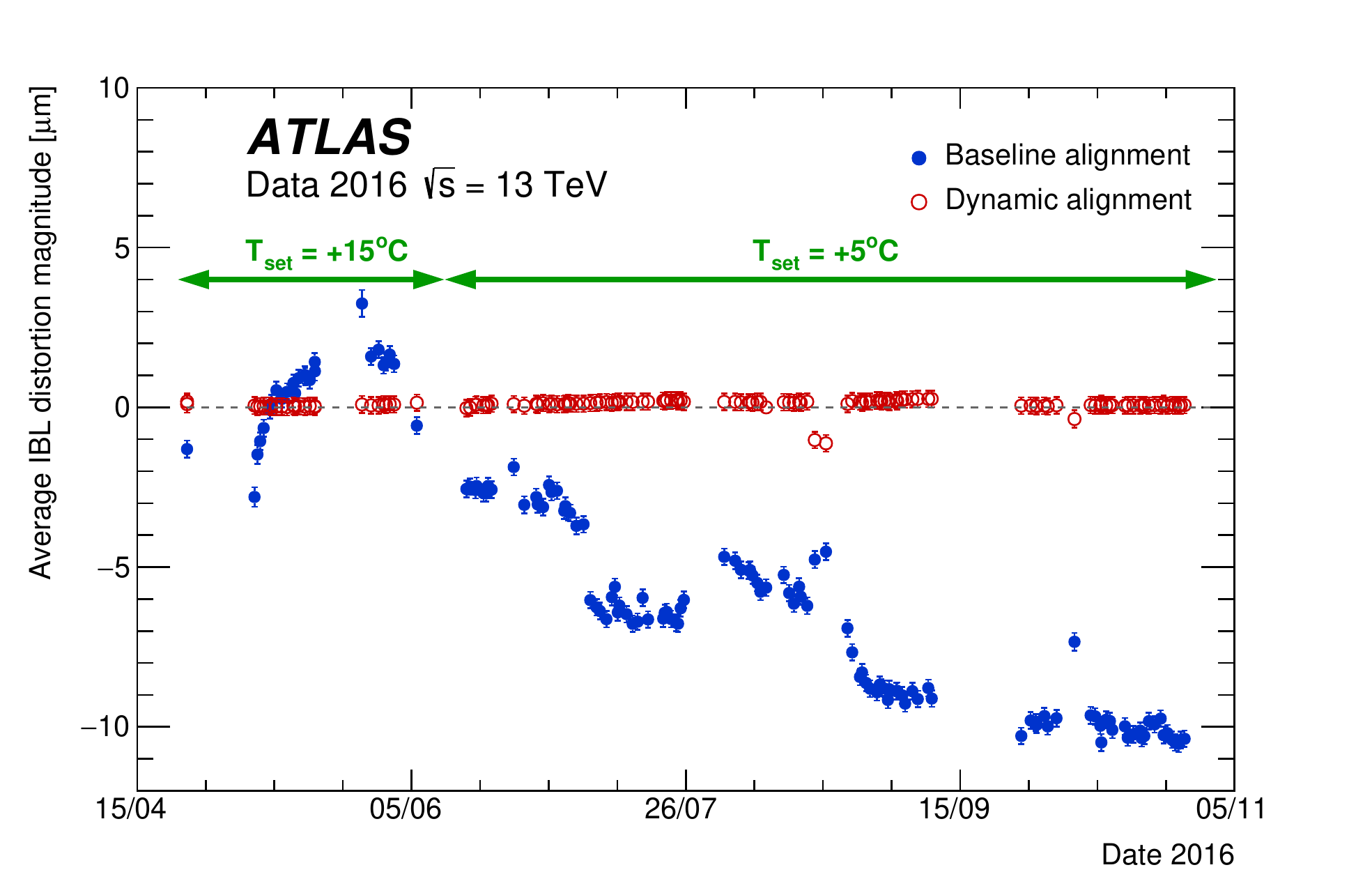}
\caption{Bowing magnitude averaged over the 14 IBL staves relative to the baseline alignment (blue full circles) and the geometry after dynamic alignment (red open circles) with its statistical uncertainty.
The IBL operation temperature (T$_{\mathrm{set}}$) for each period is shown.}
\label{fig:DynamicIBLMain}
\end{figure}

\begin{figure}[ht!]
\centering
\includegraphics[width=0.48\textwidth]{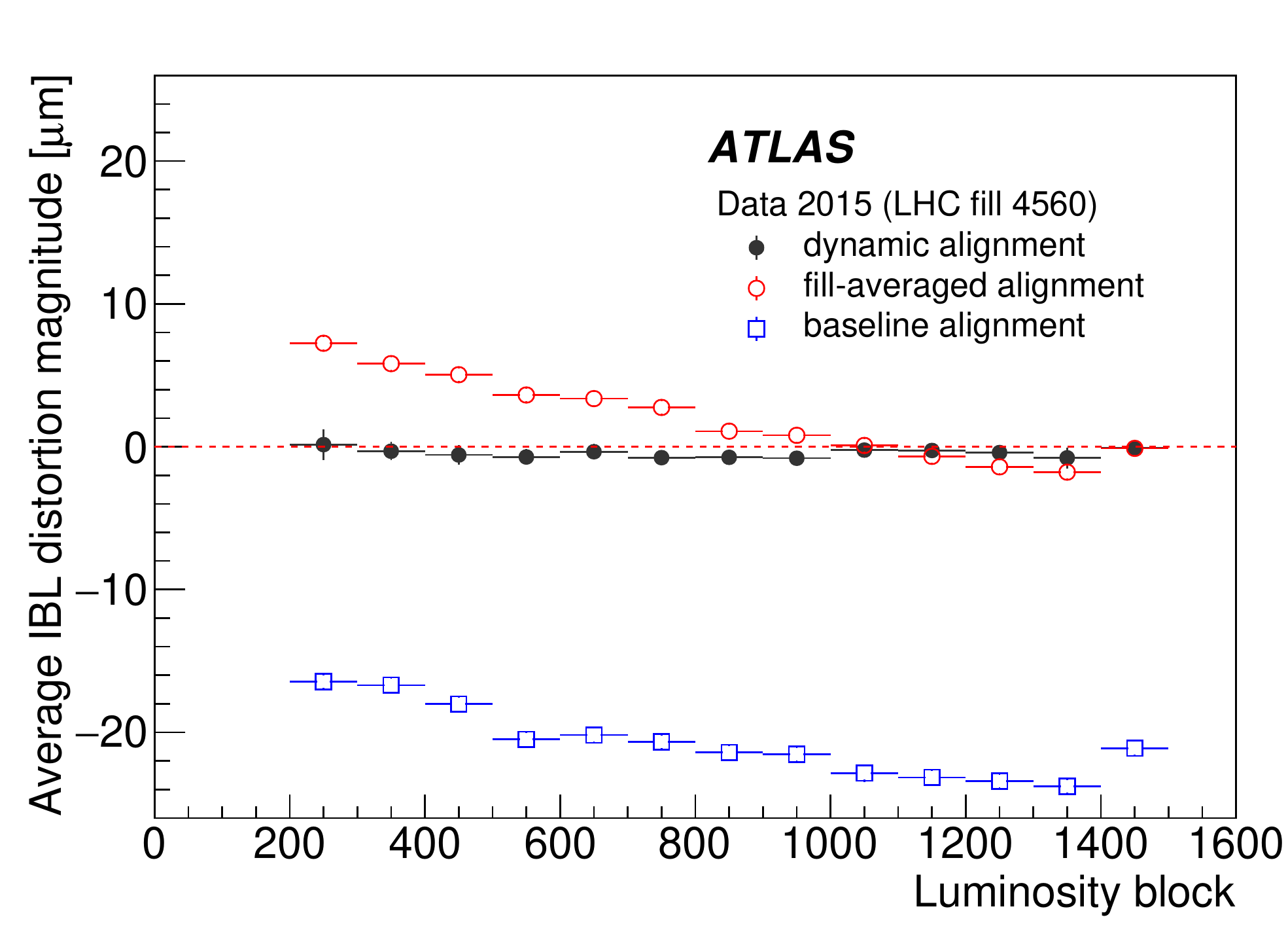}
\includegraphics[width=0.48\textwidth]{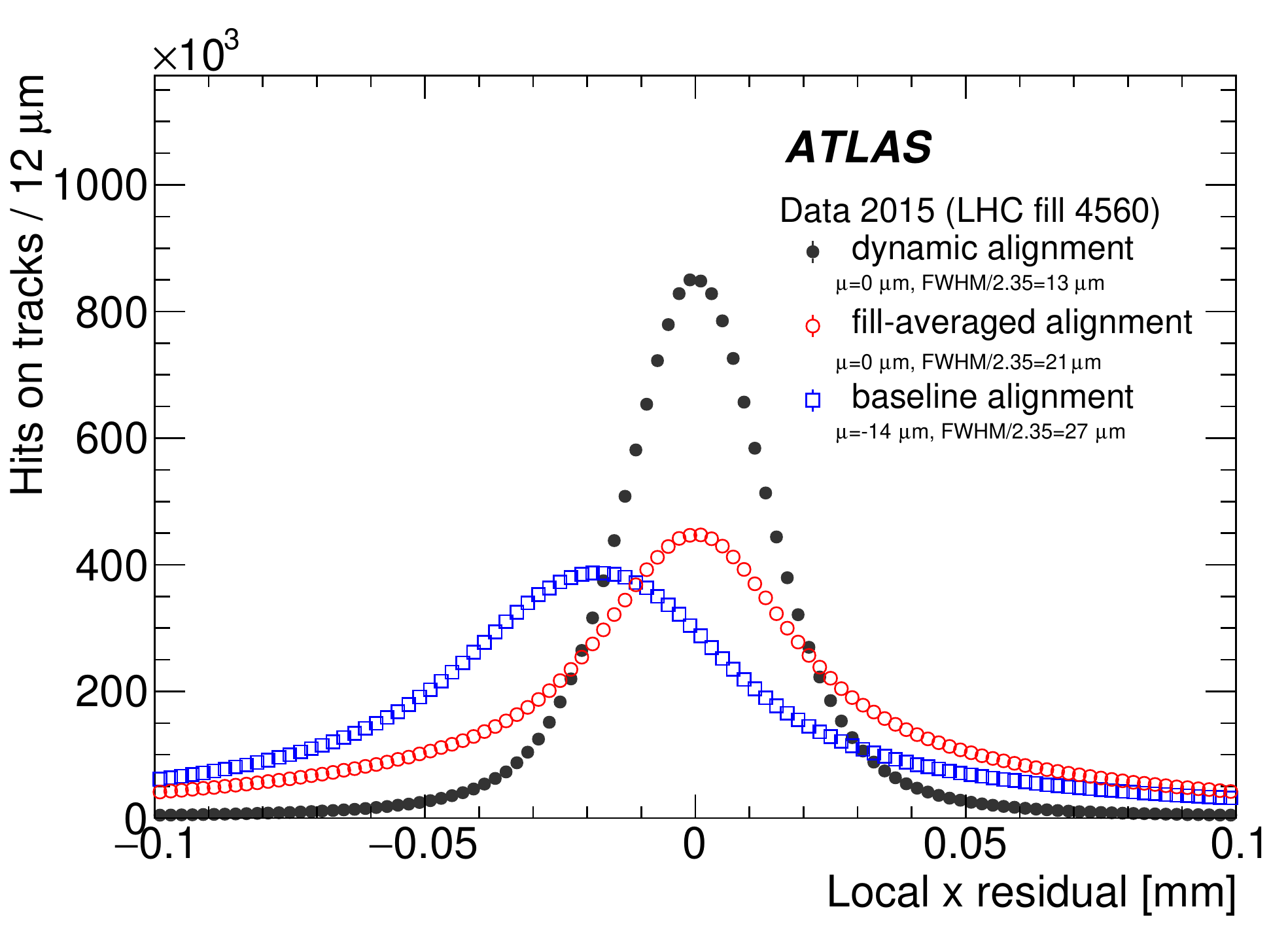}
\caption{IBL distortion magnitude in the transverse plane per luminosity block (LB) range (left) and the IBL local-$x$ unbiased residual distributions (right) for an LHC fill averaged over all 14 IBL staves. The open blue squares (baseline alignment) show the average IBL distortion in the transverse plane after the baseline ID alignment. The open red circles show the fill-averaged correction and the solid black circles show the full dynamic alignment correction. }
\label{fig:DynamicIBL_LBandRes}
\end{figure}

The performance of the dynamic alignment scheme using 2016 $pp$ collision data is shown in
\Cref{fig:DynamicIBLMain,fig:DynamicIBL_LBandRes}.
The average bowing magnitude of the 14 IBL staves relative to the baseline alignment is compared
with the results of the dynamic alignment in \Cref{fig:DynamicIBLMain}.
\Cref{fig:DynamicIBL_LBandRes} shows the average IBL distortion computed after different alignment
corrections versus time in the form of luminosity blocks (LB), which correspond to stable data-taking
conditions in periods of approximately one minute.  It also compares the unbiased local-$x$ residuals
computed using the a fill-averaged correction (for illustration only) with those obtained after computing the full
dynamic alignment correction, which is derived in short time-intervals.
A clear improvement in the residual distributions is seen after applying dynamic alignment corrections.
\Cref{fig:DynamicIBLMain,fig:DynamicIBL_LBandRes} illustrate that, averaged over an LHC fill,
even very large values of $M$ (up to $\SI{30}{\micro\metre}$) are accurately corrected for using $B_x$ as an alignment DoF.
These features
were present for all Run 2 data, although there was some saturation of the effect in the later years of
Run 2, as observed in the radiation damage studies of the IBL~\cite{IBLradiation}.
 
\begin{figure}[ht]
\centering
\begin{subfigure}[c]{.9\textwidth}
\includegraphics[width=0.97\textwidth]{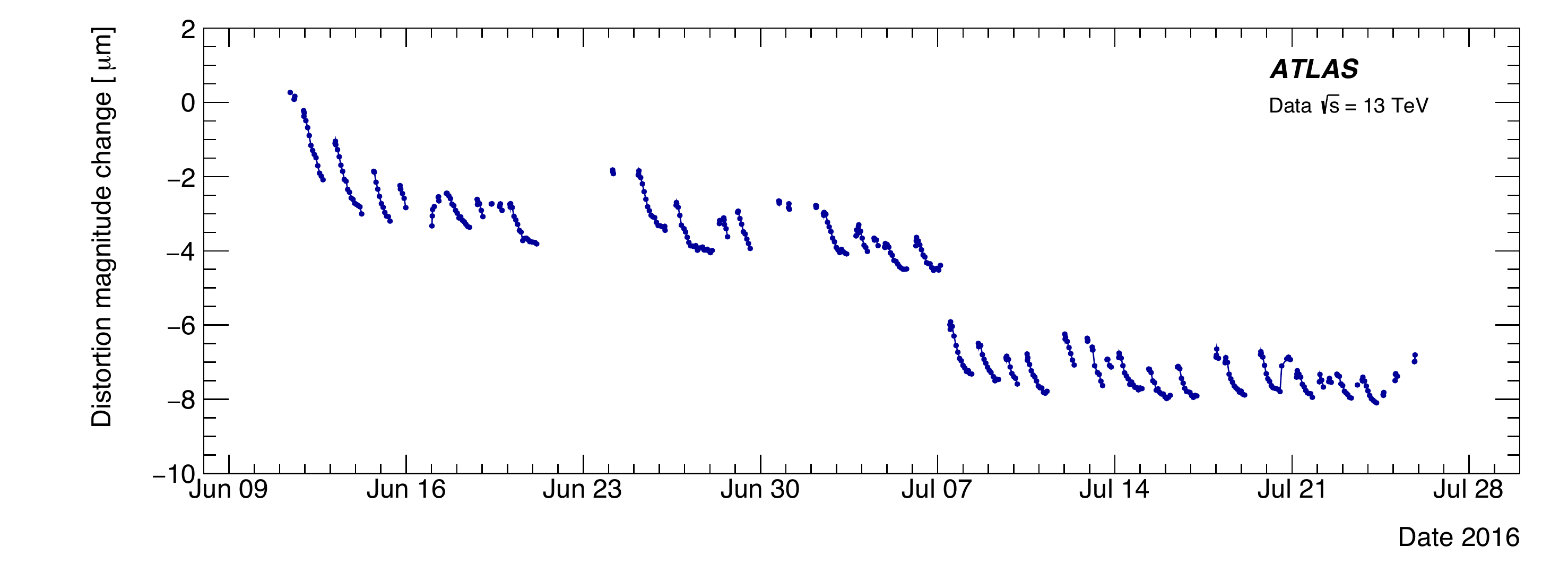}
\label{fig:dynamicBow}
\end{subfigure}
\begin{subfigure}[c]{.9\textwidth}
\includegraphics[width=0.97\textwidth]{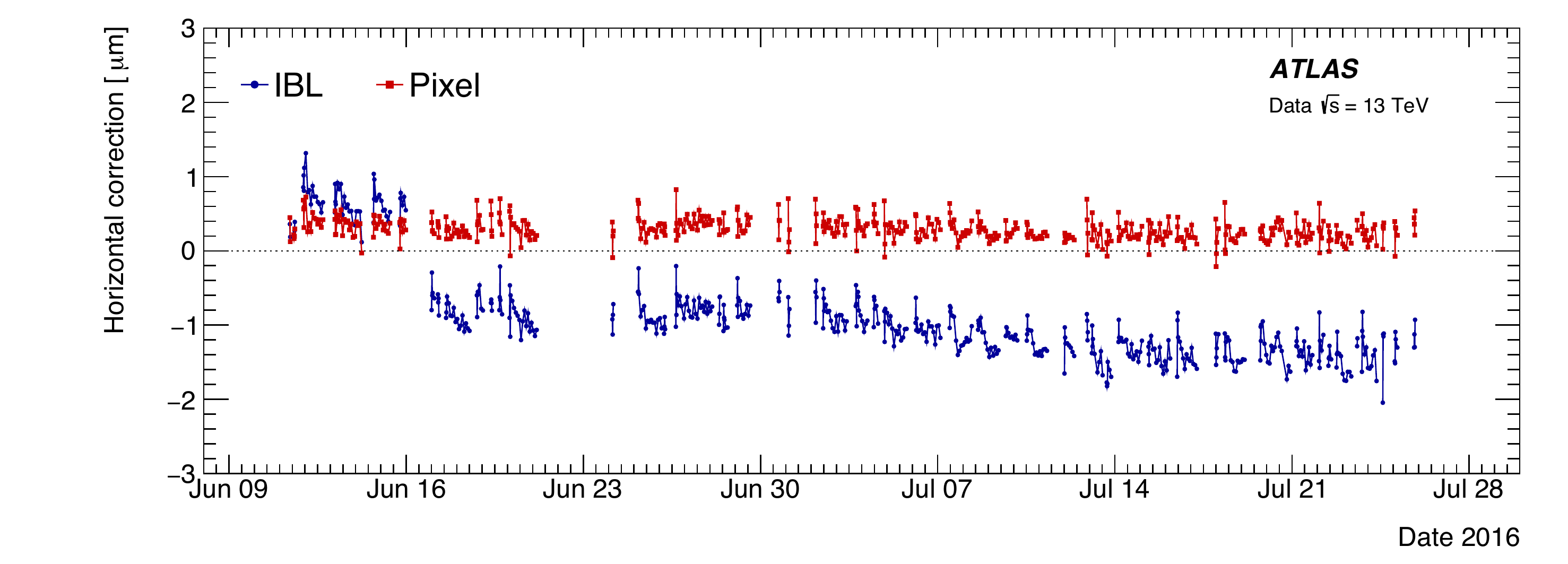}
\label{fig:dynamicX}
\end{subfigure}
\begin{subfigure}[c]{.9\textwidth}
\includegraphics[width=0.97\textwidth]{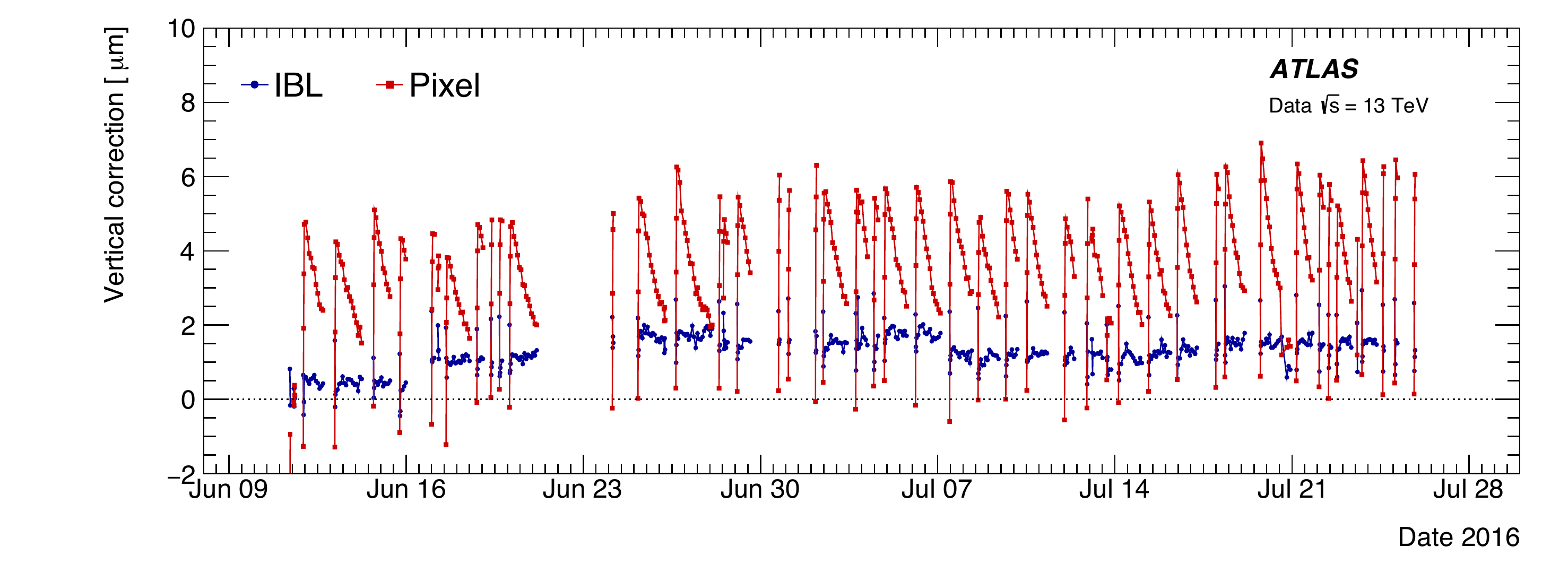}
\label{fig:pixb2_x}
\end{subfigure}
\caption{Average correction of the IBL bowing magnitude, $B_x$, (top), IBL and
Pixel detector's horizontal position, $T_x$, (middle), IBL and Pixel detector's vertical position, $T_y$,
(bottom) relative to the baseline alignment in 2016 $pp$ collision runs between
LHC technical shutdown period 1 and LHC machine development period 1. The correction is calculated every 20
minutes for the first 60 minutes of the data taking, and every 100 minutes for the rest of the
data-taking period. Each connected series of points represents a continuous data-taking period.}
\label{fig:dynamicPlots}
\end{figure}

The long-term trend of the Pixel and IBL detector movements relative to the baseline alignment
correction is shown in \Cref{fig:dynamicPlots} for the average $B_x$ correction, the global $T_x$ correction, and the global $T_y$ correction.
For the sake of clarity, the plots in \Cref{fig:dynamicPlots} show only a fraction of the Run 2 data; the
remaining data follow the same trend.

\subsubsection{Final alignment precision of each subsystem}
\label{ResultsPrecision}
 
The final alignment precision of each ID subsystem is determined from the track-hit residuals of individual silicon modules for each LHC fill in 2015 and 2016 data taking after the dynamic alignment corrections are applied.
These dynamic alignment corrections are computed either for large structures (e.g.\ the Pixel detector) as a collective movement of all modules or using a simplified parameterisation (like $B_x$).
In this context, less significant module-to-module movements remain uncorrected by the dynamic alignment.
This effect is seen as a residual time-dependent misalignment or `instability' of the modules.
This instability is estimated for each silicon layer and module $z$-position by integrating modules over $\phi$ into one group.
Results are presented for the `in-plane' translation DoFs only (local-$x$ and local-$y$).
 
For each module, the average track-hit residual, $\langle r_{x,y}\rangle$, is computed for each LHC fill, for both local-$x$ and local-$y$, on a set of calibration data, whose size is approximately independent from the fill conditions.
Its statistical uncertainty, $\sigma_{r_{x,y}}/\sqrt{N}$, where $N$ is given by the number of tracks per module and $\sigma_{r_{x,y}}$ is the standard deviation of the residuals, is computed assuming that the residual distribution is approximately Gaussian.
The dispersion $\sigma_{\langle r_{x,y}\rangle}$ of the distribution in $\langle r_{x,y}\rangle$ obtained from all LHC fills is an estimate of the total instability of the module position after all alignment corrections are applied.
This total uncertainty can be divided into a statistical component $(\sigma_{r_{x,y}}/\sqrt{N})$ and a component describing residual instability due to uncorrected time-dependent movements and stochastic fluctuations, $\sigma_{x,y}^{\mathrm{time}}$:
\begin{linenomath}
\begin{equation*}
\sigma_{\langle r_{x,y}\rangle} \sim \sigma_{x,y}^{\mathrm{time}} \oplus \frac{\sigma_{r_{x,y}}}{\sqrt{N}} \, .
\label{eq:combined}
\end{equation*}
\end{linenomath}
 
As the size of the statistical contribution per module per LHC fill is generally small, $\sigma_{x,y}^{\mathrm{time}}$ is estimated by
\begin{linenomath}
\begin{equation*}
\sigma_{x,y}^{\mathrm{time}} \equiv \sqrt{\sigma_{\langle r_{x,y}\rangle}^{2} - \left(\frac{\sigma_{r_{x,y}}}{\sqrt{N}}\right)^{2}}~.
\end{equation*}
\end{linenomath}

\begin{figure}[ht!]
\begin{center}
\includegraphics[width=0.70\textwidth]{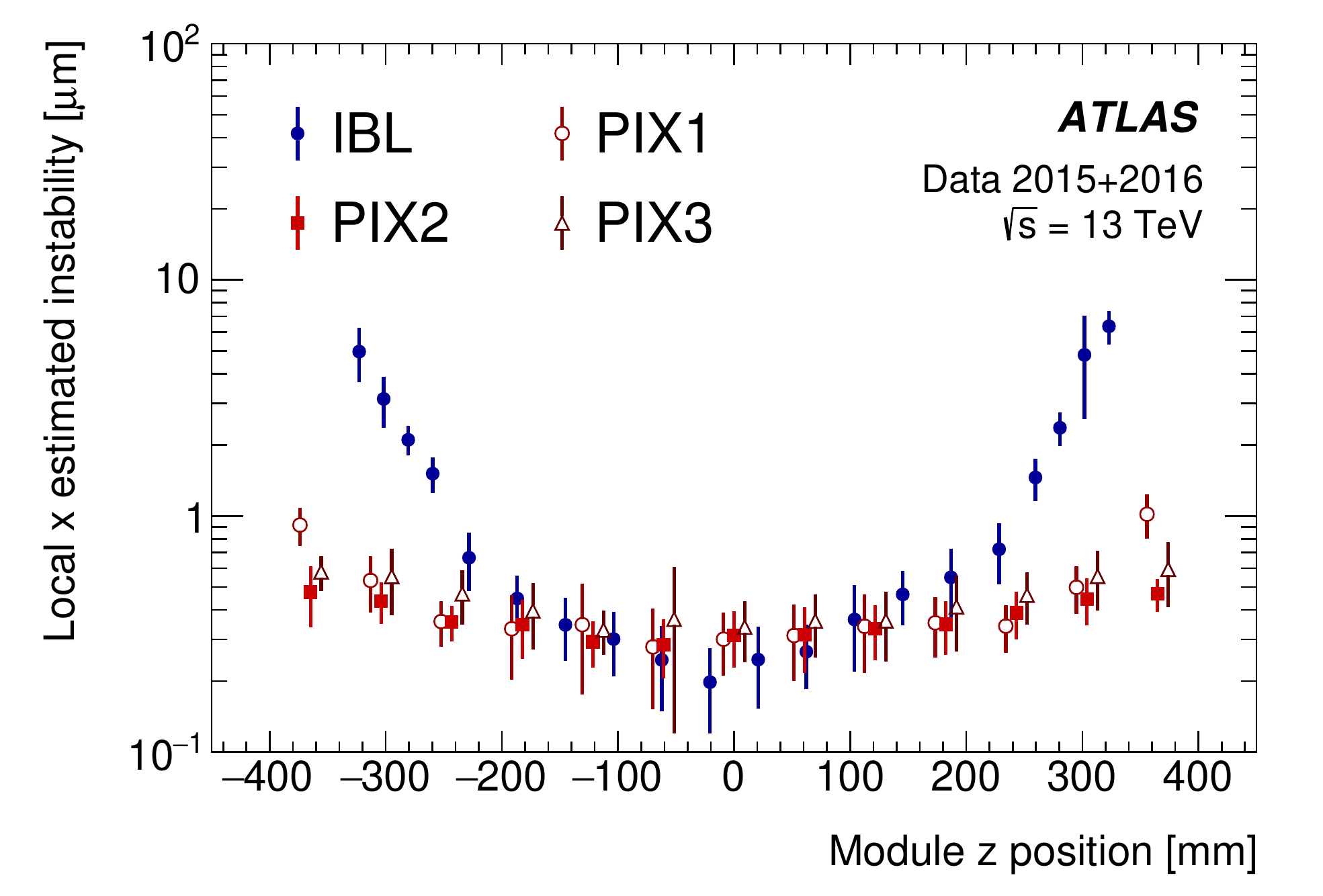}
\caption{Estimated $\sigma_{x}^{\mathrm{time}}$ as a function of the module global-$z$ position for the IBL and Pixel barrel layers. The vertical bar on each marker represents the standard deviation of the estimated value over modules at the same $z$-position along different staves. The global-$z$ position is slightly modified from its true value for visualisation purposes.}
\label{fig:pixel_barrel_inst_x}
\end{center}
\end{figure}

\begin{figure}[ht!]
\begin{center}
\includegraphics[width=0.70\textwidth]{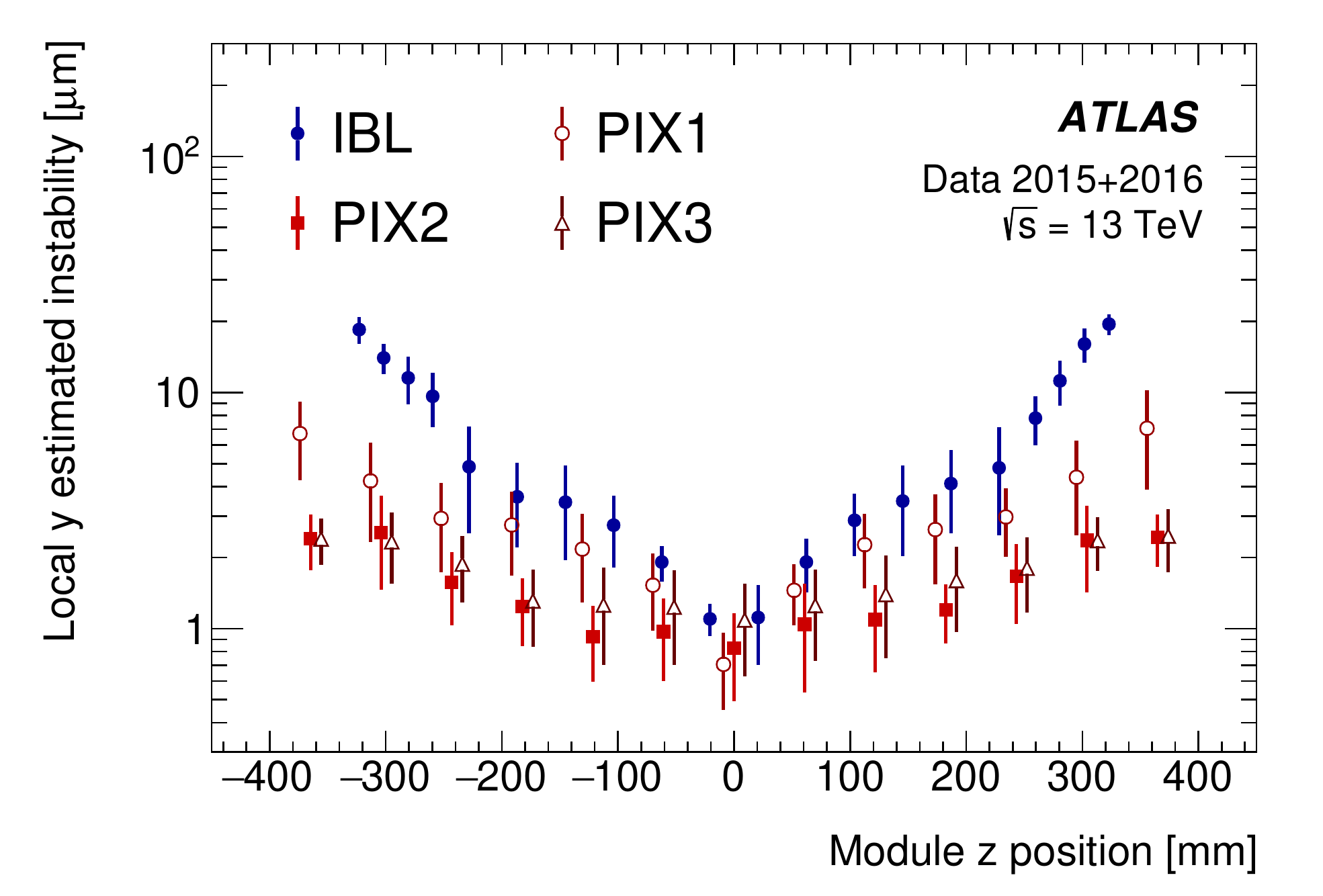}
\caption{Estimated $\sigma_{y}^{\mathrm{time}}$ as a function of the module global-$z$ position for Pixel
barrel layers. The vertical bar on each marker represents the standard deviation of the estimated value over
modules at the same $z$-position along different staves. The global-$z$ position is slightly modified for the
different Pixel layers for visualisation purposes.}
\label{fig:pixel_barrel_inst_y}
\end{center}
\end{figure}

\begin{figure}[ht!]
\begin{center}
\includegraphics[width=0.70\textwidth]{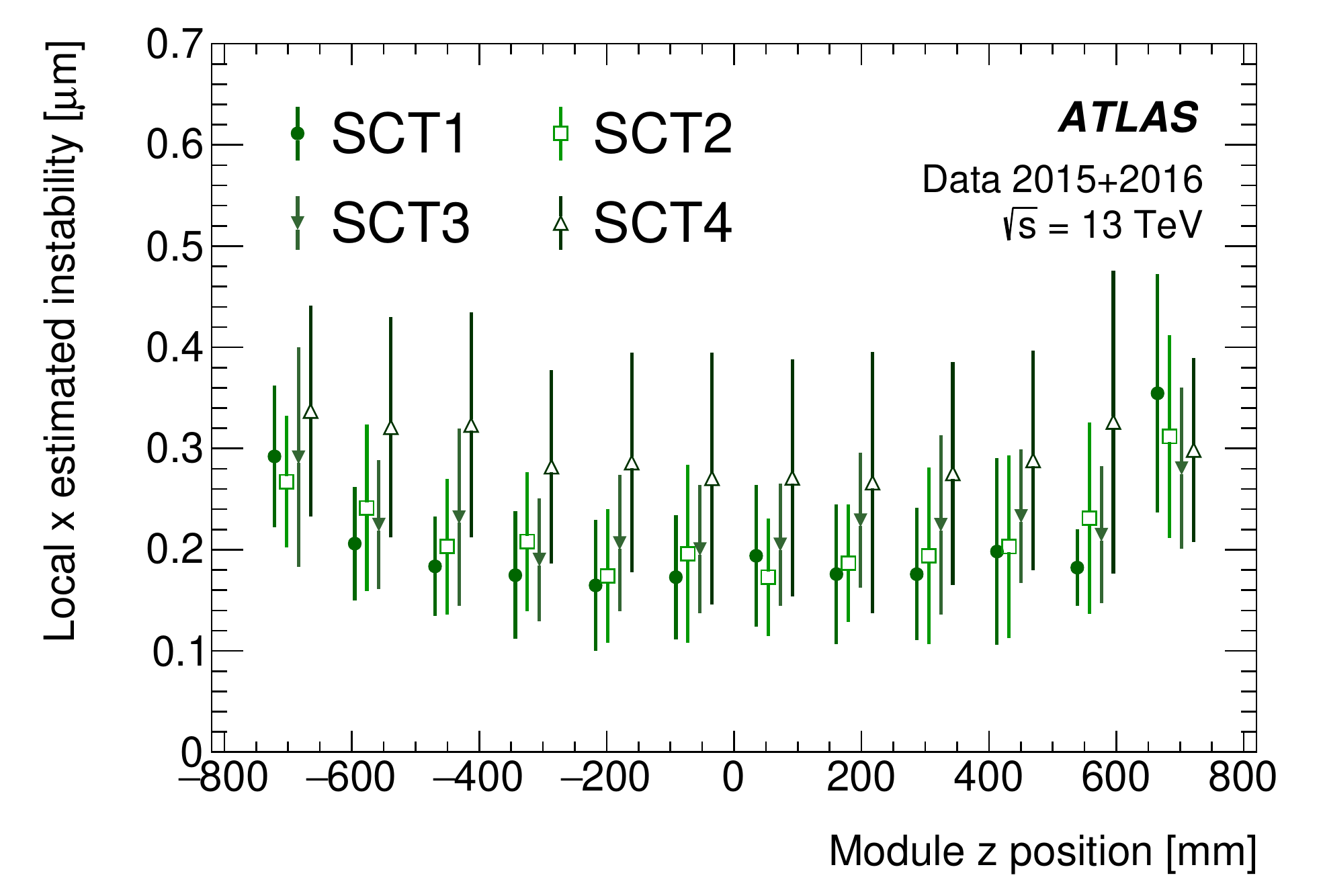}
\caption{Estimated $\sigma_{x}^{\mathrm{time}}$ as a function of module global-$z$ position for SCT
barrel layers. The vertical bar on each marker represents the standard deviation of the estimated value over
modules at the same $z$-position along different staves. The global-$z$ position is slightly modified for the
different SCT layers for visualisation purposes.}
\label{fig:SCT_barrel_inst_x}
\end{center}
\end{figure}

\begin{figure}[ht!]
\centering
\begin{subfigure}[c]{.49\textwidth}
\includegraphics[width=\textwidth]{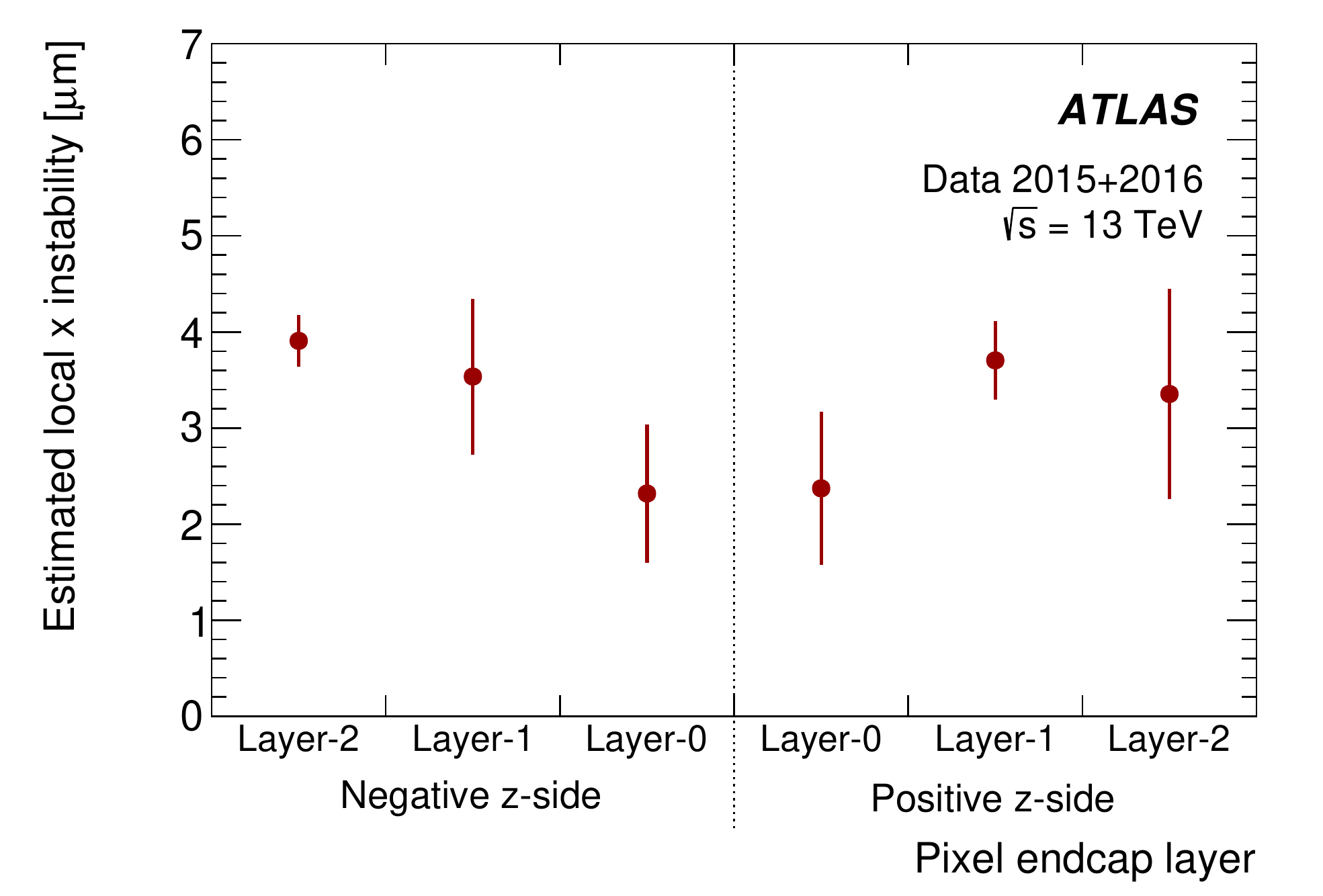}
\label{fig:pixb0_y}
\end{subfigure}
\begin{subfigure}[c]{.49\textwidth}
\includegraphics[width=\textwidth]{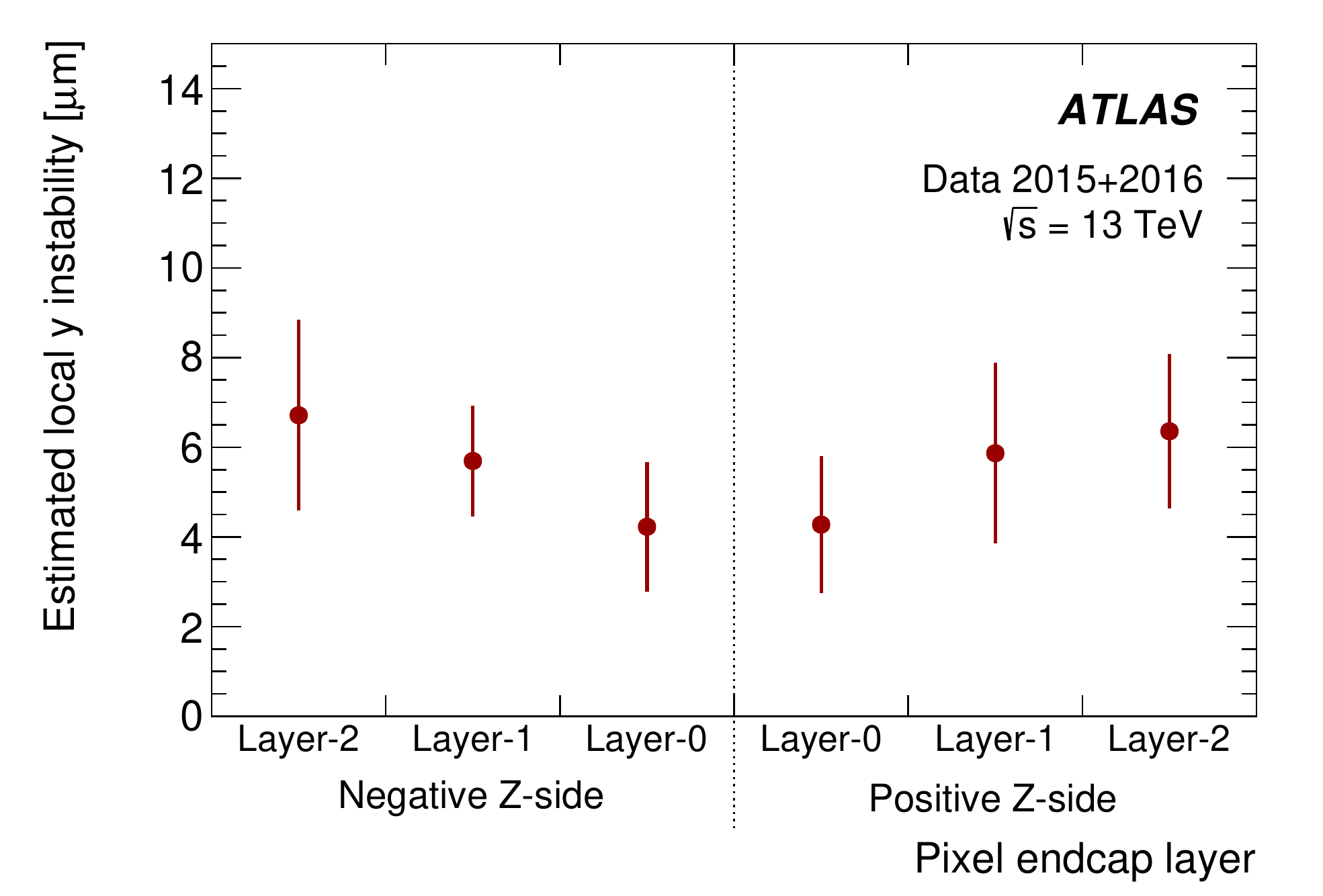}
\label{fig:pixb1_y}
\end{subfigure}
\caption{Estimated $\sigma_{x,y}^{\mathrm{time}}$ as a function of module $\eta$ index for Pixel endcap layers. The vertical bar on each marker represents the standard deviation of the estimated value over modules of the same layer.}
\label{fig:pixel_inst_endcap}
\end{figure}

\Crefrange{fig:pixel_barrel_inst_x}{fig:SCT_barrel_inst_x} show the estimated instability in local-$x$ and local-$y$
of the Pixel and SCT barrel layers as a function of module $z$-position.
\Cref{fig:pixel_inst_endcap} shows the local-$x$ and local-$y$ instabilities of the modules in the Pixel endcap layers.
The alignment in the central pseudorapidity range for both the Pixel and SCT barrel modules is controlled to a
precision better than \SI{0.5}{\micro\metre} and \SI{2}{\micro\metre} in local-$x$ and local-$y$, respectively.
This level of control is considered to be very good given the time-dependent corrections of
$\mathcal{O}(\SI{10}{\micro\metre})$ due to the IBL distortion and the vertical movement of the Pixel detector.
The same level of precision is not achieved for the outermost IBL modules (3D sensors) corresponding to
the range $|\eta|>2.5$.
There the alignment precision in local-$x$ (local-$y$) is measured with an uncertainty better than \SI{3}{\micro\metre} (\SI{15}{\micro\metre}).
This region is particularly challenging due to the low number of tracks.
Moreover, the tracks in this region have only small overlap with other ID tracking layers.
Furthermore, the large IBL local-$y$ uncertainty may be related to a deformation of IBL staves in local-$z$ over time,
which is not corrected for in the prompt calibration loop.
 
The instability for Pixel endcap modules is larger than for barrel modules;
the local-$x$ and local-$y$ instabilities are 2--4\,\si{\micro\metre} and 4--7\,\si{\micro\metre}, respectively.
This instability corresponds to the size of the movements of the Pixel endcap modules relative to the baseline
alignment over time. The precision achieved is nearly one order of magnitude better than the required precision~\cite{ATLAS-TDR-4}. This required precision was defined in order to limit the degradation of the resolution of the track parameters for high-momentum tracks to less than 20\% in comparison with a perfectly aligned detector.
While succeeding in its primary goal, these results, specifically the residual Pixel endcap movements, also imply that the current dynamic alignment scheme, which allows time-dependent alignment of the entire Pixel detector as one unit, is not optimal.
A higher level of precision might be achieved if the Pixel endcap disks were aligned individually.
This improvement is under study for LHC Run 3,
including detailed cross-checks for new weak modes that may be introduced due to the additional DoFs within the calibration loop.
 
\FloatBarrier

% End of text imported from the .//chapters/TimeDependence.tex input file

\section{Momentum biases}
\label{sec:MomentumBias}
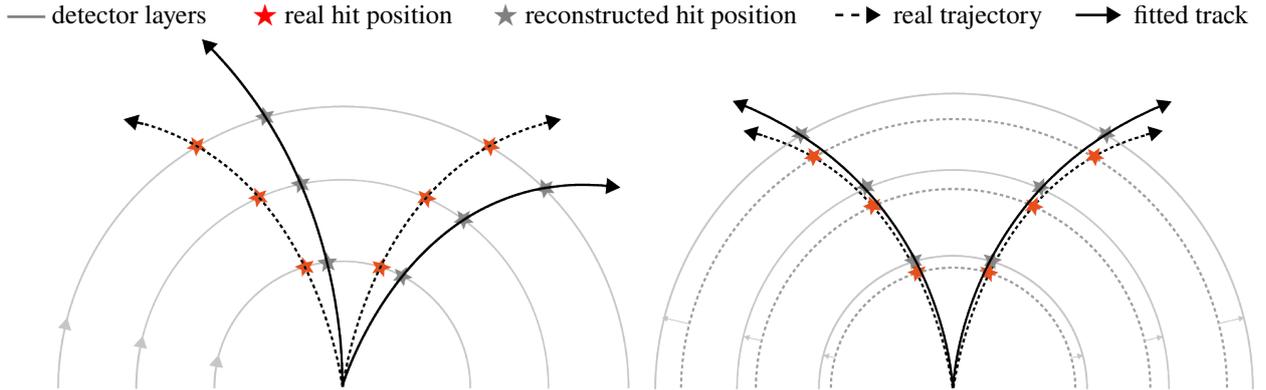
\begin{figure}[ht!]
\centering
\includegraphics[width=0.49\textwidth]{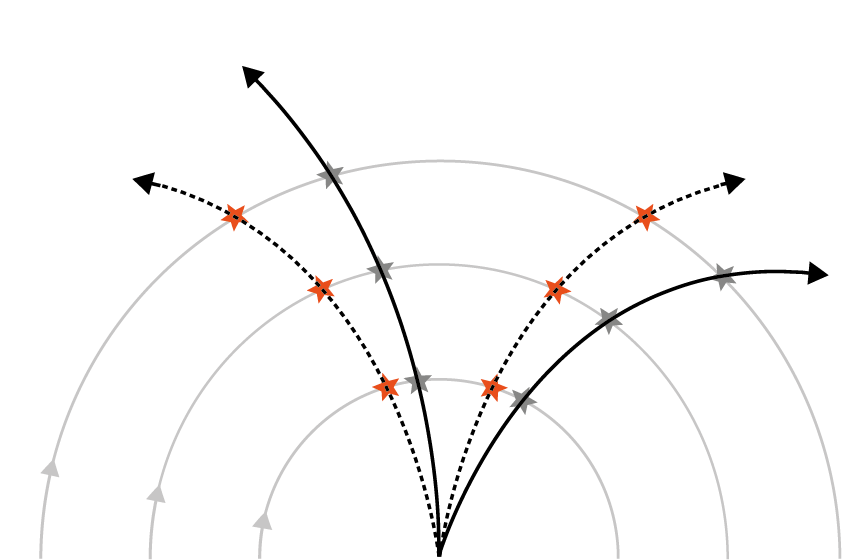}\hfill
\includegraphics[width=0.49\textwidth]{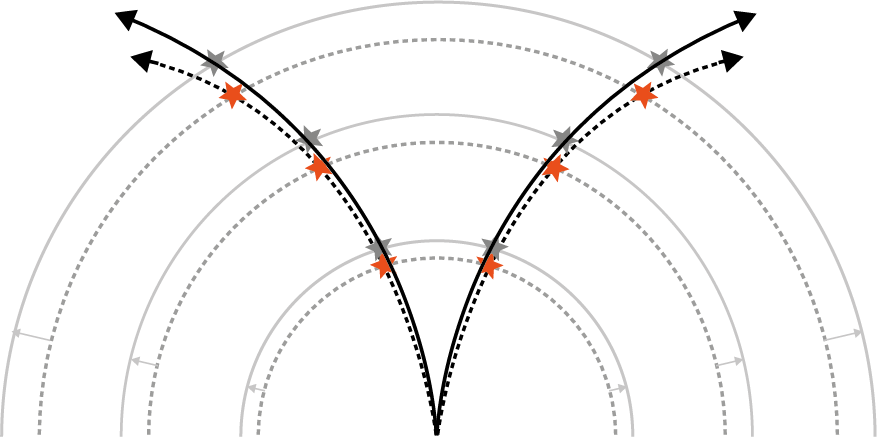}
\put(-470., 140.){
\begin{tikzpicture}
\draw[line width=1pt, color=gray] (0,0) -- (0.5,0);
\end{tikzpicture}
}
\put(-450., 138.){\small detector layers}
\put(-375., 138.){\small {\color{red} {$\bigstar$}} real hit position}
\put(-285., 138.){\small {\color{gray} {$\bigstar$}} reconstructed hit position}
\put(-160., 138.){
\begin{tikzpicture}
\draw[line width=1pt, arrows=->, >=Triangle, dashed] (0,0) -- (0.6,0);
\end{tikzpicture}
}
\put(-135., 138.){\small real trajectory}
\put(-70., 138.){
\begin{tikzpicture}
\draw[line width=1pt, arrows=->, >=Triangle] (0,0) -- (0.6,0);
\end{tikzpicture}
}
\put(-45., 138.){\small fitted track}
\caption{A simplified representation of two common weak modes that bias the track momentum. A \emph{sagitta} bias (left) is caused by a deformation in the bending plane of the tracks, e.g. a rotation of the detector layers depending linearly on the radius.  A \emph{length scale} bias (right) caused by a deformation along the track trajectory, e.g. a radial expansion of the detector layers depending linearly on the radius. The real (dashed black line) and fitted (solid black line) particle trajectories are shown. Red stars indicate real measurement positions and grey stars show the reconstructed hit positions (biased measurements).}
\label{fig:cartoons_weak_modes}
\end{figure}

% The next lines are included from the .//chapters/MomentumBias.tex input file
The alignment corrections described in \Cref{sec:IDalign,,sec:TimeAlign} target misalignments that change the $\chi^2$ of the track fit  in \cref{eq:basicTrackChi2}.
In contrast, correlated geometrical distortions referred to as \emph{weak modes} leave the $\chi^2$ of the fitted tracks virtually unchanged and can systematically bias the reconstructed track parameters.
Momentum biases induced by correlated detector misalignments can generally be classified into two categories:
\begin{itemize}
\item \emph{Sagitta} deformations consist of detector geometry distortions in the bending plane that affect the reconstructed track curvature differently for positively and negatively charged particles (\Cref{fig:cartoons_weak_modes} left).
\item  \emph{Length scale} biases are characterised by detector geometry distortions along the track trajectory and affect the reconstructed curvature identically for positively and negatively charged particles (\Cref{fig:cartoons_weak_modes} right).
\end{itemize}
 
These biases can be mitigated through the use of constraints either on track parameters (\Cref{sec:constraints}) or on alignment parameters (\Cref{sec:alignconst}), or on both simultaneously.
Residual momentum biases, after corrections to the detector alignment have been made, are sufficiently small that they can be accounted for by directly correcting track parameters.

\subsection{Sagitta bias}
\label{sec:sagittabias}
 
Displacements of the reconstructed hits in the bending plane orthogonal to the track path result in a charge-antisymmetric alteration of the track curvature, which is parameterised as
\begin{linenomath}
\begin{equation}
p' = p ~(1+q ~p_{\mathrm{T}} ~\delta_{\mathrm{sagitta}})^{-1} \label{eq:sagittaeffect} \, ,
\end{equation}
\end{linenomath}
where the un-primed quantities correspond to the true values, the primed quantities correspond to the reconstructed values,
$q$ refers to the sign of the electric charge of the particle and
$\delta_{\mathrm{sagitta}}$ is a bias parameter common to all measured momenta and uniquely defines the detector geometry deformation.
 
Two iterative methods are used to determine the sagitta biases.
The first method uses $Z\rightarrow\mu^+\mu^-$ decays.
The second uses the electromagnetic calorimeter as a reference, and utilises the ratio of the measured energy deposited in the calorimeter ($E$)
to the momentum  ($p$) measured by the ID for electrons and positrons.
Both methods allow the detector to be segmented arbitrarily in $\eta$ and $\phi$, allowing the study of localised sagitta biases.
Sagitta biases have, to a great extent, been corrected for during the determination of the alignment constants by adding constraints to the parameters of the tracks used to perform the detector alignment,
as given by \cref{eq:constrained_chi2,eq:alignconst_chi2} and
explained in \Cref{sec:constraints,sec:alignconst},
and also in Refs.~\cite{ATLAS-CONF-2012-141,ATL-SOFT-PUB-2014-003}.
The methods used to calculate the constraints are described below,
and the residual sagitta biases after alignment corrections are shown.
 
\subsubsection{Measuring sagitta biases using $Z \rightarrow \mu^{+} \mu^{-}$ decays}
\label{sec:sagittabias_Zmumu}
 
The invariant mass, $m$, of two highly relativistic opposite-charge particles is given approximately by
\begin{linenomath}
\begin{equation*}
m^2 = 2 p^+ p^- (1-\cos\gamma) \, ,
\end{equation*}
\end{linenomath}
where $p^+$ and $p^-$ are the magnitudes of the momenta of the positively and negatively charged particles,
and $\gamma$ is defined as their opening angle.
In the following, $+$ and $-$ superscripts refer to the properties of the positively and negatively charged muons respectively.
Sagitta biases can be measured 
using any particle (of reasonably narrow width) that decays into pairs of stable particles.
In LHC conditions,
resonances that decay into pairs of muons (such as $J/\psi$, $\Upsilon$ and $Z$) present the advantage that the dimuon signature can be clearly distinguished from the large hadronic background.
For $\delta_{\mathrm{sagitta}}$ studies,
$Z\rightarrow\mu^+\mu^-$ decays are preferred due to the high momentum of the \Zboson decay products.
Data quality selection criteria, summarised in~\Cref{Tab:Selection}, are applied to both the selected muon candidates and the dimuon system.
In total, more than 70~million $Z\rightarrow \mu^{+} \mu^{-}$ candidate events were selected.
 
In general, geometrical distortions that bias sagitta measurements can be localised in specific regions of the detector.
As a result, the sagitta bias parameter explicitly depends on the path of the track, which can be approximated by the direction of the track at the $pp$ interaction point, given by $\eta$ and $\phi$: $\delta_{\mathrm{sagitta}} \to \delta_{\mathrm{sagitta}} (\eta,\phi)$.
The difference at leading order in $\delta_{\mathrm{sagitta}} (\eta,\phi)$ between the reconstructed dimuon invariant mass using the uncorrected geometry ($m_{\mu\mu}$) and the expected mass ($m_{Z}$) for each event is given by:
\begin{linenomath}
\begin{equation}
m_{\mu\mu}^{2}- m^{2}_{Z}  \approx m^{2}_{Z}~\left( p'^{+}_{\mathrm{T}} ~\delta_{\mathrm{sagitta}} (\eta^+, \phi^+) -
p'^{-}_{\mathrm{T}}~ \delta_{\mathrm{sagitta}}(\eta^-, \phi^-) \right) \, .
\label{eq:MassCurl}
\end{equation}
\end{linenomath}
where $m^{2}_{Z}$ is a reference mass (in this case the world average mass for the Z boson~\cite{PDG2018}).
 
An iterative procedure is used to determine $\delta_{\mathrm{sagitta}} (\eta,\phi)$.
For the $i$-th iteration, $\delta_{\mathrm{sagitta},i}(\eta,\phi)$ is computed for every muon in the $Z\rightarrow\mu^+\mu^-$ sample with:
\begin{linenomath}
\begin{equation}
\Delta \delta_{\mathrm{sagitta},i} (\eta,\phi)  =  -q~\frac{m_{\mu\mu}^{2}- m^{2}_{Z}}{2~m^{2}_{Z}}~
\frac{\left(1+~q~p'_{\mathrm{T}} ~ \langle\delta_{\mathrm{sagitta},{i-1}} (\eta,\phi)\rangle\right)}{\pt'},
\label{eq:deltaSagittaIter}
\end{equation}
\end{linenomath}
where $\langle\delta_{\mathrm{sagitta},{i-1}} (\eta,\phi)\rangle$ is the result of the previous iteration.
The corrections are updated by adding the average of the current iteration to the result of the previous iteration:
\begin{linenomath}
\begin{equation}
\langle\delta_{\mathrm{sagitta},{i}} (\eta,\phi)\rangle =  \langle\delta_{\mathrm{sagitta},{i-1}} (\eta,\phi)\rangle +  \langle\Delta \delta_{\mathrm{sagitta},i} (\eta,\phi) \rangle,
\end{equation}
\end{linenomath}
where  $\langle\Delta \delta_{\mathrm{sagitta},i} (\eta,\phi) \rangle $ is the average bias in a $(\eta,\phi)$ region.
The value of $m_{\mu\mu}^{2}$ is computed using \cref{eq:MassCurl} with the values of $\delta_{\mathrm{sagitta}}$ from the previous iteration.
The iterations are repeated until convergence is reached.
 
The method, as described by \cref{eq:MassCurl}, is only sensitive to relative sagitta biases in different sectors of the detector.
An alternative method, comparing the $\pt$ spectrum of the $\mu^+$ and $\mu^-$ ~\cite{CMS-TRK-11-002} was also tested.
This method is sensitive to global sagitta biases,
although it is also subject to detector acceptance effects and requires more data to achieve the same statistical precision as the mass-based method.
 
\begin{table}[t!]
\begin{center}
\caption{Event selection criteria for $Z\rightarrow \mu^{+} \mu^{-}$ (\Cref{sec:sagittabias_Zmumu}) and $Z\rightarrow e^{+} e^{-}$ (\Cref{sec:sagittabias_ep}) candidate events for the analyses of the sagitta biases in data. Events triggered by the lowest-threshold unprescaled single and double electron and muon triggers are used to select $Z\rightarrow \mu^{+} \mu^{-}$ and $Z\rightarrow e^{+} e^{-}$ event candidates. $\gamma(\mu^{+}, \mu^{-})$ is the opening angle between the muons. $\Delta d_0(\mu^{+}, \mu^{-})$  and $\Delta z_0(\mu^{+}, \mu^{-})$ are defined as the difference in $d_0$ and $z_0$ between the two muons.}
\begin{tabular}{l | c | c }
\toprule
Selection criteria & $Z\rightarrow \mu^{+} \mu^{-}$ & $Z\rightarrow e^{+} e^{-}$ \\
\midrule
Lepton selection	&   Two muons$^{1}$ associated                                                                          	& Two electrons$^{5}$ associated   		 \\
& with the primary vertex$^{2}$						                                        & with the primary vertex$^{2}$  		 \\
&   $p_{\mathrm{T}} > \SI{12}{\GeV}$   					& $E_\mathrm{T} > \SI{25}{\GeV}$ 			  \\
&   $|\eta|<2.5$		& $\abs{\eta} < 2.47 \,^{3}$   		 \\
&   $d_0$ significance$^{4}$ < 4&\\
\midrule
Dilepton selection &   $\SI{70}{\GeV} < m_{\mu^{+} \mu^{-}}< \SI{110}{\GeV}$ 		& $\abs{m_{e^{+}e^{-}} - m_{Z}} < \SI{30}{\GeV}$   \\
&   $|\Delta d_0(\mu^{+}, \mu^{-})|< \SI{0.1}{mm}$   		&     			  \\
&   $|\Delta z_0(\mu^{+}, \mu^{-})|< \SI{0.6}{mm}$   		&     			  \\
&   $\gamma(\mu^{+}, \mu^{-})> 45^\circ$		   		&     			  \\
\bottomrule
\multicolumn{3}{l}{$^{1}$\footnotesize{Both muon candidates are required to satisfy the `medium' quality criteria as defined in Ref.~\cite{atlas_MCPpaper}.}} \\
\multicolumn{3}{l}{$^{2}$\footnotesize{The reconstructed vertex with the largest $\sum p_{\mathrm{T}}^{2}$ of its tracks.}}\\
\multicolumn{3}{l}{$^{3}$\footnotesize{Excluding the transition region between the barrel and forward calorimeters.}}\\
\multicolumn{3}{l}{$^{4}$\footnotesize{The significance is defined as $|d_0|/\sigma(d_0)$, where $\sigma(d_0)$ is the uncertainty on the $d_0$ from the track fit.}} \\
\multicolumn{3}{l}{$^{5}$\footnotesize{Both electron candidates are required to satisfy the `loose' quality criteria as defined in Ref.~\cite{PERF-2017-01}.}} \\
\end{tabular}
\label{Tab:Selection}
\end{center}
\end{table}
 
\begin{figure}
\centering
\begin{subfigure}[c]{.49\textwidth}
\includegraphics[width=\textwidth]{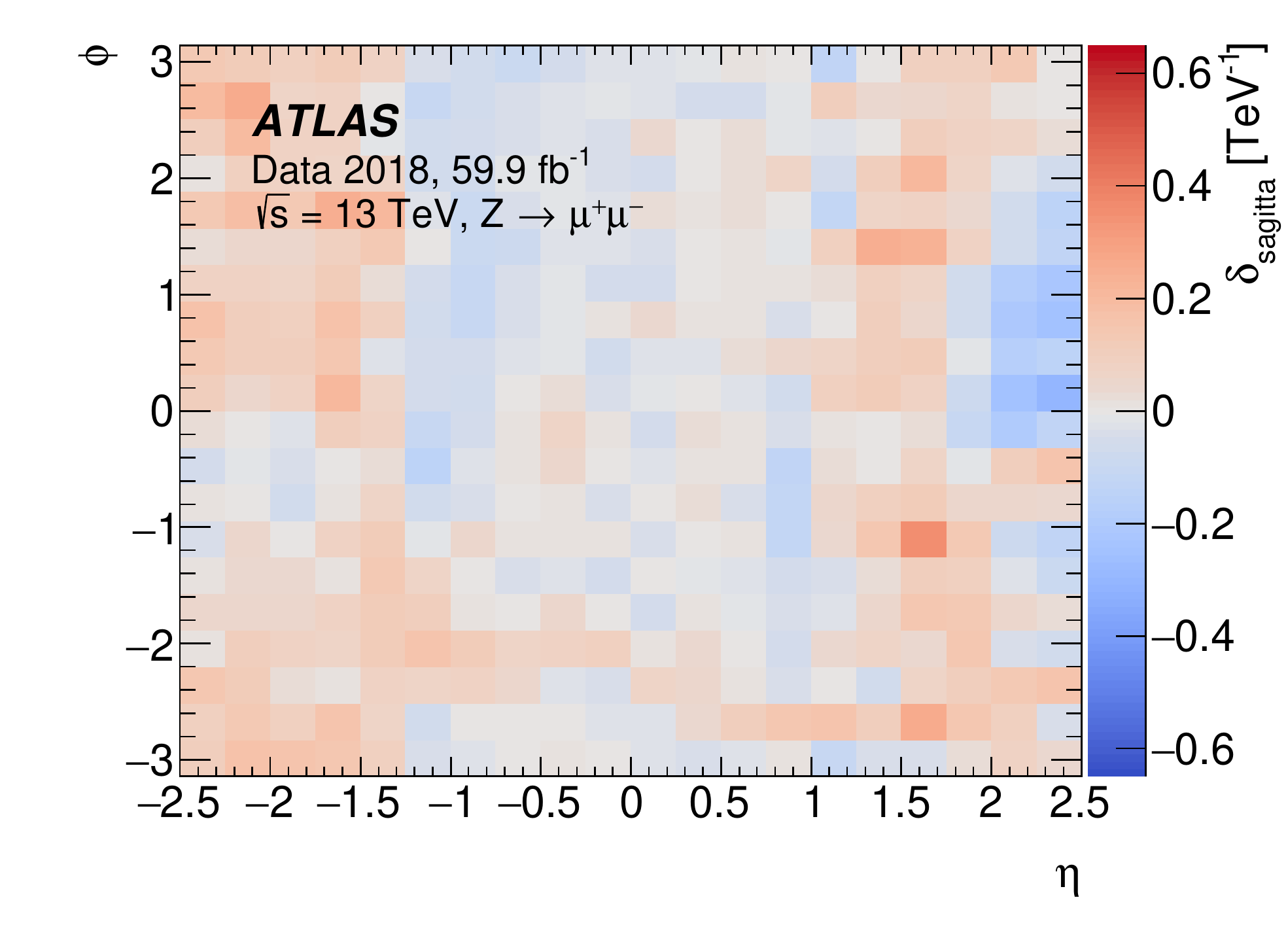}
\label{fig:SagittaVsEtaPhi}
\end{subfigure}
\begin{subfigure}[c]{.49\textwidth}
\includegraphics[width=\textwidth]{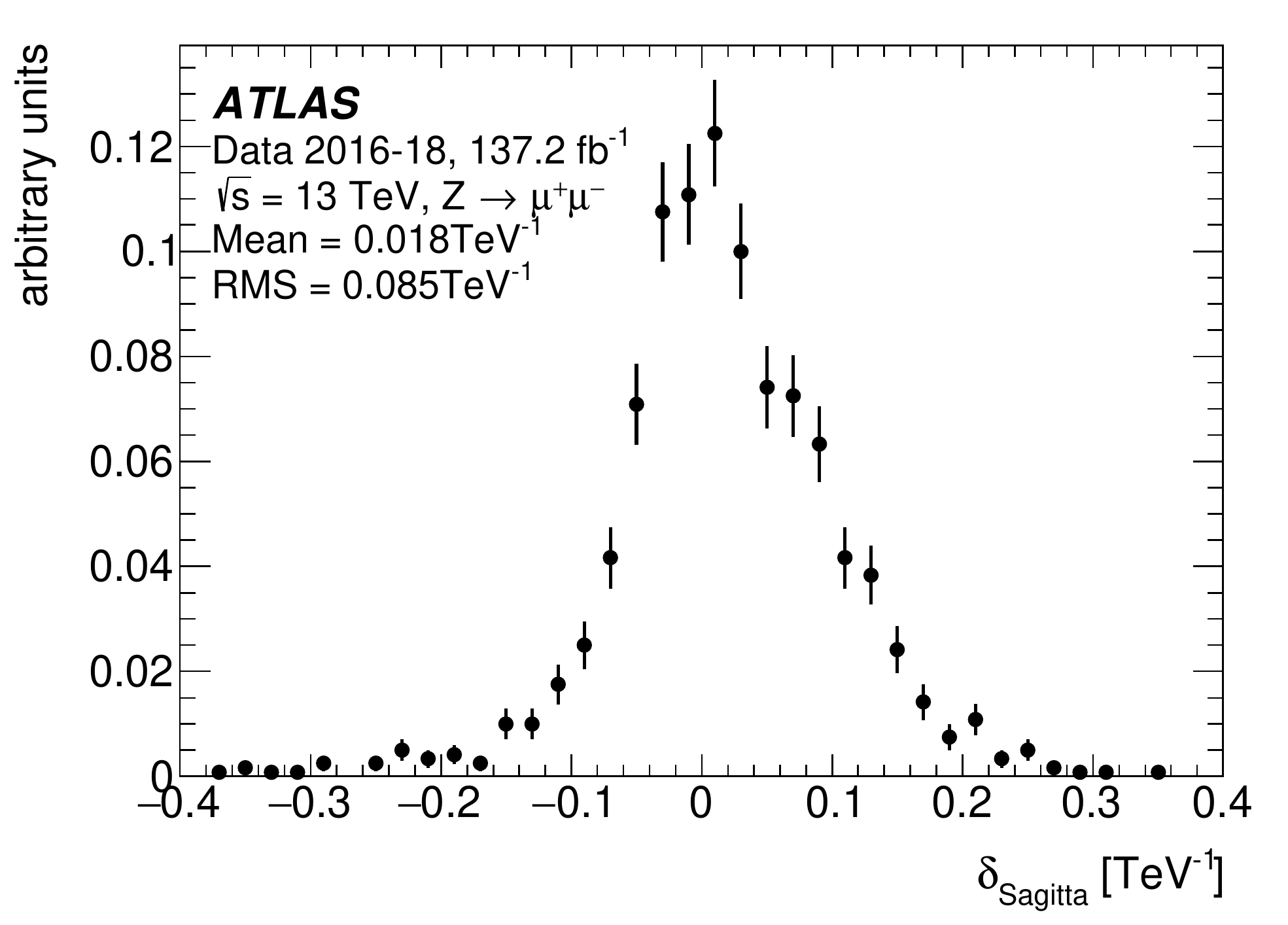}
\label{fig:SagittaOverall}
\end{subfigure}
\caption{Sagitta biases versus $\eta$ and $\phi$ (left) for 2018 data and the overall sagitta biases
in the Run 2 data (right) for the $Z\rightarrow\mu^+\mu^-$ method.
The error bars represent the statistical uncertainty.}
\label{fig:zmumusagitta2018}
\end{figure}

\begin{figure}
\centering
\begin{subfigure}[c]{.49\textwidth}
\includegraphics[width=\textwidth]{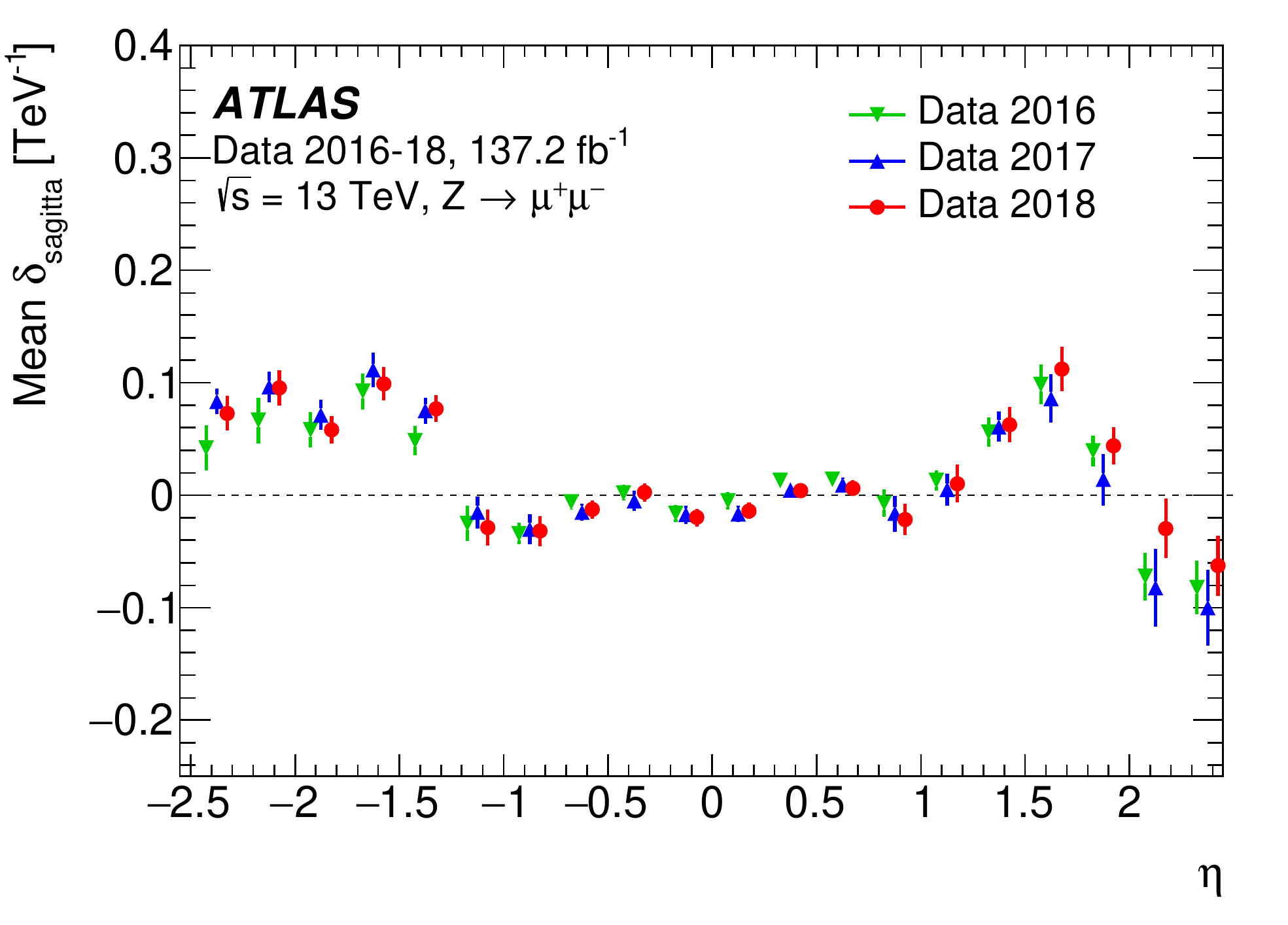}
\label{fig:SagittaVsEtaMean}
\end{subfigure}
\begin{subfigure}[c]{.49\textwidth}
\includegraphics[width=\textwidth]{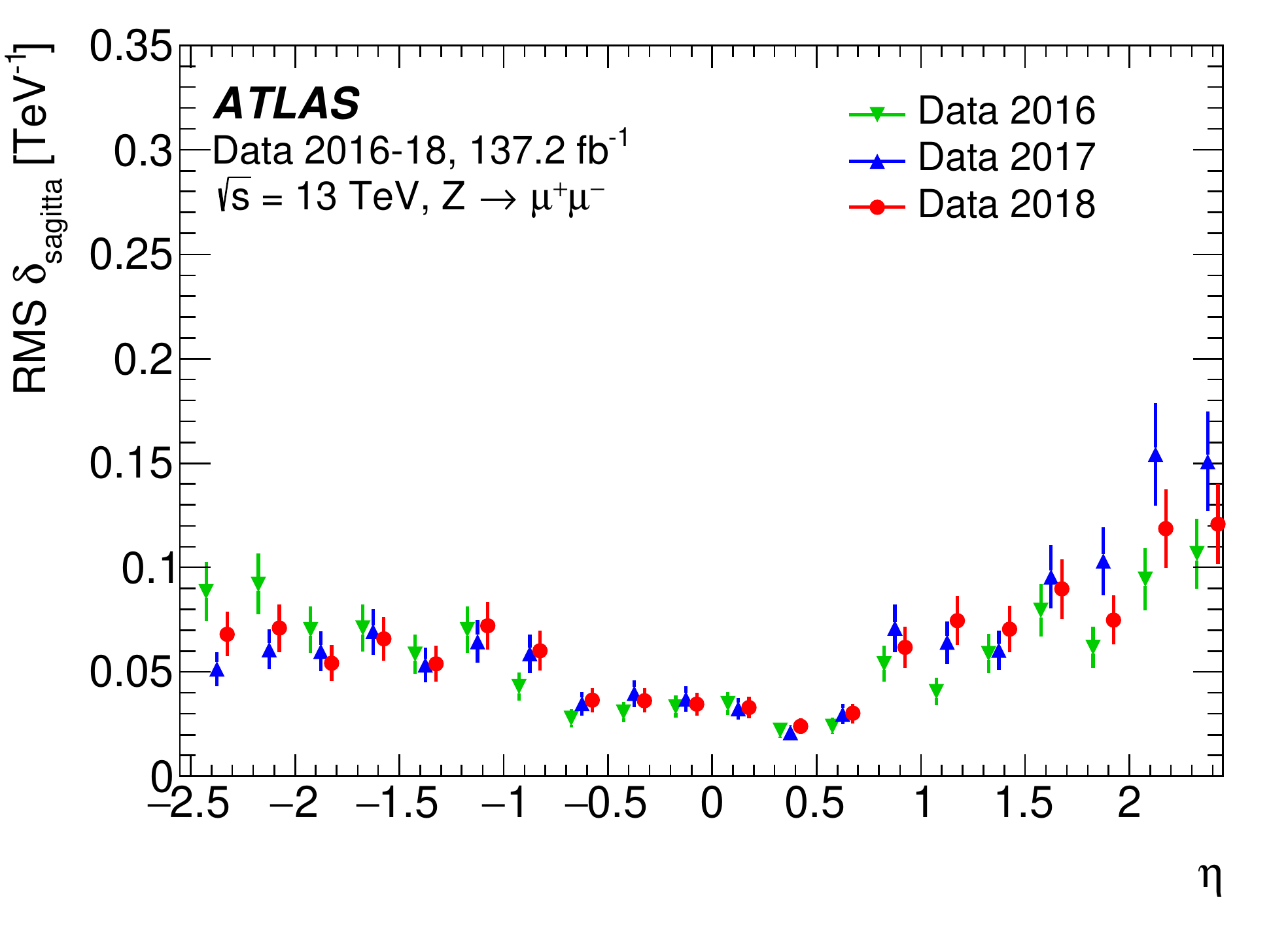}
\label{fig:SagittaVsEtaRMS}
\end{subfigure}\\
\begin{subfigure}[c]{.49\textwidth}
\includegraphics[width=\textwidth]{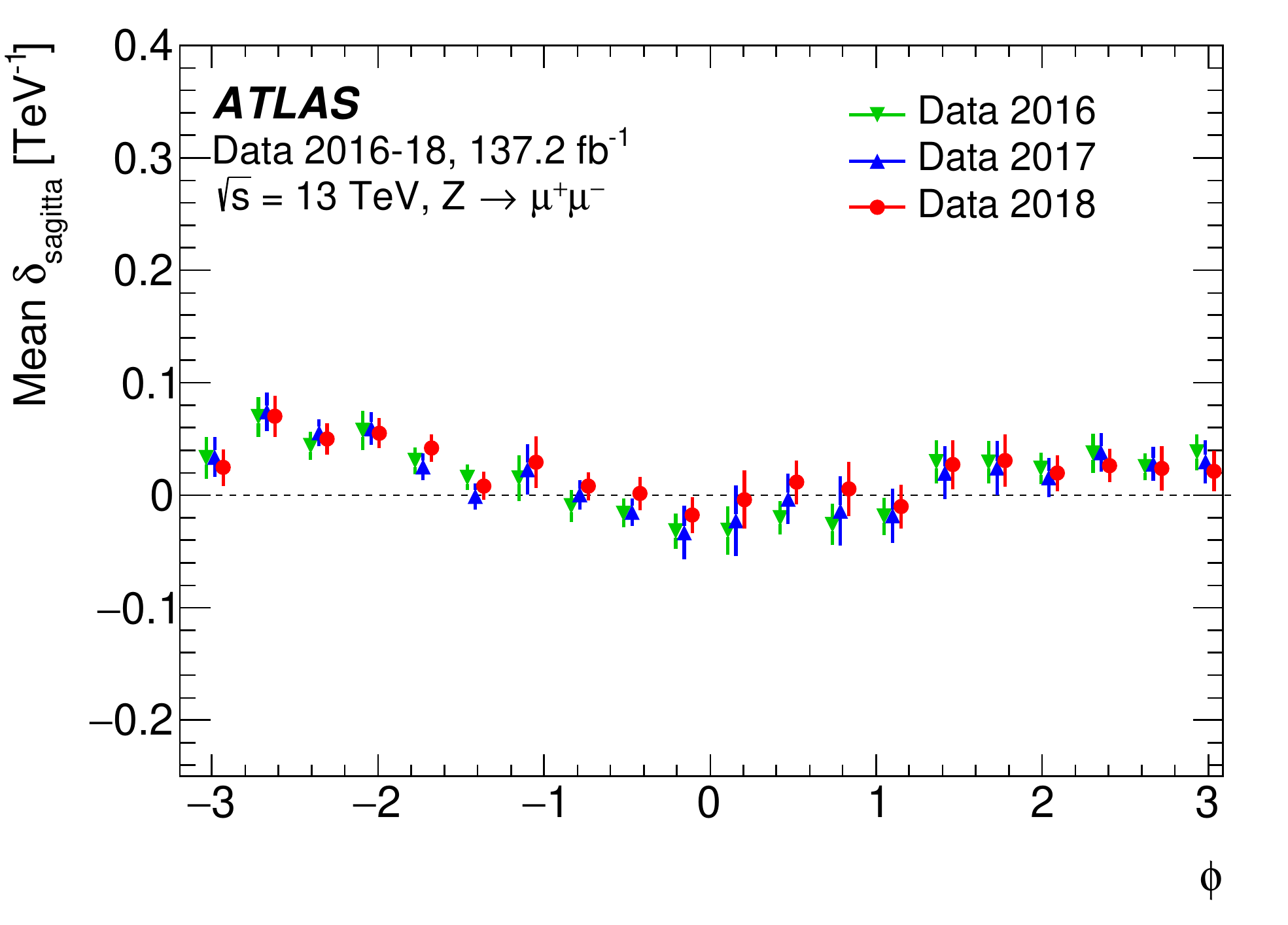}
\label{fig:SagittaVsPhiMean}
\end{subfigure}
\begin{subfigure}[c]{.49\textwidth}
\includegraphics[width=\textwidth]{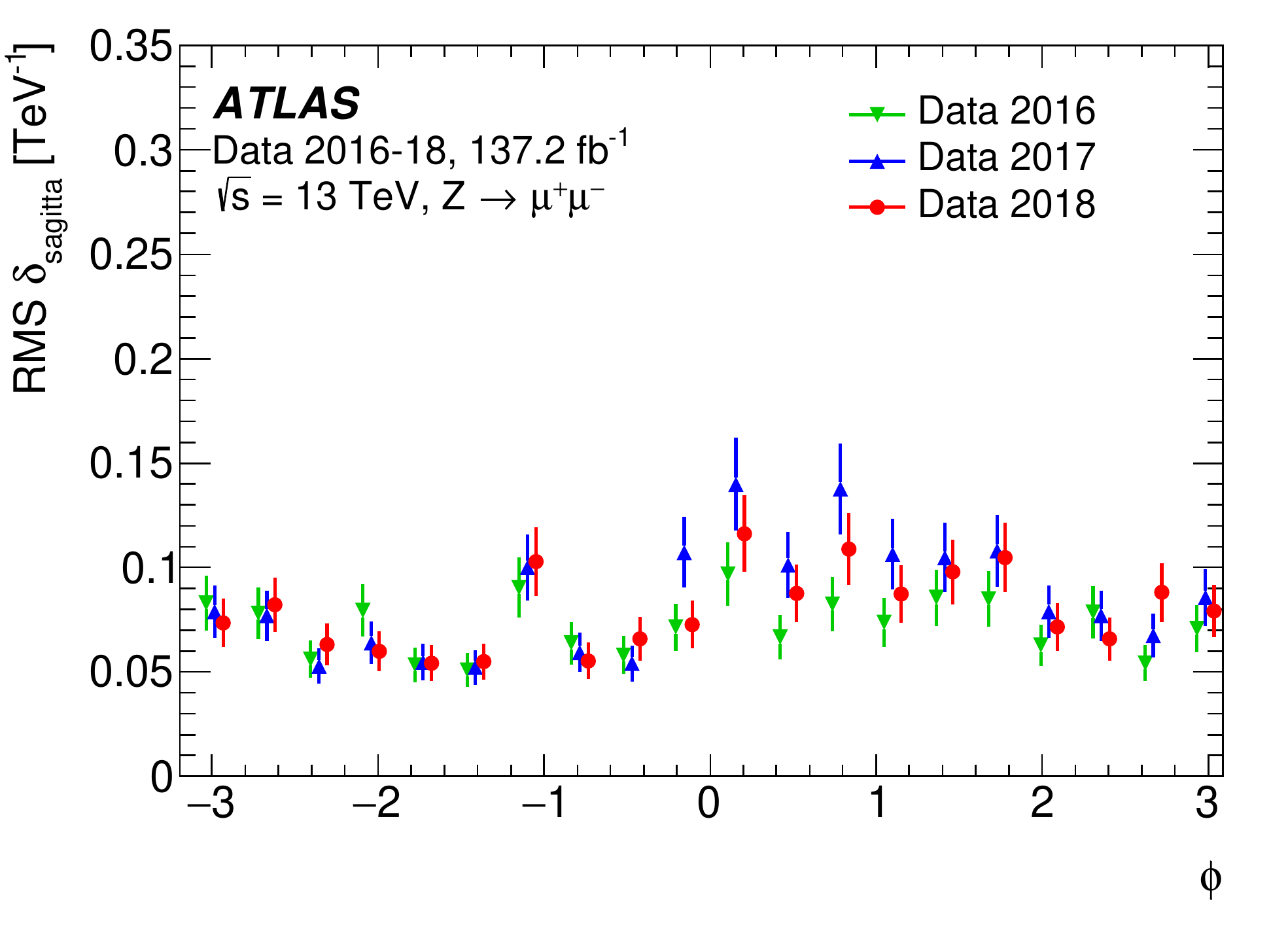}
\label{fig:SagittaVsPhiRMS}
\end{subfigure}
\caption{Measured sagitta biases versus $\eta$ (top) and $\phi$ (bottom) using the $Z\rightarrow\mu^+\mu^-$ method. The average (left) and the RMS (right) of the sagitta bias is shown.
The markers of the data points of the different years are slightly shifted in $\eta$ and $\phi$ for better visibility.
The error bars represent the statistical uncertainty.}
\label{fig:SagittaVsEtaAndVsPhi}
\end{figure}

\Cref{fig:zmumusagitta2018} shows the measured sagitta distortions depending on the track direction using this technique. The central barrel region of the detector is largely free of sagitta bias, while the endcap regions exhibit some areas of small residual sagitta bias.
The distribution of $\delta_{\mathrm{sagitta}}$ for the full Run 2 data is shown in \Cref{fig:zmumusagitta2018}.
\Cref{fig:SagittaVsEtaAndVsPhi} shows the average $\delta_{\mathrm{sagitta}}$ versus $\eta$ and $\phi$, as well as its RMS.
The distributions, split by data-taking year, have compatible shapes indicating a consistent and stable detector geometry during Run 2.
 
\subsubsection{Measuring sagitta biases using the $E/p$ ratio of electrons and positrons}
\label{sec:sagittabias_ep}
 
Assuming that the calorimeter response is independent of the charge of the incoming particle
and that a perfectly aligned detector reconstructs the momentum of charged particles correctly,
charge-dependent momentum biases are expected to result in differences in the $E/p$ ratio of positive and negative particles.
This ratio is defined as the ratio of the calorimeter energy measurement ($E$) to the track momentum measurement ($p$).
This technique is mainly suitable for electrons and positrons.
In the presence of a sagitta bias,
the $\langle E/p \rangle$ ratio would be modified as $\langle E/p' \rangle =  \langle E/p \rangle + q ~\langle E_\mathrm{T} \rangle ~\delta_{\mathrm{sagitta}}$,
where $E_\mathrm{T}  \equiv E/\cosh{\eta}$ is referred to as the transverse energy.
Assuming that the average transverse momentum of positrons and electrons is equal, the sagitta bias can be estimated~\cite{ATLAS-CONF-2012-141} as
\begin{linenomath}
\begin{equation*}
\delta_{\mathrm{sagitta}} = \frac{ \left \langle E/p' \right \rangle^{+} - \left \langle E/p' \right \rangle^{-} }{2 \left \langle E_\mathrm{T}\right \rangle }.
\end{equation*}
\end{linenomath}
To take into account any biases introduced by the aforementioned assumptions the value of $\delta_{\mathrm{sagitta}}$ is determined iteratively, correcting the momentum using \cref{eq:sagittaeffect} at each iteration.
It should be noted that biases in the calorimeter energy scale cancel out to first order and any residual dependence would be reduced by this iterative procedure.
In addition, this method is, by construction, sensitive to global sagitta biases.
Data quality selection criteria are applied to both the selected electron candidates and the electron--positron system and are summarised in~\Cref{Tab:Selection}.
 
\begin{figure}
\centering
\begin{subfigure}[c]{.49\textwidth}
\includegraphics[width=\textwidth]{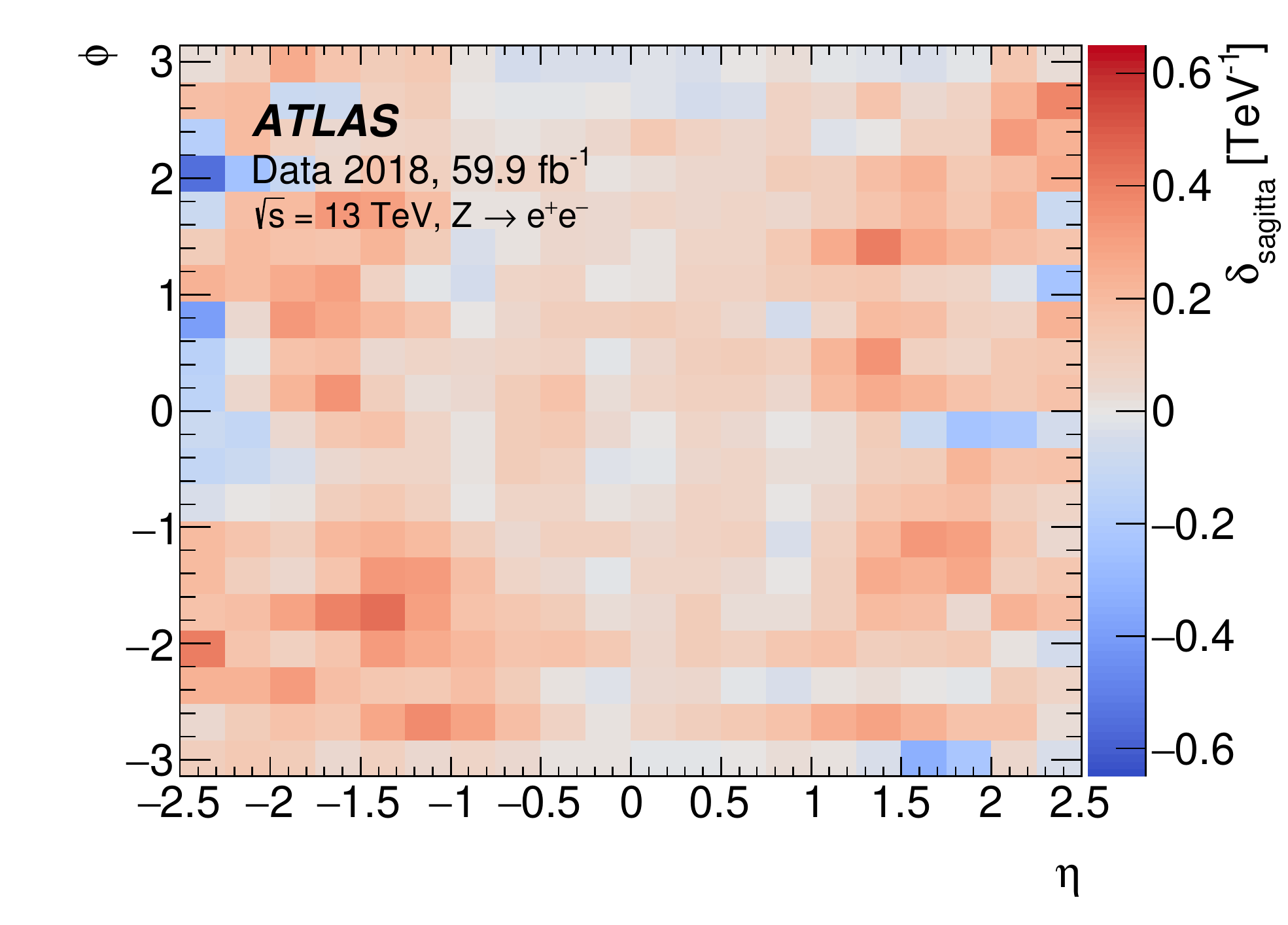}
\label{fig:SagittaEPVsEtaPhi}
\end{subfigure}
\begin{subfigure}[c]{.49\textwidth}
\includegraphics[width=\textwidth]{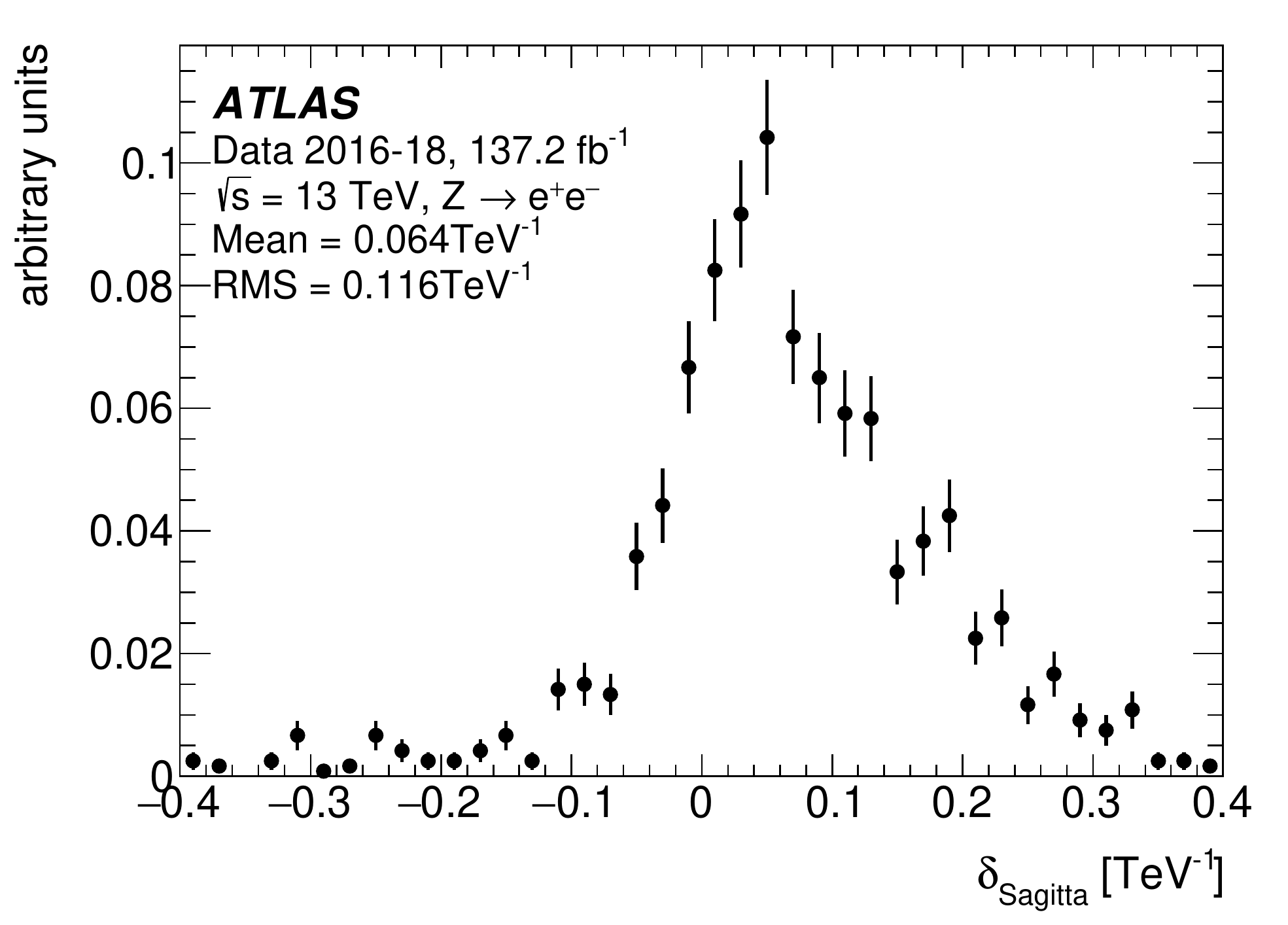}
\label{fig:SagittaEPOverall}
\end{subfigure}
\caption{Sagitta biases versus $\eta$ and $\phi$ (left) for 2018 data and the overall sagitta biases
of the Run 2 data (right) for the $E/p$ method.
The error bars represent the statistical uncertainty.}
\label{fig:zeesagitta2018}
\end{figure}

\begin{figure}
\centering
\begin{subfigure}[c]{.49\textwidth}
\includegraphics[width=\textwidth]{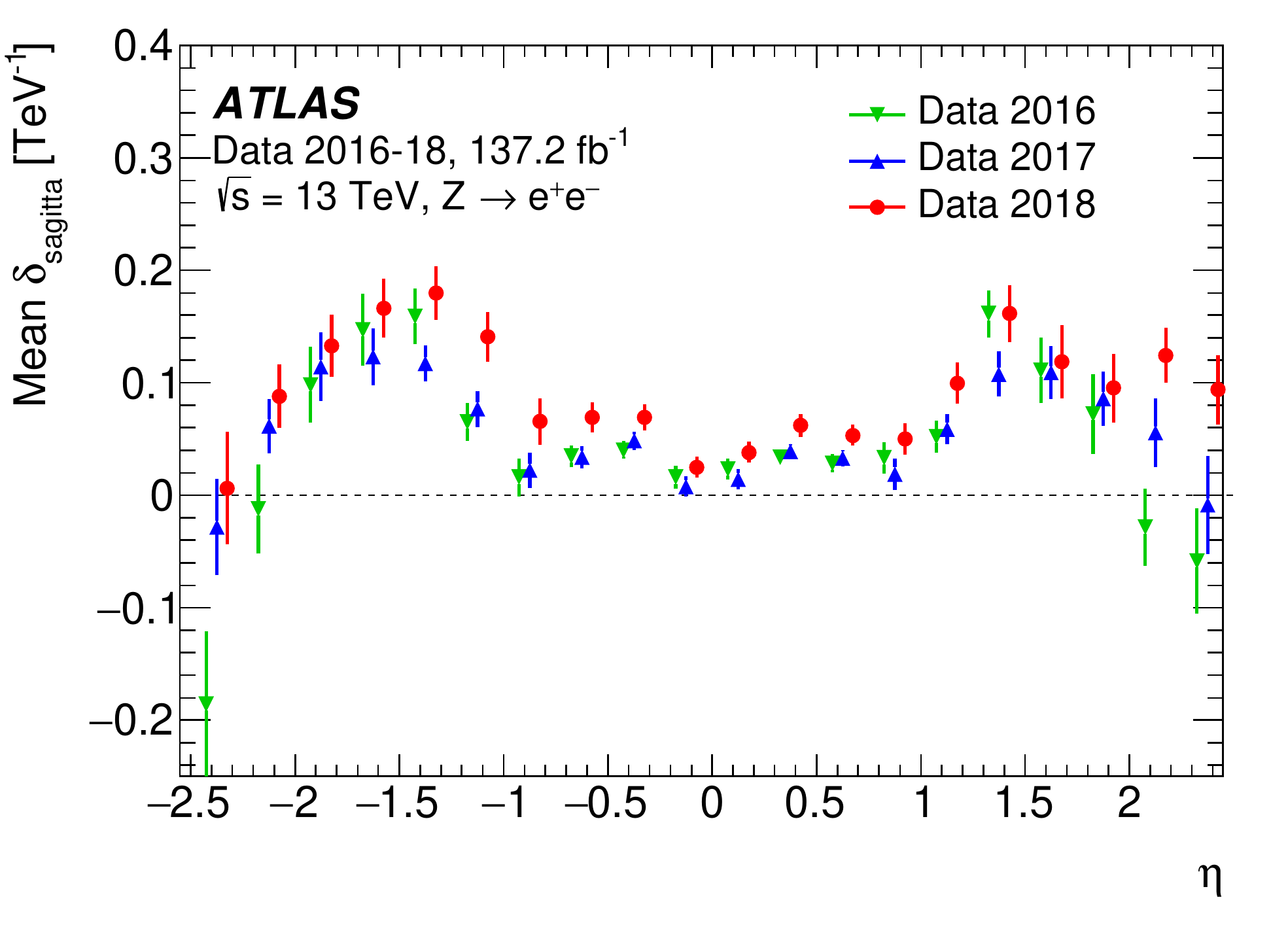}
\label{fig:SagittaEPVsEtaMean}
\end{subfigure}
\begin{subfigure}[c]{.49\textwidth}
\includegraphics[width=\textwidth]{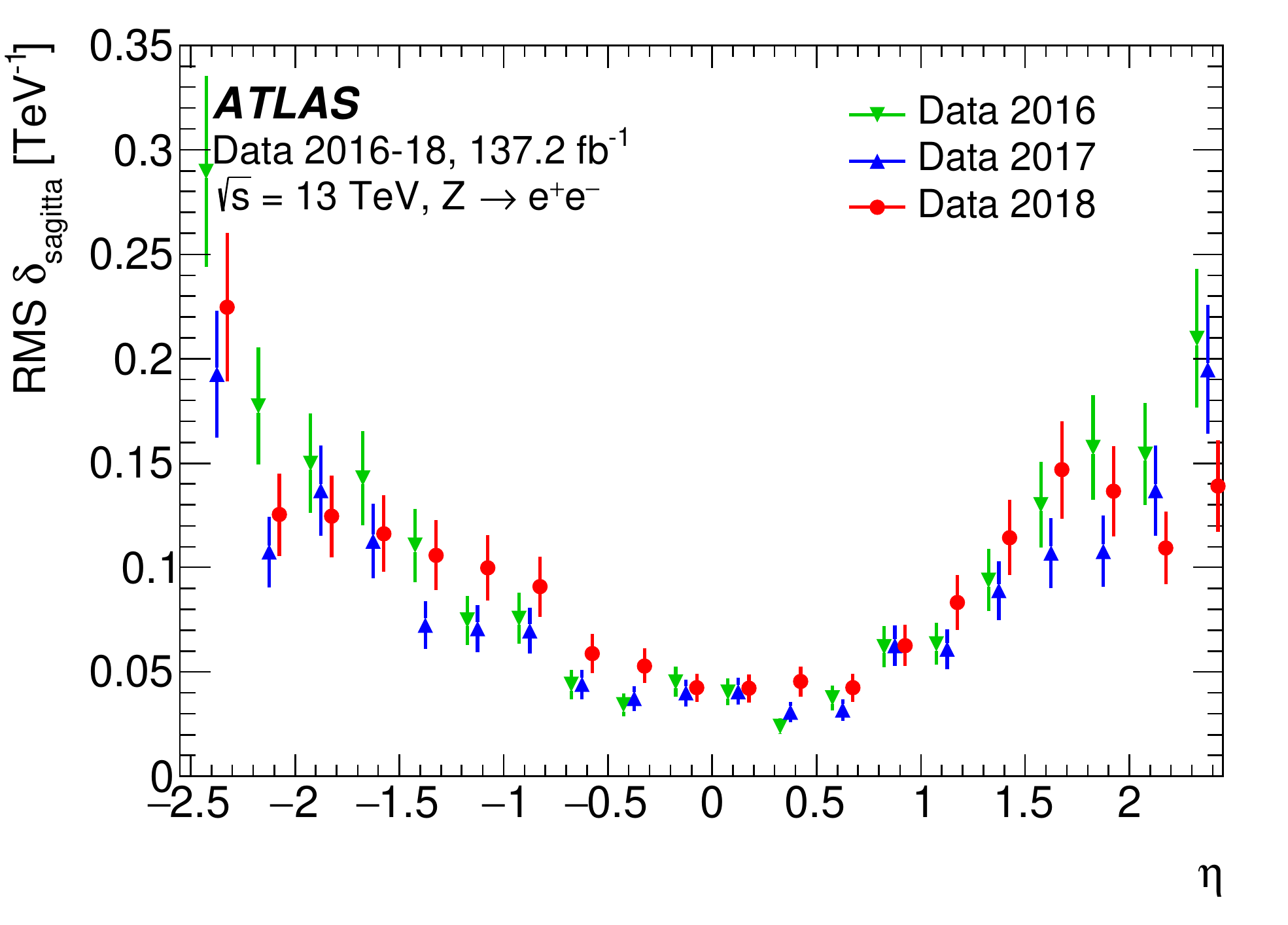}
\label{fig:SagittaEPVsEtaRMS}
\end{subfigure}\\
\begin{subfigure}[c]{.49\textwidth}
\includegraphics[width=\textwidth]{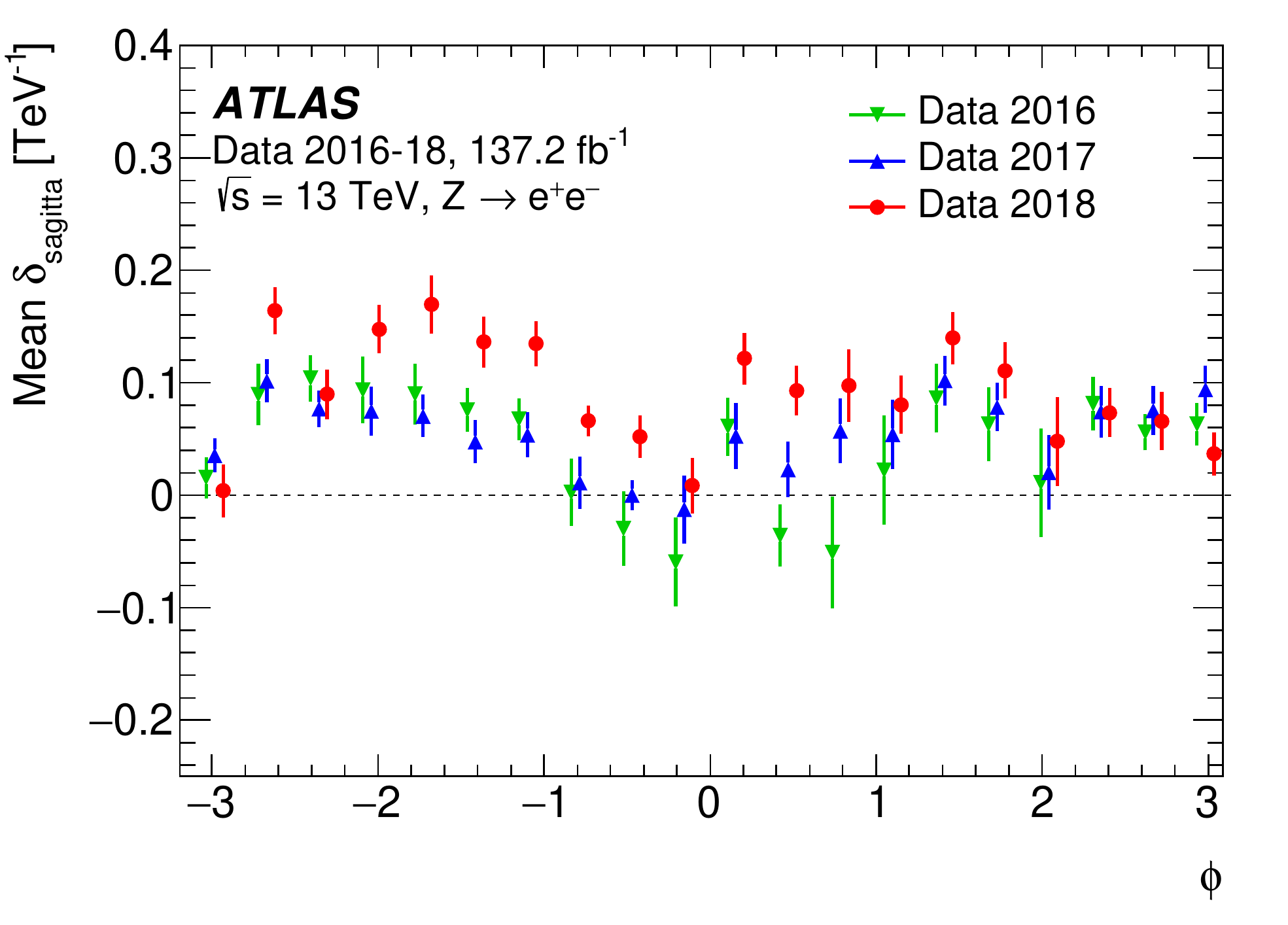}
\label{fig:SagittaEPVsPhiMean}
\end{subfigure}
\begin{subfigure}[c]{.49\textwidth}
\includegraphics[width=\textwidth]{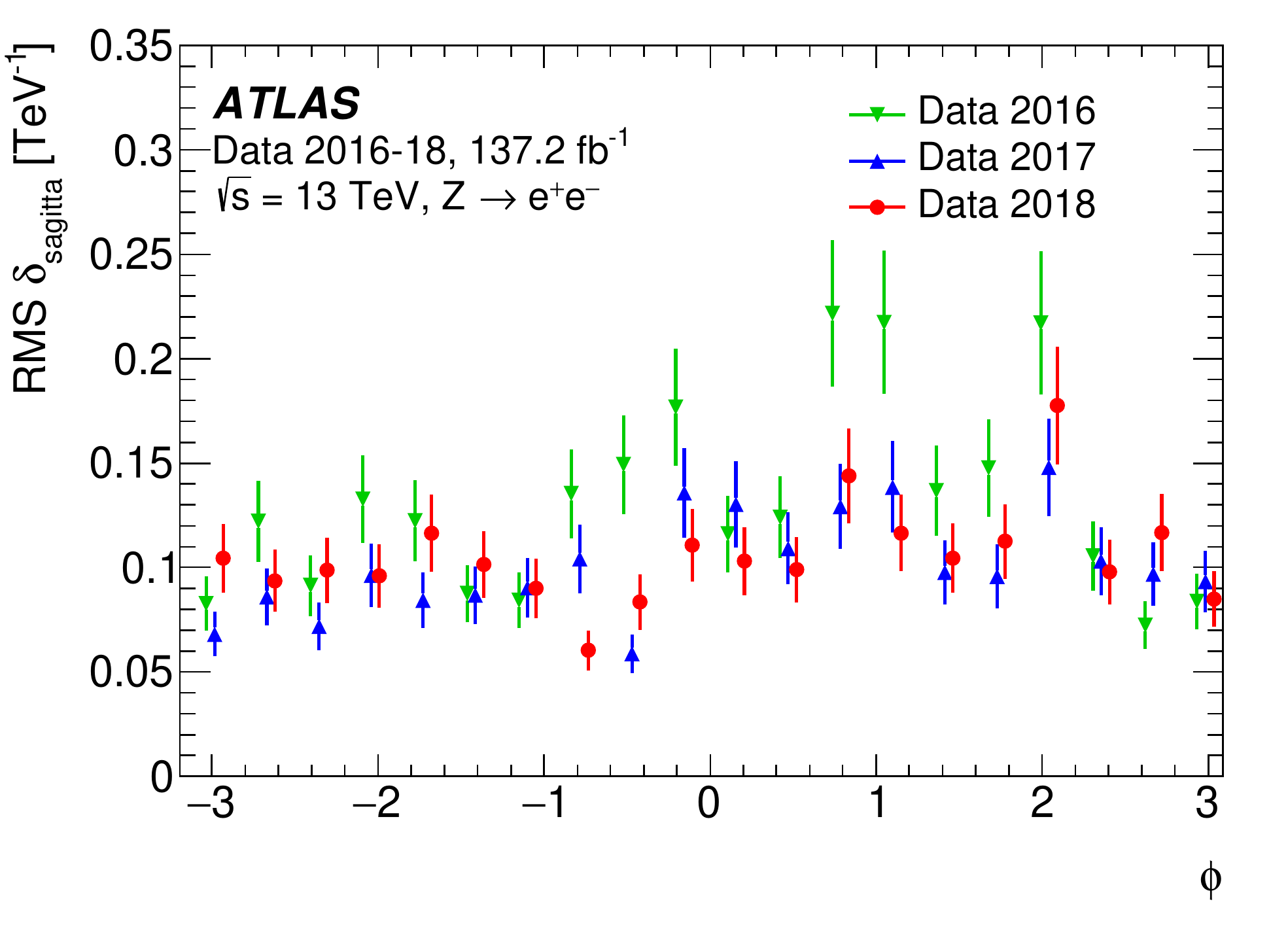}
\label{fig:SagittaEPVsPhiRMS}
\end{subfigure}
\caption{Sagitta biases versus $\eta$ (top) and $\phi$ (bottom) for the $E/p$ method. The average (left) and the RMS (right) of the sagitta bias is shown.
The markers of the data points of the different years are slightly shifted in $\eta$ and $\phi$ for better visibility.
The error bars represent the statistical uncertainty.}
\label{fig:SagittaEPVsEtaAndVsPhi}
\end{figure}

\Cref{fig:zeesagitta2018} shows the $\delta_{\mathrm{sagitta}}$ as obtained from the $E/p$ method.
These results support the observations from \Cref{fig:SagittaVsEtaPhi}: the central barrel region of the detector is largely free of sagitta bias, while the endcap regions exhibit regions of small residual sagitta bias.
Compared to \Cref{fig:zmumusagitta2018}~(right) a global offset of $\sim$\SI{0.05}{\TeV^{-1}} can be seen in \Cref{fig:zeesagitta2018}~(right) indicating the presence of a small global sagitta bias.
\Cref{fig:SagittaEPVsEtaAndVsPhi} shows the average $\delta_{\mathrm{sagitta}}$ versus $\eta$ and $\phi$, as well as its RMS.
The $\delta_{\mathrm{sagitta}}$ distributions from the $E/p$ method split by data-taking year have comparable shape to those obtained from the $Z\rightarrow\mu^+\mu^-$ mass method, further supporting the observation of a consistent and stable detector geometry during Run 2.
The change in position due to the residual sagitta bias ($\sim$\SI{0.1}{\TeV^{-1}}) when extrapolating a track from the detector origin to the outermost SCT endcap disk (radius of 500\,mm and a $z$-axis distance of 2720\,mm from the detector origin) is less than \SI{10}{\micro\metre}.
 
\subsection{Length scale biases\label{sec:RadialDistortion}}
 
Displacements of the reconstructed hits parallel to the track direction result in a charge-symmetric alteration of the measured track curvature.
In a tracker with a solenoidal magnetic field these can be induced by changes in the radial or longitudinal length scale of the detector with little impact on the track fit quality.
 
If the actual radius of a detector module, $R$, is assumed to be $R~(1+\varepsilon_{r})$,
then for small distortions ($|\varepsilon_{r}| \ll 1$), the reconstructed momentum will be:
\begin{linenomath}
\begin{equation}\label{eq:radialdisteffect}
\begin{split}
p'_{\mathrm{T}} & = p_{\mathrm{T}} \left(1+ \varepsilon_{r}\right) \\
p'_{\mathrm{z}} & = p_{\mathrm{z}} \, .
\end{split}
\end{equation}
\end{linenomath}
 
\Cref{eq:radialdisteffect} assumes that the length scale in the bending plane also expands by a factor $(1+\varepsilon_{r})$, which implies that dimensions of sensitive detector modules would also expand by the same factor.
If it is assumed that detector modules do not expand in the bending plane then the reconstructed transverse momentum will be biased by a factor of $(1+2\varepsilon_{r})$.  
 
Similarly, if the actual longitudinal dimension of a detector module, $z$, is assumed to be $z~(1+\varepsilon_{z})$,
the reconstructed momentum will be :
\begin{linenomath}
\begin{equation}\label{eq:longitudinaldisteffect}
\begin{split}
p'_{\mathrm{T}} & = p_{\mathrm{T}} \\
p'_{\mathrm{z}} & = p_{\mathrm{z}} \left(1+ \varepsilon_{z}\right)\, .
\end{split}
\end{equation}
\end{linenomath}
 
Linear combinations of both the radial and longitudinal biases are also considered.
It is worth noting that there is a degeneracy between the effects of a bias in the magnetic field
and a global scaling of the detector (radial and longitudinal: $\varepsilon_{s}$),
as both lead to a momentum bias of the form $\pmb{p}~(1+ \varepsilon_{s})$.
Consequently, if $B$ is assumed to be $B~(1+\varepsilon_s)$ the particle momentum scales as $\pmb{p}~(1+ \varepsilon_{s})$.
 
The relationship between the reconstructed invariant mass of a particle decaying into two muons
(${m'}_{\mu\mu}$),
and the true mass ($m_{\mu\mu}$),
assuming that the radial and longitudinal biases in \cref{eq:radialdisteffect,eq:longitudinaldisteffect} are both small, is given by:
 
\begin{linenomath}
\begin{equation*}
\begin{split}
{m'}_{\mu\mu}^2 \approx m_{\mu\mu}^2 &+ 2 E^+E^- \left[ \left(\beta_{\mathrm{T}}^+ \right)^2 - \pmb{\beta}_{\mathrm{T}}^+ \cdot \pmb{\beta}_{\mathrm{T}}^- \right] \varepsilon_{r}(\eta^+,\phi^+) \\
&+ 2 E^+E^- \left[ \left(\beta_{\mathrm{T}}^- \right)^2 - \pmb{\beta}_{\mathrm{T}}^+ \cdot \pmb{\beta}_{\mathrm{T}}^- \right] \varepsilon_{r}(\eta^-,\phi^-) \\
&+ 2 E^+E^- \left[ \left(\beta_{\mathrm{z}}^+ \right)^2 - \pmb{\beta}_{\mathrm{z}}^+ \cdot \pmb{\beta}_{\mathrm{z}}^- \right] \varepsilon_{z}(\eta^+,\phi^+) \\
&+ 2 E^+E^- \left[ \left(\beta_{\mathrm{z}}^- \right)^2 - \pmb{\beta}_{\mathrm{z}}^+ \cdot \pmb{\beta}_{\mathrm{z}}^-\right] \varepsilon_{z}(\eta^-,\phi^-) \, ,
\end{split}
\end{equation*}
\end{linenomath}
where the $\pmb{\beta} = \pmb{p}/E$ is the velocity of the particle.
This approximation is valid to first order in $\varepsilon$.

In a simpler case, where only a global radial and longitudinal bias are present, the reconstructed mass is:
\begin{linenomath}
\begin{equation*}
\begin{split}
{m'}_{\mu\mu}^2 \approx m_{\mu\mu}^2 &+ 2 E^+E^- \left[ \pmb{\beta}_{\mathrm{T}}^+ - \pmb{\beta}_{\mathrm{T}}^- \right]^2  \varepsilon_{r} \\
&+ 2 E^+E^- \left[  \pmb{\beta}_{\mathrm{z}}^+ - \pmb{\beta}_{\mathrm{z}}^- \right]^2  \varepsilon_{z} \,,
\end{split}
\end{equation*}
\end{linenomath}
which, in the limit where the muon mass is ignored leads to 
\begin{eqnarray*}
{m'}_{\mu\mu}^2 & \approx  m_{\mu\mu}^2 + 2 m_{\mu\mu}^2\varepsilon_{r}~\sin^2\alpha + 2 m_{\mu\mu}^2~\varepsilon_{z}~\cos^2\alpha  \\
& \approx m_{\mu\mu}^2 + 2 m_{\mu\mu}^2 \left(  \varepsilon_{s}\ + \varepsilon_{r'}~\sin^2\alpha \right) \, ,
\end{eqnarray*}
where
\begin{equation}
\sin^2\alpha = E^+E^- \left[ \pmb{\beta}_{\mathrm{T}}^+ - \pmb{\beta}_{\mathrm{T}}^- \right]^2 /m_{\mu\mu}^2\,,
\label{eq:def_sin2alpha}
\end{equation}
$\varepsilon_{s} = \varepsilon_{z}$, and $\varepsilon_{r'} = \varepsilon_{r}-\varepsilon_{z}$ is the difference
between the radial and longitudinal components of the momentum scale.\footnote{For the massless case, defining $\cos^2\alpha = E^+E^- \left[ \pmb{\beta}_{\mathrm{z}}^+ - \pmb{\beta}_{\mathrm{z}}^- \right]^2 /m_{\mu\mu}$ one obtains $\sin^2\alpha + \cos^2\alpha = 1$.}
 
\begin{figure}[ht!]
\centering
\includegraphics[width=0.50\textwidth]{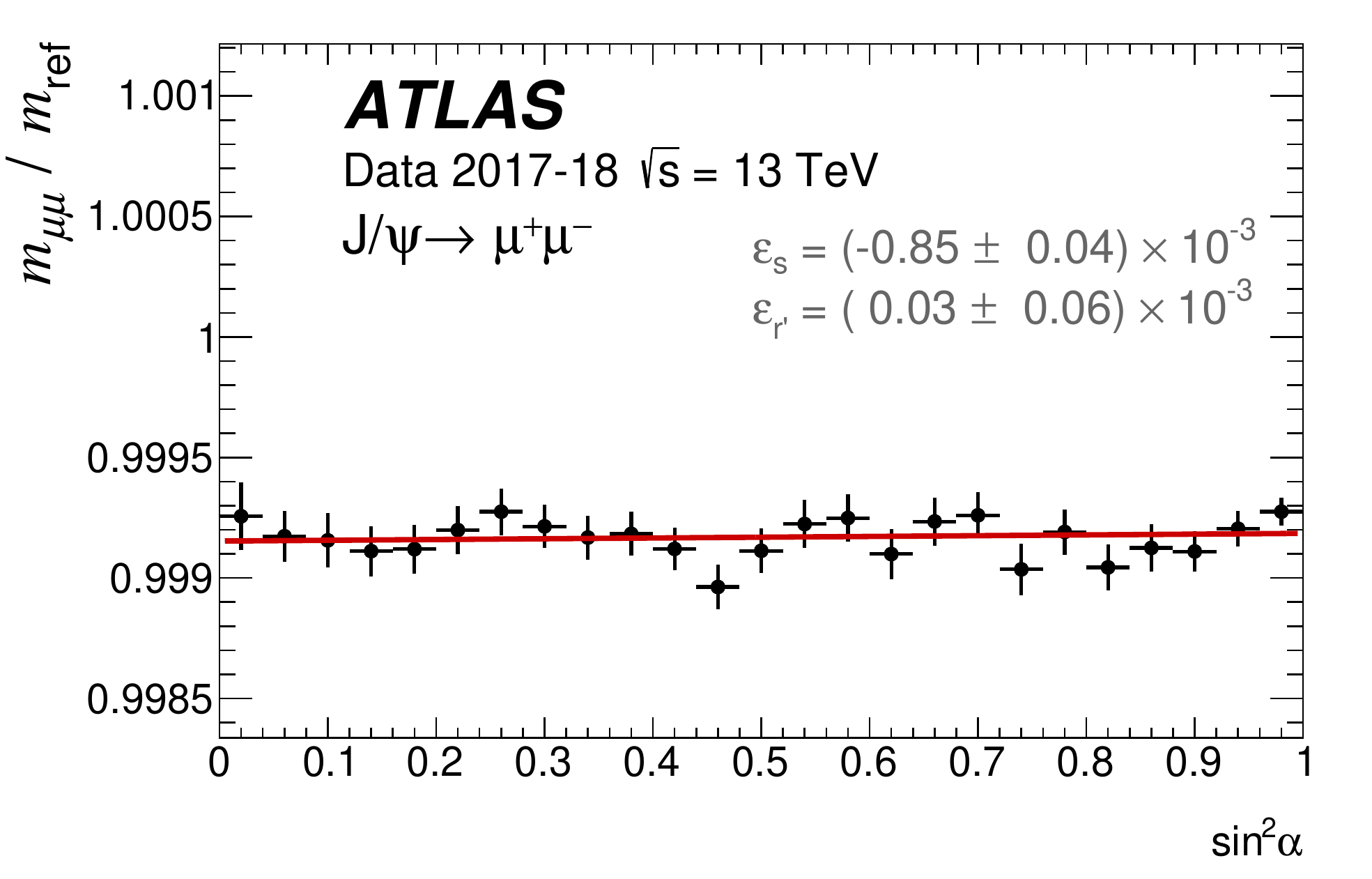}\hfill
\includegraphics[width=0.50\textwidth]{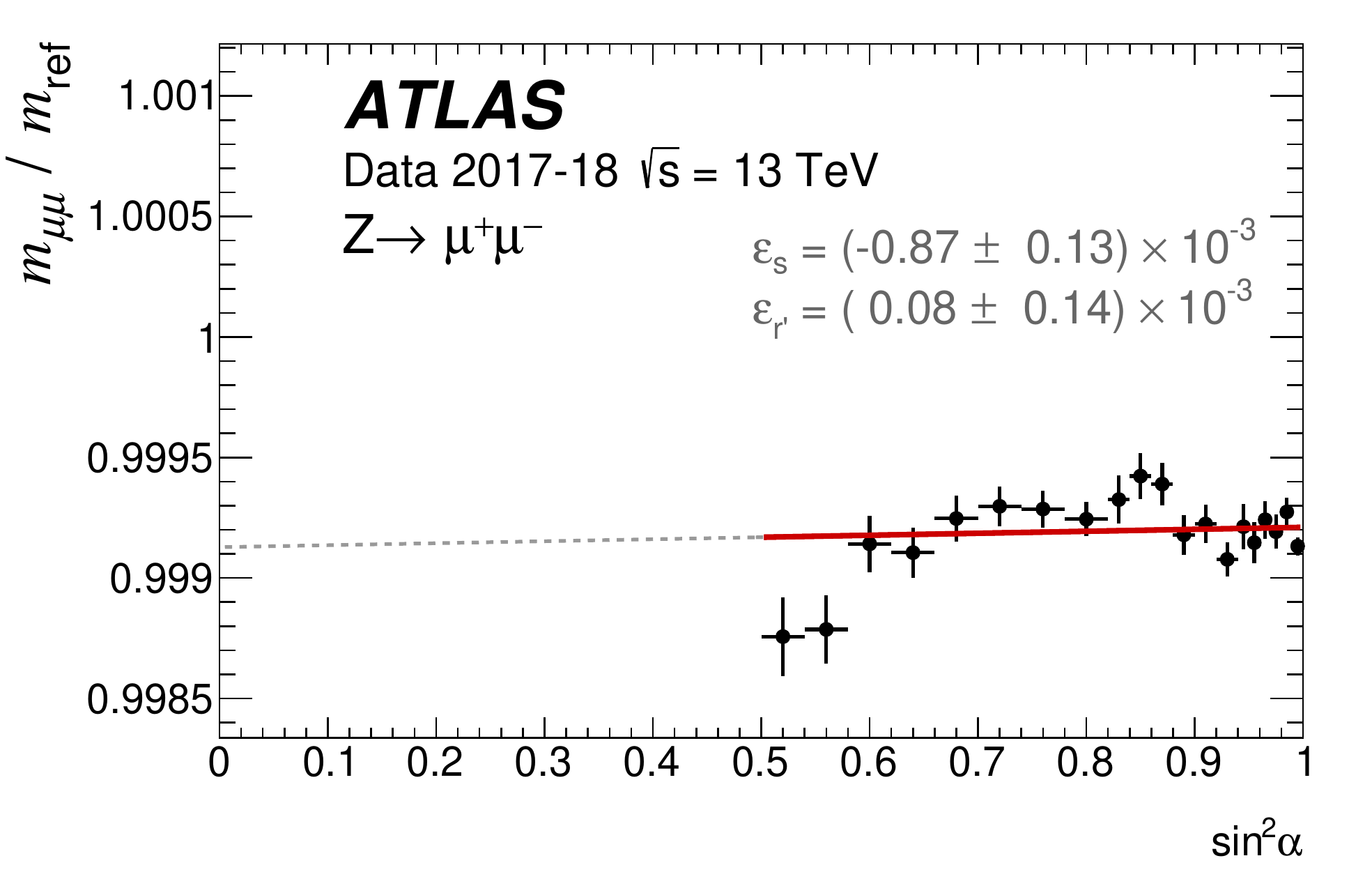}
\caption{Ratio of the measured mass to the reference as a function of $\sin^2\alpha$. Due to event kinematics, $J/\psi\rightarrow\mu^+\mu^-$ events (left) cover the entire $\sin^2\alpha$ range while $Z\rightarrow\mu^+\mu^-$ events (right) cover a smaller range. Error bars represent the statistical uncertainty. The red lines show the fit to \cref{eq:def_sin2alpha} from which the values of $\varepsilon_s$ and $\varepsilon_{r'}$ are extracted.}
\label{fig:radialDistortion_sin2alpha}
\end{figure}

Thus, by measuring the mass as function of $\sin^2\alpha$ it is possible to differentiate between radial and scale biases.
\Cref{fig:radialDistortion_sin2alpha} shows the measured scale using $J/\psi$ and $Z$ decays into $\mu^+\mu^-$ in the barrel of the ID.
The results show a clear momentum scale bias but no significant radial scale ($\varepsilon_{r'}$) as the reconstructed mass is constant as a function of $\sin^2\alpha$.

\begin{figure}[ht!]
\centering
\includegraphics[width=0.50\textwidth]{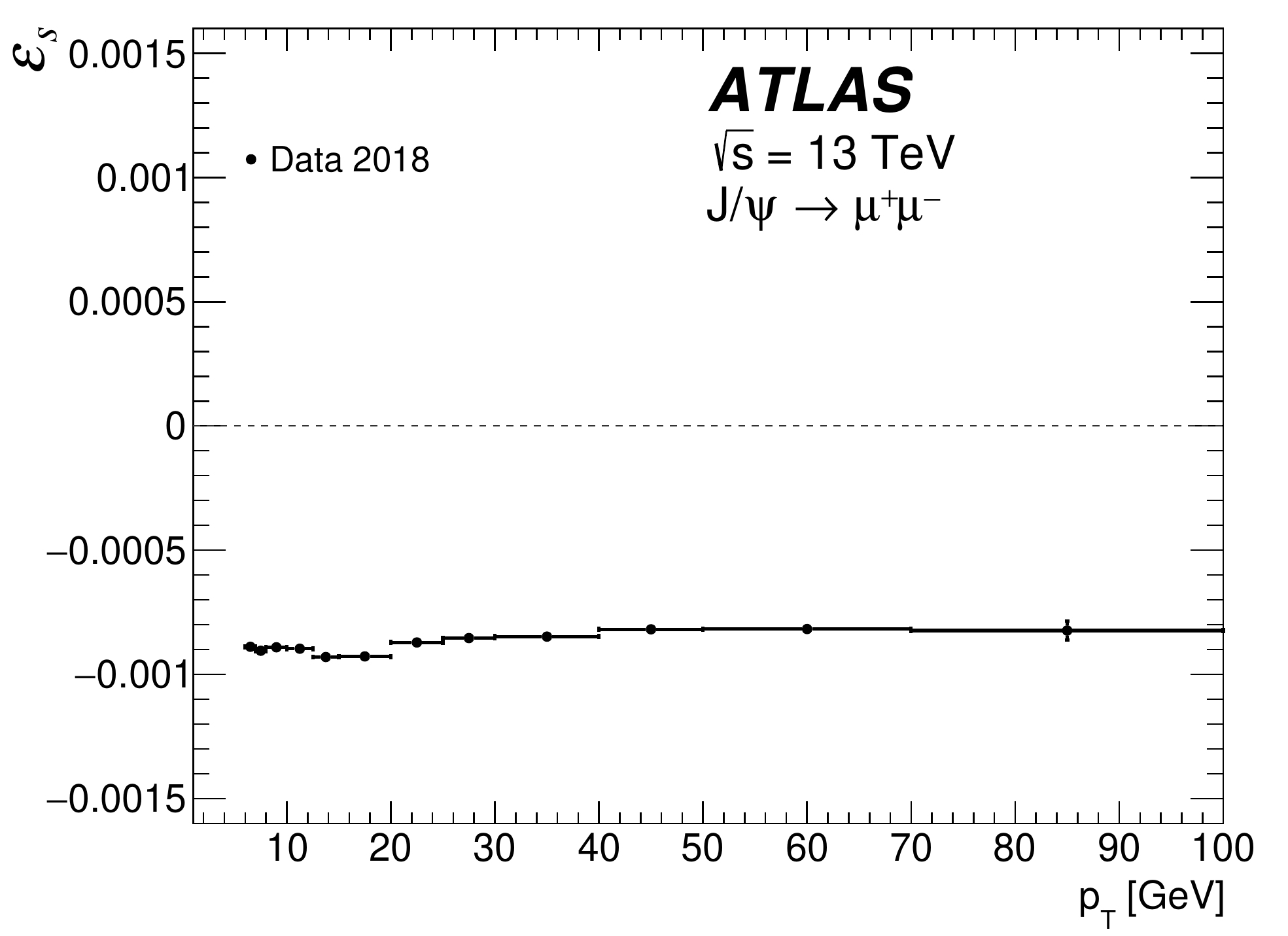}\hfill
\includegraphics[width=0.50\textwidth]{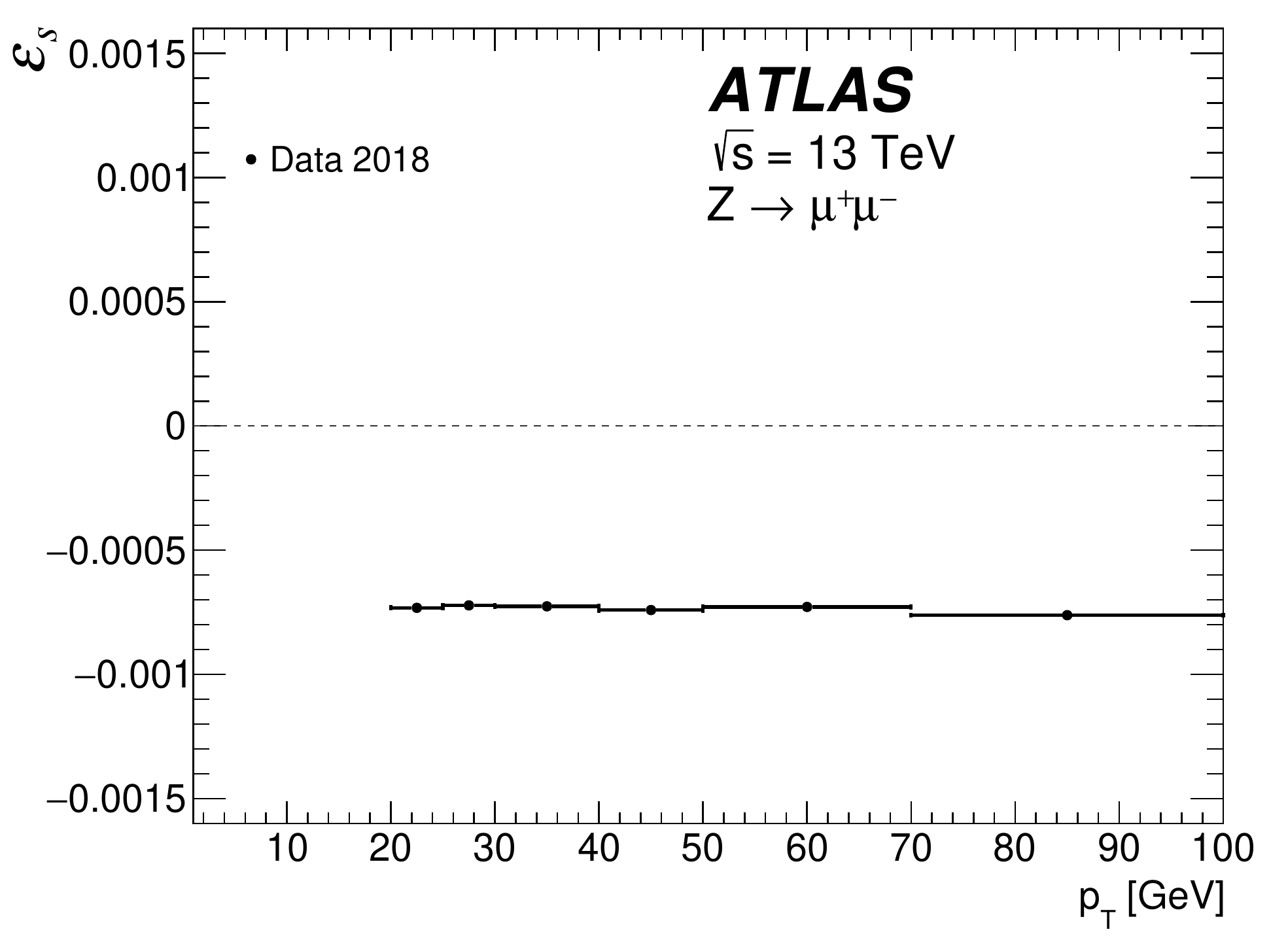}
\caption{The measured momentum scale bias $\varepsilon_{s}$ as a function of track \pt. Error bars represent the statistical uncertainty. Left: $J/\psi\rightarrow\mu^+\mu^-$ decays; right: $Z\rightarrow\mu^+\mu^-$ decays.}
\label{fig:repsilon_vs_pt}
\end{figure}

An analysis using an iterative procedure similar to the $\delta_{\mathrm{sagitta}}$ method, \cref{eq:deltaSagittaIter}, is also performed.
Here, the momentum scale factor ($\varepsilon_s$) is computed and consequently used to update the momentum of the tracks at the next iteration.
This method allows biases to be measured as a function of any kinematic or geometric parameter.
The results as a function of the track \pt are presented in \Cref{fig:repsilon_vs_pt}.
The magnitude of the momentum scale bias is observed to be constant as a function of track \pt as expected from a length scale or magnetic field strength bias.
 
The magnitude of the measured scale bias is consistent for the two studies, demonstrating that there is a global momentum scale bias of  $\varepsilon_s \approx - 0.9 \times 10^{-3}$.
This result is in agreement with the momentum scale at the ID for muons~\cite{atlas_MCPpaper}.
As previously highlighted, the origin of such a global momentum scale bias cannot be unambiguously resolved by these studies.
It should be noted that: the measurement of the absolute scale of the magnetic field has an uncertainty,
which is about four times smaller than the observed scale bias~\cite{ATLAS_solenoid_paper}.
 
% End of text imported from the .//chapters/MomentumBias.tex input file
\newpage
\section{Impact parameter biases}
\label{sec:IPbias}
% The next lines are included from the .//chapters/IPBias.tex input file
The weak modes of the alignment can also lead to a bias in the transverse ($d_0$) and longitudinal ($z_0$) impact parameters.
For example, a rotation of the IBL or radial distortions of the Pixel layers can lead to transverse impact parameter biases.
The quality of the detector alignment can be assessed by analysing impact parameter biases as a function of track $p_{\mathrm{T}}$ and $\eta$.
For this study, events are selected using a combination of single-jet triggers with several jet $p_{\mathrm{T}}$ thresholds starting at \SI{100}{\GeV}.
The standard ATLAS event cleaning selection is applied, ensuring that all detectors were fully operational.
In order to disentangle the biases due to residual misalignment from those originating from the track reconstruction algorithms,
recorded data are compared with a dijet Monte Carlo simulation sample generated with \PYTHIA~\cite{pythia}.
The primary vertex of each selected event must have at least three tracks associated with it.
Tracks are selected by requiring them to be assigned to jets using ghost association~\cite{ghost}, a procedure that treats them as four-vectors of infinitesimal momentum magnitude
during the jet reconstruction and assigns them to the jet with which they are clustered. Jets are reconstructed using the anti-$k_t$ algorithm~\cite{antikt} with radius parameter $R = 0.4$.
In addition, tracks are required to have at least 9 silicon (Pixels + SCT) hits for $|\eta|\le1.65$,
at least 11 silicon hits for $|\eta|>1.65$,
a maximum of 2 SCT holes\footnote{A hole is defined as a missing hit in a module where a hit is expected, based on the extrapolation of the particle trajectory to the module surface.}, no Pixel hole, $p_{\mathrm{T}}>\SI{3}{\GeV}$, $|\eta|<2.5$, and an opening angle $\Delta R({\text{track},\text{jet}})<0.4$ relative to the reconstructed jet axis.
A track $p_{\mathrm{T}}$ of $\SI{3}{\GeV}$ corresponds to the lowest momentum threshold typically used within the alignment to reduce MCS effects (see~\Cref{sec:trackfitting}).
The impact parameters are obtained relative to the primary vertex by extrapolating the particle trajectory to its position.
This is particularly relevant for the longitudinal impact parameter,
as the width of the luminous region in the $z$ direction is very broad.
The impact parameter biases are extracted by iteratively\footnote{This procedure is adopted as the impact parameter distributions have long tails.} fitting the distribution of impact parameters relative to the primary vertex with a Gaussian function within a $\pm 2\sigma$ range until the fitted $\mu$ and $\sigma$ are stable within 1\%.
The resulting value of the Gaussian mean ($\mu$) represents the estimate of the impact parameter bias.
 
\begin{figure}[ht!]
\centering
\begin{subfigure}[t]{0.5\textwidth}
\centering
\includegraphics[width=1\textwidth]{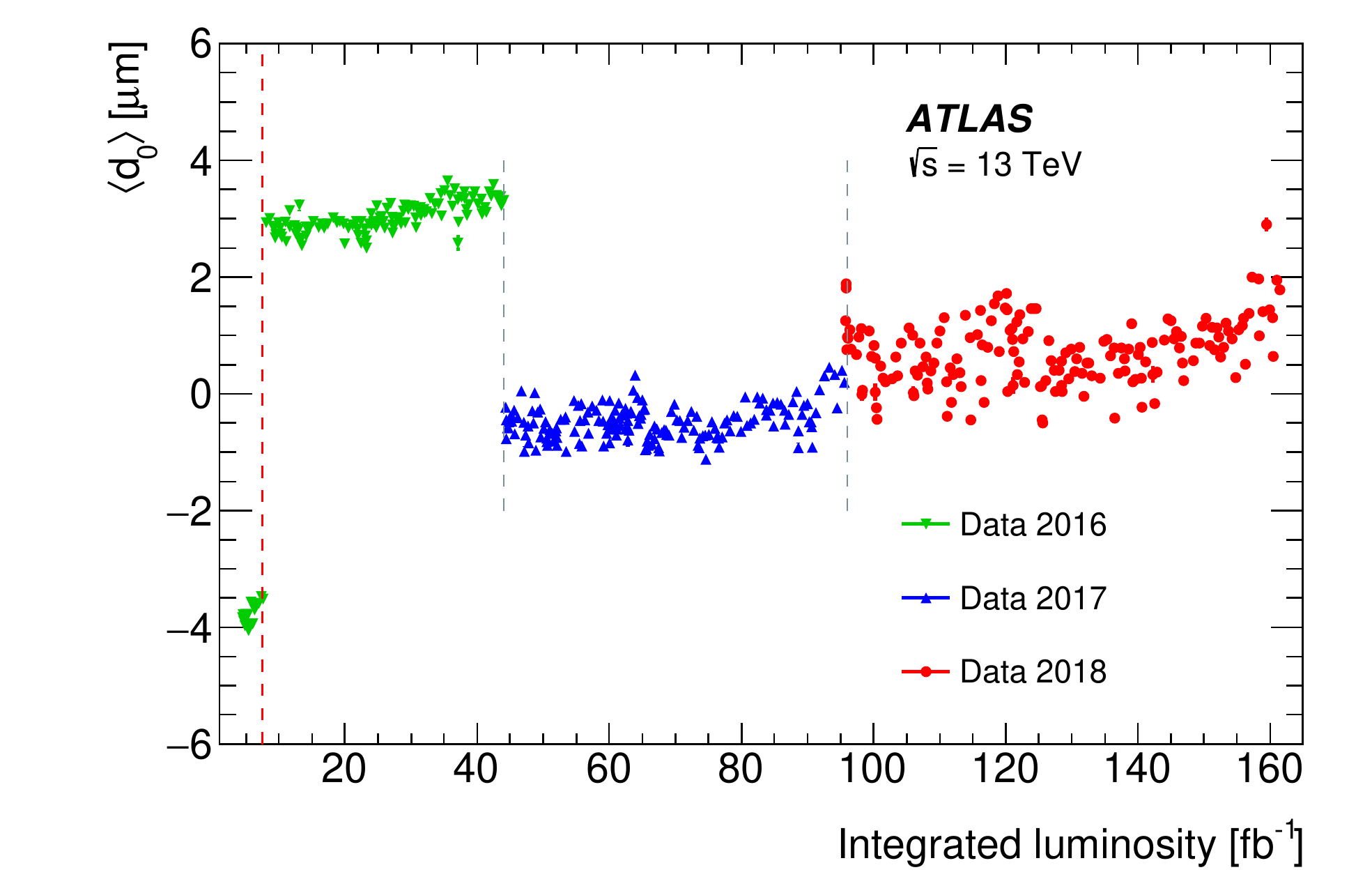}
\end{subfigure}
~
\begin{subfigure}[t]{0.5\textwidth}
\centering
\includegraphics[width=1\textwidth]{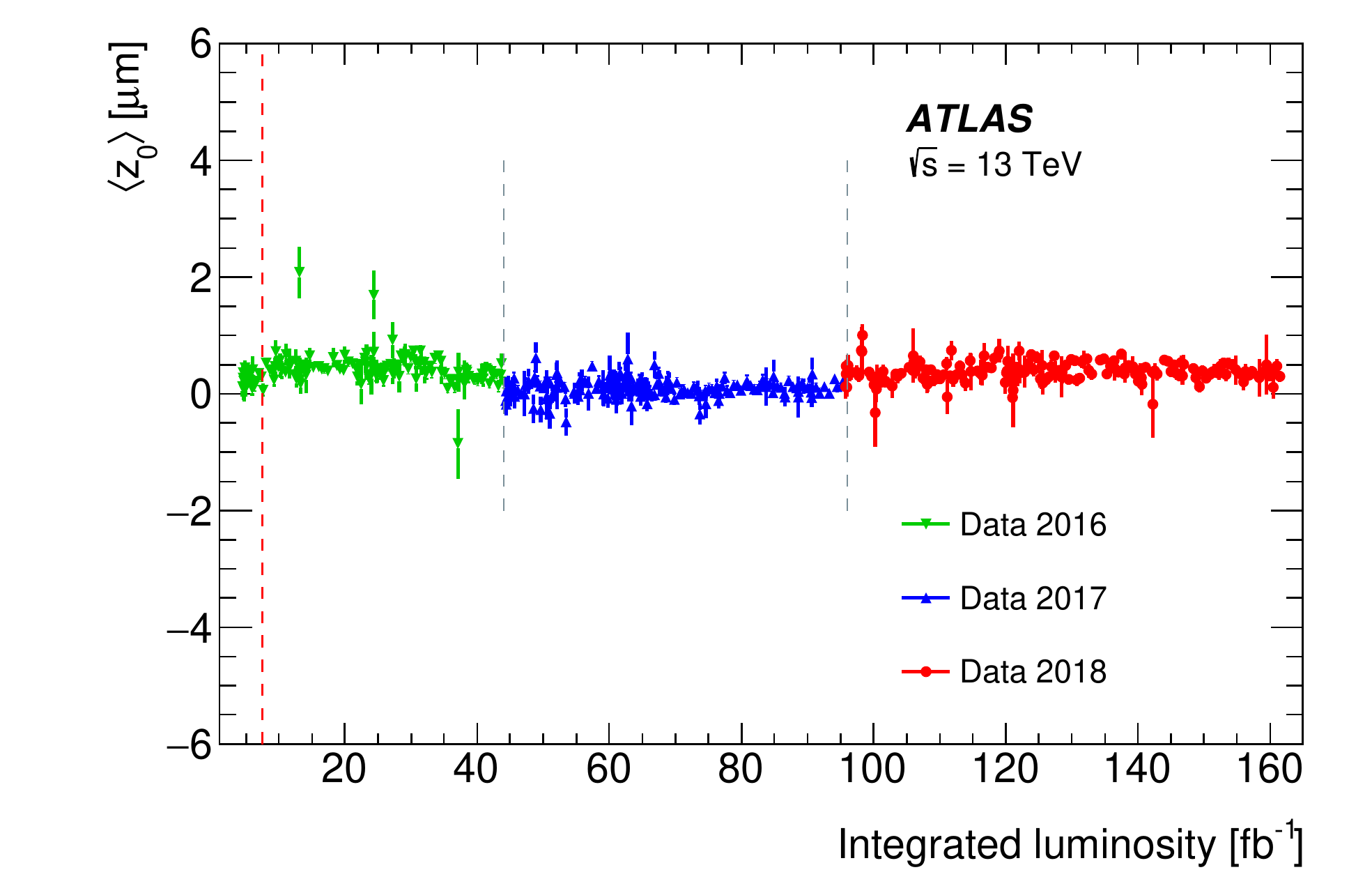}
\end{subfigure}
\caption{The transverse (left) and longitudinal (right) impact parameter biases as function of the Run 2
delivered luminosity.
The red dotted line indicate the change in the underlying ATLAS ID alignment geometry description.
This splits the 2016 data in two periods.
The grey dotted lines indicate the change of the data-taking years.
The $\sim\SI{4}{fb^{-1}}$ corresponding to 2015 data are not shown in this plot.
Only statistical uncertainties are shown.}
\label{fig:ipbiasestime}
\end{figure}

\begin{figure}[ht!]
\centering
\begin{subfigure}[t]{0.5\textwidth}
\centering
\includegraphics[width=1\textwidth]{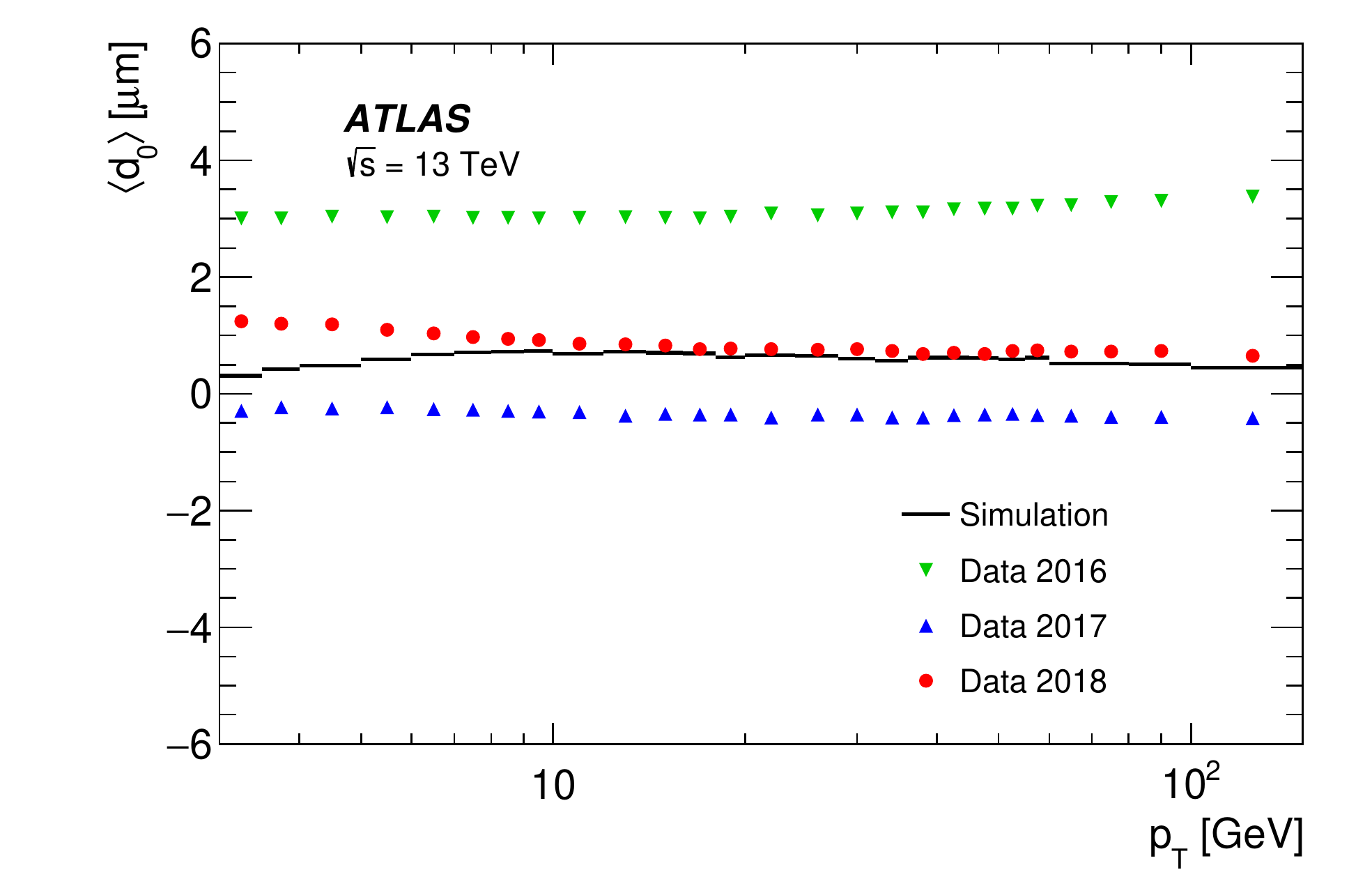}
\end{subfigure}
~
\begin{subfigure}[t]{0.5\textwidth}
\centering
\includegraphics[width=1\textwidth]{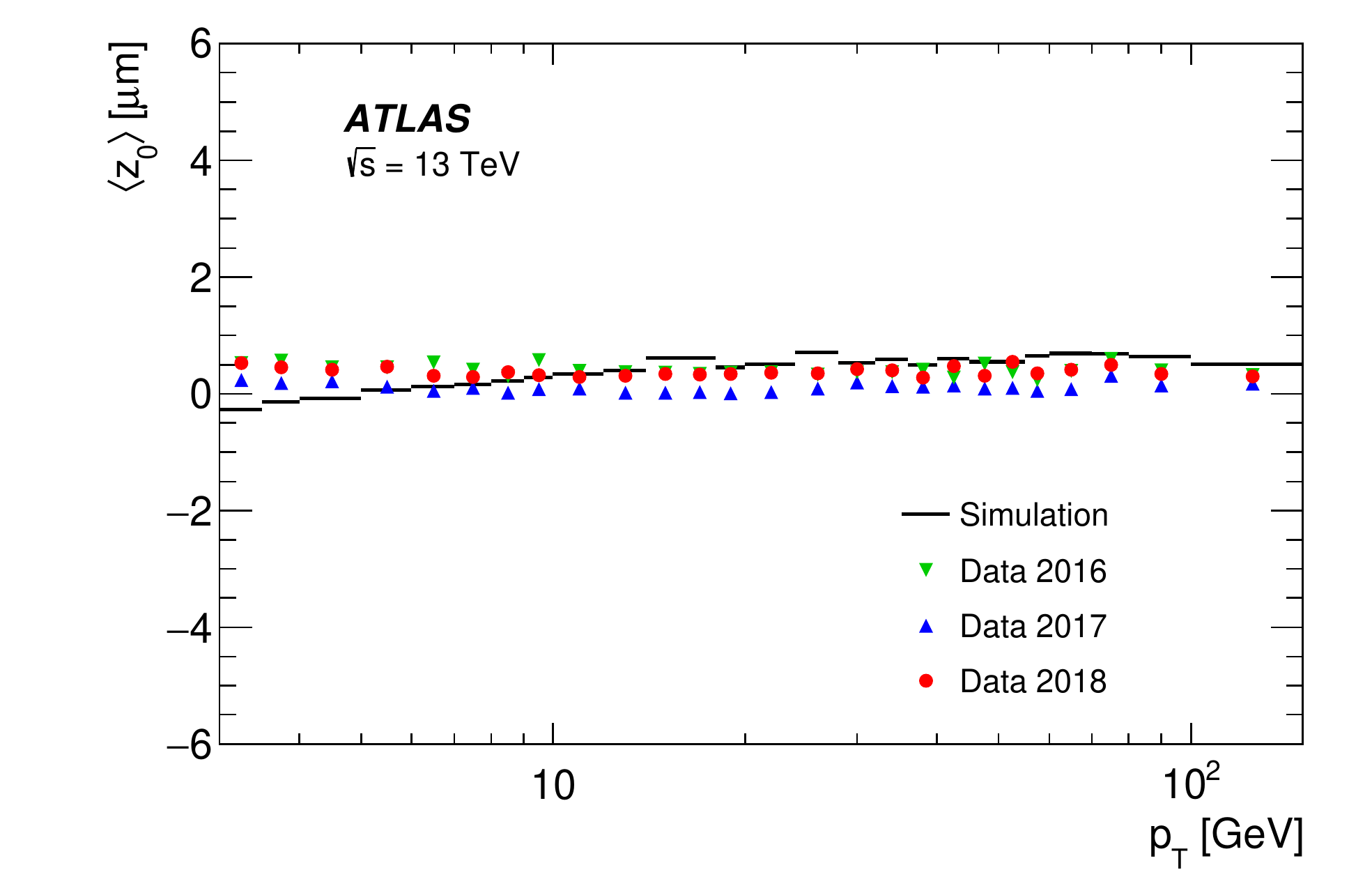}
\end{subfigure}
\caption{The transverse (left) and longitudinal (right) impact parameter biases as function of the track $p_{\mathrm{T}}$.
The 2016 data entries in this figure are taken from the second part of the 2016 data visible in \Cref{fig:ipbiasestime};
the first part of the 2016 data also show no impact parameter dependence on track $p_{\mathrm{T}}$.}
\label{fig:ipbiasespt}
\end{figure}

\begin{figure}[ht!]
\centering
\begin{subfigure}[t]{0.5\textwidth}
\centering
\includegraphics[width=1\textwidth]{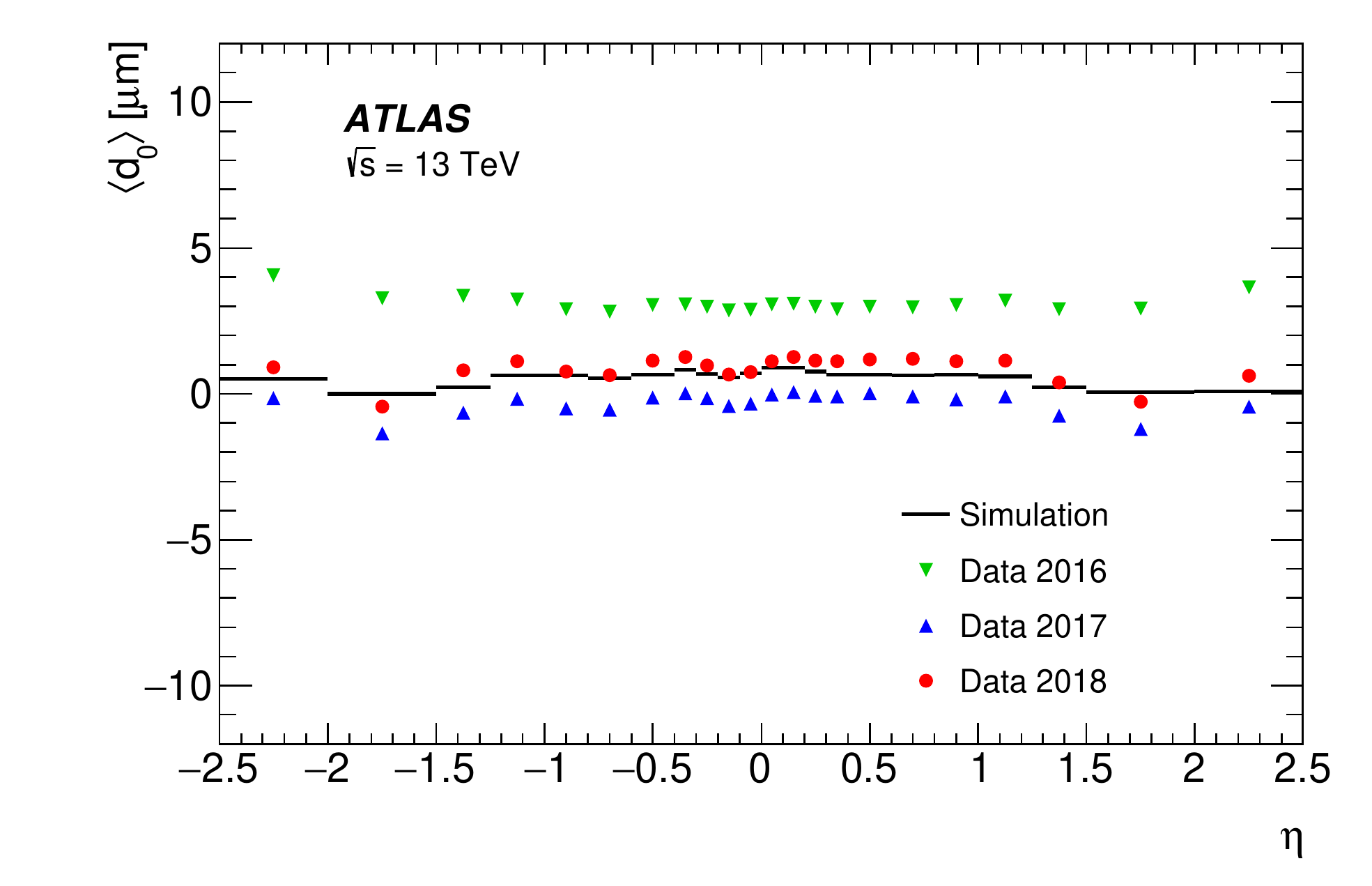}
\end{subfigure}
~
\begin{subfigure}[t]{0.5\textwidth}
\centering
\includegraphics[width=1\textwidth]{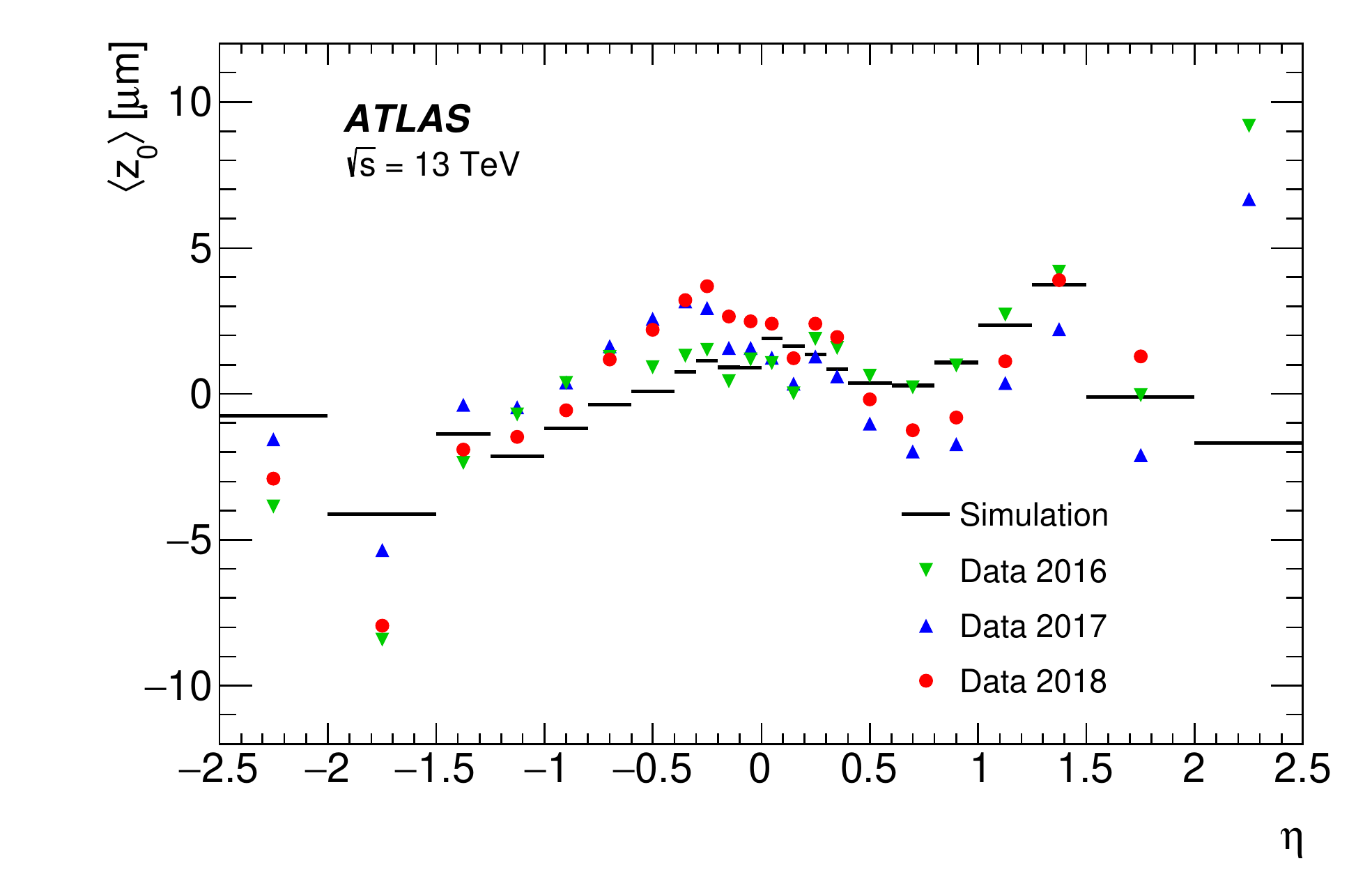}
\end{subfigure}
\caption{The transverse (left) and longitudinal (right) impact parameter biases as function of the track $\eta$.
The 2016 data entries in this figure are taken from the second part of the 2016 data visible in \Cref{fig:ipbiasestime}; the first part of 2016 data also shows no impact parameter dependence on track $\eta$.}
\label{fig:ipbiaseseta}
\end{figure}

\Cref{fig:ipbiasestime} shows the transverse and longitudinal impact parameter biases as a function of the delivered luminosity in Run 2.
Data collected in 2016 have a period-dependent $d_0$ bias of $\SI{-4}{\micro\metre}$ (early 2016) and $+\SI{3}{\micro\metre}$ (late 2016).
This bias was introduced by a change in the underlying geometry description of the ATLAS ID and a misconfiguration of the beam-spot constraint.\footnote{The impact parameters of tracks used in ATLAS physics analyses are corrected by a time-dependent constant to remove the observed biases. In addition, this bias will be corrected in future data processing campaigns.}
Data collected in 2017 and 2018 show overall $d_0$ biases of less than $\SI{1}{\micro\metre}$.
The longitudinal impact parameter bias is negligible and constant across the years (below $\SI{0.5}{\micro\metre}$).
In \Cref{fig:ipbiasespt,,fig:ipbiaseseta}
the transverse and longitudinal impact parameter biases are shown as function of the track transverse momentum and track $\eta$, respectively. The small bias in the longitudinal impact parameter as a function of track $\eta$ is present in simulation and data and is consequently not introduced by the track-based alignment because it is not applied to simulation (where perfect alignment is assumed). The resulting bias has no significant effect on the ATLAS tracking performance as the longitudinal impact parameter resolution is on the order of $\SI{100}{\micro\metre}$ for tracks with $p_{\mathrm{T}} > 10\,\GeV$.
 
% End of text imported from the .//chapters/IPBias.tex input file
 
\FloatBarrier
 
\section{Conclusion}
\label{sec:conclusion}
% The next lines are included from the .//chapters/Conclusion.tex input file
This paper describes the precision alignment of the ATLAS Inner Detector (ID) for Run 2 and quantifies the impact
of alignment uncertainties on track parameter biases.
The alignment procedure consists of a track-based algorithm that minimises track-hit residuals.
It calculates the track parameters at each measurement surface and encodes the relationship between
track-hit residuals and the alignment parameters of each alignable structure.
To resolve ambiguities, it imposes externally determined constraints on track parameters, e.g.\ using tracks from resonance decays.
The alignment procedure is performed at different hierarchical levels,
starting from the largest physical structures and proceeding to individual detector modules or sensor elements.
The number of degrees of freedom increases for each subsequent alignment level.
In total, more than \num{36000} degrees of freedom are considered when aligning all silicon modules (IBL,
Pixel and SCT) and more than \num{700000} degrees of freedom are added for the TRT.
 
It has been observed that operational conditions affect the positions of ID elements.
The Pixel detector moves rapidly upwards every time the data acquisition is activated. 
The staves of the IBL bow depending on the temperature; the degree of variation depends on the thermal load and is a function of the accumulated radiation dose and of the luminosity.
The remaining detector structures are quite stable during an LHC fill;
the movements of individual modules in the barrel have an RMS at the micrometer level while those
in the endcap regions range from 2\,\si{\micro\metre} to $\mathcal{O}(\SI{10}{\micro\metre}$).
 
An automated alignment procedure that corrects for relatively rapid movements of the Pixel detector and IBL and the relative positions of all of other subdetectors is executed for every LHC fill for which the ID collects data.
The detailed alignment of all the other structures (subdetectors, barrel, endcaps, layers, disks, modules or wires) is determined in dedicated alignment campaigns.
The impact of alignment weak modes,
namely distortions that leave the track fit quality largely unchanged and can bias the measured track parameters,
is minimised during these campaigns by employing external constraints on track parameters.
Independent measurements are performed to quantify potential biases, enabling them to be largely removed.
The residual sagitta bias and momentum scale bias after the full Run 2 alignment are reduced to less than $\sim$\SI{0.1}{TeV^{-1}} and 0.9$\times10^{-3}$, respectively. Remaining track parameter biases do not significantly impact ATLAS physics analyses.
 
% End of text imported from the .//chapters/Conclusion.tex input file
 
\FloatBarrier
\clearpage
\appendix
\part*{Appendix}
\addcontentsline{toc}{part}{Appendix}
\section{Track fitting with multiple Coulomb scattering effects}
\label{sec:trackfittingMCS}
% The next lines are included from the .//chapters/Appendix_AdditionalFormalism.tex input file
The track fit can be improved by considering the charged particle scattering as it passes through material in the
detector.
Knowing the detector material description, one can estimate and fit the scattering angles in the sensor planes~\cite{ThijsThesis}.
To include the effects of multiple scattering, terms are added directly to the track $\chi^2$ as it is done in \cref{eq:TrackChi2Scatter}, which can be written as:
\begin{linenomath}
\begin{equation*}
\XTrk = \sum_i \left( \frac{r_i(\per,\tet )}{\sigma_i} \right)^2 + \sum_j \frac{(\hat{\theta} - \theta_j)^2}{\Theta_{jj}}~.
\end{equation*}
\end{linenomath}
It should be noted that the residuals now also depend on the scattering angles, \tet .
The scattering expectation value, $\hat{\theta}$, is zero and its variance, $\Theta_{jj}$, depends on the particle momentum and amount of material traversed.
The uncertainty of the $i$-th measurement is denoted by $\sigma_i$.
 
The $\chi^2$ has to be minimised for \per\ and \tet\ simultaneously.
Defining the derivative of residuals with respect to track and scattering parameters to be:
\begin{linenomath}
\begin{equation*}
G \equiv \frac{\partial \rer}{\partial \per} \hspace{48 pt} S \equiv \frac{\partial \rer}{\partial \tet }
\end{equation*}
\end{linenomath}
the derivatives of $\chi^2$ with respect to the track and scattering parameters are:
\begin{linenomath}
\begin{equation*}
\begin{split}
\frac{1}{2} \frac{\full \chi^2}{\full \per } &= G^\top \Omega^{-1} \rer , \\
\frac{1}{2} \frac{\full \chi^2}{\full \tet } &= S^\top \Omega^{-1} \rer + \Theta^{-1}\tet .
\end{split}
\end{equation*}
\end{linenomath}
 
Neglecting second-order derivatives of residuals, the second derivatives of $\chi^2$ with respect to perigee and scattering parameters are:
\begin{linenomath}
\begin{equation*}
\begin{split}
\frac{1}{2} \frac{\full ^2\chi^2}{\full \per ^2} &= G^\top \Omega^{-1} G ,\\
\frac{1}{2} \frac{\full ^2\chi^2}{\full \tet ^2} &= S^\top \Omega^{-1} S + \Theta^{-1} ,\\
\frac{1}{2} \frac{\full ^2\chi^2}{\full \tet \full \per } &= G^\top \Omega^{-1} S .
\end{split}
\end{equation*}
\end{linenomath}

The above can be written down in a compact form using \cref{ref:auxdef}:
\begin{linenomath}
\begin{equation*}
\begin{split}
\frac{1}{2} \frac{\full \chi^2}{\full \trk } &= H^\top V^{-1} \res ,\\
\frac{1}{2} \frac{\full ^2\chi^2}{\full \trk ^2} &= H^\top V^{-1} H.
\end{split}
\end{equation*}
\end{linenomath}
% End of text imported from the .//chapters/Appendix_AdditionalFormalism.tex input file
\FloatBarrier

\section*{Acknowledgements}
 
% The next lines are included from the .//acknowledgements/Acknowledgements.tex input file

We thank CERN for the very successful operation of the LHC, as well as the
support staff from our institutions without whom ATLAS could not be
operated efficiently.
 
We acknowledge the support of ANPCyT, Argentina; YerPhI, Armenia; ARC, Australia; BMWFW and FWF, Austria; ANAS, Azerbaijan; SSTC, Belarus; CNPq and FAPESP, Brazil; NSERC, NRC and CFI, Canada; CERN; CONICYT, Chile; CAS, MOST and NSFC, China; COLCIENCIAS, Colombia; MSMT CR, MPO CR and VSC CR, Czech Republic; DNRF and DNSRC, Denmark; IN2P3-CNRS and CEA-DRF/IRFU, France; SRNSFG, Georgia; BMBF, HGF and MPG, Germany; GSRT, Greece; RGC and Hong Kong SAR, China; ISF and Benoziyo Center, Israel; INFN, Italy; MEXT and JSPS, Japan; CNRST, Morocco; NWO, Netherlands; RCN, Norway; MNiSW and NCN, Poland; FCT, Portugal; MNE/IFA, Romania; MES of Russia and NRC KI, Russia Federation; JINR; MESTD, Serbia; MSSR, Slovakia; ARRS and MIZ\v{S}, Slovenia; DST/NRF, South Africa; MINECO, Spain; SRC and Wallenberg Foundation, Sweden; SERI, SNSF and Cantons of Bern and Geneva, Switzerland; MOST, Taiwan; TAEK, Turkey; STFC, United Kingdom; DOE and NSF, United States of America. In addition, individual groups and members have received support from BCKDF, CANARIE, Compute Canada and CRC, Canada; ERC, ERDF, Horizon 2020, Marie Sk{\l}odowska-Curie Actions and COST, European Union; Investissements d'Avenir Labex, Investissements d'Avenir Idex and ANR, France; DFG and AvH Foundation, Germany; Herakleitos, Thales and Aristeia programmes co-financed by EU-ESF and the Greek NSRF, Greece; BSF-NSF and GIF, Israel; CERCA Programme Generalitat de Catalunya and PROMETEO Programme Generalitat Valenciana, Spain; G\"{o}ran Gustafssons Stiftelse, Sweden; The Royal Society and Leverhulme Trust, United Kingdom.
 
The crucial computing support from all WLCG partners is acknowledged gratefully, in particular from CERN, the ATLAS Tier-1 facilities at TRIUMF (Canada), NDGF (Denmark, Norway, Sweden), CC-IN2P3 (France), KIT/GridKA (Germany), INFN-CNAF (Italy), NL-T1 (Netherlands), PIC (Spain), ASGC (Taiwan), RAL (UK) and BNL (USA), the Tier-2 facilities worldwide and large non-WLCG resource providers. Major contributors of computing resources are listed in Ref.~\cite{ATL-SOFT-PUB-2020-001}.
 
% End of text imported from the .//acknowledgements/Acknowledgements.tex input file

\printbibliography

@article{PDG2018,
  title = {Review of Particle Physics},
  author = {Tanabashi, M. and others},
  collaboration = {Particle Data Group},
  journal = {Phys. Rev. D},
  volume = {98},
  issue = {3},
  pages = {030001},
  numpages = {1898},
  year = {2018},
  month = {Aug},
  publisher = {American Physical Society},
  doi = {10.1103/PhysRevD.98.030001}
}

@article{ghost,
	Archiveprefix = {arXiv},
	Author = {Matteo Cacciari and Gavin P. Salam},
	Doi = {https://doi.org/10.1016/j.physletb.2007.09.077},
	Eprint = {0707.1378},
	Journal = {Phys. Lett. B},
	Number = {1},
	Pages = {119 - 126},
	Primaryclass = {hep-ph},
	Title = {Pileup subtraction using jet areas},
	Volume = {659},
	Year = {2008}}

@article{Wollrath:2019rrw,
	Archiveprefix = {arXiv},
	Author = {Wollrath, J.},
	Booktitle = {{Connecting the Dots and Workshop on Intelligent Trackers (CTD/WIT 2019) Valencia, Valencia, Spain, April 2-5, 2019}},
	Eprint = {1910.06060},
	Primaryclass = {physics.ins-det},
	Reportnumber = {ATL-PHYS-PROC-2019-039, PROC-CTD19-134},
	Slaccitation = {%%CITATION = ARXIV:1910.06060;%%},
	Title = {{Sensor shapes and weak modes of the ATLAS Inner Detector track-based alignment}},
	Year = {2019}}

@article{ATLAS_solenoid_paper,
	Author = {Aleksa, M. and others},
	Doi = {10.1088/1748-0221/3/04/P04003},
	Journal = {JINST},
	Pages = {P04003},
	Slaccitation = {%%CITATION = JINST,3,P04003;%%},
	Title = {{Measurement of the ATLAS solenoid magnetic field}},
	Volume = {3},
	Year = {2008}}

@article{ATLASRun2DataQuality,
	Archiveprefix = {arXiv},
	Author = {{{ATLAS Collaboration}}},
	Collaboration = {ATLAS},
	Doi = {10.1088/1748-0221/15/04/P04003},
	Eprint = {1911.04632},
	Journal = {JINST},
	Pages = {P04003},
	Primaryclass = {physics.ins-det},
	Reportnumber = {CERN-EP-2019-207},
	Slaccitation = {%%CITATION = ARXIV:1911.04632;%%},
	Title = {{ATLAS data quality operations and performance for 2015-2018 data-taking}},
	Volume = {15},
	Year = {2020}}

@article{IBLradiation,
	Author = {{ATLAS Collaboration}},
	Institution = {CERN},
	Journal = {ATL-INDET-PUB-2017-001},
	Reportnumber = {ATL-INDET-PUB-2017-001},
	Title = {Radiation induced effects in the ATLAS Insertable B-Layer readout chip},
	Url = {http://cds.cern.ch/record/2291800},
	Year = {2017}}

@article{IBL-3d-perf,
	Archiveprefix = {arXiv},
	Author = {Darbo, G.},
	Booktitle = {{Proceedings, 7th International Workshop on Semiconductor Pixel Detectors for Particles and Imaging (PIXEL 2014): Niagara Falls, Ontario, Canada, September 1-5, 2014}},
	Collaboration = {ATLAS},
	Doi = {10.1088/1748-0221/10/05/C05001},
	Eprint = {1411.6937},
	Journal = {JINST},
	Number = {05},
	Pages = {C05001},
	Primaryclass = {physics.ins-det},
	Reportnumber = {ATL-INDET-PROC-2014-017},
	Slaccitation = {%%CITATION = ARXIV:1411.6937;%%},
	Title = {{Experience on 3D Silicon Sensors for ATLAS IBL}},
	Volume = {10},
	Year = {2015}}

@article{IBL-3d-sensors,
	Author = {Da Via, Cinzia and others},
	Doi = {10.1016/j.nima.2012.07.058},
	Journal = {Nucl. Instrum. Meth. A},
	Pages = {321-330},
	Slaccitation = {%%CITATION = NUIMA,A694,321;%%},
	Title = {{3D silicon sensors: Design, large area production and quality assurance for the ATLAS IBL pixel detector upgrade}},
	Volume = {694},
	Year = {2012}}

@article{trt-barrel-det,
	Author = {Abat, E. and others},
	Collaboration = {ATLAS TRT},
	Doi = {10.1088/1748-0221/3/02/P02014},
	Journal = {JINST},
	Pages = {P02014},
	Slaccitation = {%%CITATION = JINST,3,P02014;%%},
	Title = {{The ATLAS TRT barrel detector}},
	Volume = {3},
	Year = {2008}}

@article{trt-ec-det,
	Author = {Abat, E. and others},
	Doi = {10.1088/1748-0221/3/10/P10003},
	Journal = {JINST},
	Pages = {P10003},
	Slaccitation = {%%CITATION = JINST,3,P10003;%%},
	Title = {{The ATLAS TRT end-cap detectors}},
	Volume = {3},
	Year = {2008}}

@article{sct-ec-modules,
	Author = {Abdesselam, A. and others},
	Doi = {10.1016/j.nima.2007.02.019},
	Journal = {Nucl. Instrum. Meth. A},
	Pages = {353-389},
	Title = {The ATLAS semiconductor tracker end-cap module},
	Volume = {575},
	Year = {2007}}

@article{sct-barrel-modules,
	Author = {Abdesselam, A. and others},
	Doi = {10.1016/j.nima.2006.08.036},
	Journal = {Nucl. Instrum. Meth. A},
	Pages = {642-671},
	Title = {The barrel modules of the ATLAS semiconductor tracker},
	Volume = {568},
	Year = {2006}}

@article{pixels-det-sensors,
	Author = {G. Aad and others},
	Doi = {10.1088/1748-0221/3/07/P07007},
	Journal = {JINST},
	Number = {3},
	Pages = {P07007},
	Title = {ATLAS pixel detector electronics and sensors},
	Volume = {3},
	Year = {2008}}

@article{antikt,
	Archiveprefix = {arXiv},
	Author = {Cacciari, Matteo and Salam, Gavin P. and Soyez, Gregory},
	Doi = {10.1088/1126-6708/2008/04/063},
	Eprint = {0802.1189},
	Journal = {JHEP},
	Number = {10.1088/1126-6708/2008/04/063},
	Pages = {63},
	Primaryclass = {hep-ph},
	Title = {{The anti-$k_t$ jet clustering algorithm}},
	Volume = {04},
	Year = {2008}}

@article{pythia,
	Archiveprefix = {arXiv},
	Author = {{T. Sj\"ostrand, S. Mrenna and P. Z. Skands}},
	Doi = {10.1016/j.cpc.2008.01.036},
	Eprint = {0710.3820},
	Journal = {Comput. Phys. Commun.},
	Pages = {852},
	Primaryclass = {hep-ex},
	Title = {A Brief Introduction to PYTHIA 8.1},
	Volume = {178},
	Year = {2008}}

@article{atlas_MCPpaper,
	Archiveprefix = {arXiv},
	Author = {{ATLAS Collaboration}},
	Doi = {10.1140/epjc/s10052-016-4120-y},
	Eprint = {1603.05598},
	Journal = {Eur. Phys. J. C},
	Pages = {292},
	Primaryclass = {hep-ex},
	Reportnumber = {CERN-EP-2016-033},
	Title = {{Muon reconstruction performance of the ATLAS detector in proton--proton collision data at \(\sqrt{s} = 13\,\text{TeV}\)}},
	Volume = {76},
	Year = {2016}}

@misc{eigenweb,
	Author = {Ga\"{e}l Guennebaud and Beno\^{i}t Jacob and others},
	Howpublished = {http://eigen.tuxfamily.org},
	Title = {Eigen v3},
	Year = {2010}}

@article{ATL-DAQ-PUB-2016-001,
	Author = {{ATLAS Collaboration}},
	Journal = {ATL-DAQ-PUB-2016-001},
	Month = {Mar},
	Reportnumber = {ATL-DAQ-PUB-2016-001},
	Title = {{2015 start-up trigger menu and initial performance assessment of the ATLAS trigger using Run-2 data}},
	Url = {https://cds.cern.ch/record/2136007},
	Year = {2016}}

@booklet{IBLDistTaskForce,
	Author = {{ATLAS Collaboration}},
	Howpublished = {{ATL-INDET-PUB-2015-001}},
	Title = {{Study of the mechanical stability of the ATLAS Insertable B-Layer}},
	Url = {https://cds.cern.ch/record/2022587},
	Year = {2015}}

@article{ATLASTechnicalPaper,
	Author = {{ATLAS Collaboration}},
	Doi = {10.1088/1748-0221/3/08/S08003},
	Journal = {JINST},
	Pages = {S08003},
	Slaccitation = {%%CITATION = JINST,3,S08003;%%},
	Title = {{The ATLAS Experiment at the CERN Large Hadron Collider}},
	Volume = {3},
	Year = {2008}}

@article{ATLAS-COM-CONF-2014-046,
	Author = {{ATLAS Collaboration}},
	Collaboration = {ATLAS},
	Journal = {ATLAS-CONF-2014-047},
	Title = {{Alignment of the ATLAS Inner Detector and its Performance in 2012}},
	Url = {https://cds.cern.ch/record/1741021},
	Year = {2014}}

@article{ATL-SOFT-PUB-2014-003,
	Author = {{ATLAS Collaboration}},
	Collaboration = {ATLAS},
	Journal = {ATL-SOFT-PUB-2014-003},
	Title = {{Common Framework Implementation for the Track-Based Alignment of the ATLAS Detector}},
	Url = {https://cds.cern.ch/record/1670354},
	Year = {2014}}

@booklet{IBLref,
	Author = {{ATLAS Collaboration}},
	Howpublished = {ATLAS-TDR-19},
	Related = {ATLAS-TDR-2010-19-add},
	Relatedstring = {Addendum:},
	Title = {{ATLAS Insertable B-Layer Technical Design Report}},
	Url = {https://cds.cern.ch/record/1291633},
	Year = {2010}}

@article{JimenezPenaPhd,
	Author = {Jim{\'e}nez Pe{\~n}a, Javier},
	Journal = {PhD thesis: Universitat de Valencia (Spain)},
	Title = {ATLAS Inner Detector alignment and analysis of the $Wtb$ vertex structure with single top quarks},
	Url = {https://cds.cern.ch/record/2644636/},
	Year = {2018}}

@article{TRTAlignment,
	Author = {A. Bocci and W. Hulsbergen},
	Publisher = {ATL-INDET-PUB-2007-009},
	Title = {{TRT alignment for SR1 cosmics and beyond}},
	Url = {http://cds.cern.ch/record/1039585},
	Year = 2007}

@book{LAPACK,
	Address = {Philadelphia, PA},
	Author = {Anderson, E. and Bai, Z. and Bischof, C. and Blackford, S. and Demmel, J. and Dongarra, J. and Du Croz, J. and Greenbaum, A. and Hammarling, S. and McKenney, A. and Sorensen, D.},
	Edition = {Third edition},
	Isbn = {0-89871-447-8},
	Publisher = {Society for Industrial and Applied Mathematics},
	Title = {{LAPACK} Users' Guide},
	Year = {1999}}

@article{Siena_proc,
	Author = {P. Bruckman},
	Doi = {10.1016/j.nuclphysbps.2009.10.057},
	Journal = {Nucl. Phys. B (Proc. Suppl.)},
	Pages = {158-162},
	Title = {{Alignment of the ATLAS Inner Detector Tracking System - Solving the Problem}},
	Volume = {197},
	Year = 2009}

@article{TrackingEDM,
	Author = {P. F. Akesson and others},
	Note = {ATL-SOFT-PUB-2006-004},
	Title = {{ATLAS Tracking Event Data Model}},
	Url = {http://cds.cern.ch/record/973401},
	Year = 2006}

@article{root,
	Author = {{R. Brun and F. Rademakers}},
	Doi = {10.1016/S0168-9002(97)00048-X},
	Journal = {Nucl. Instrum. Meth. A},
	Note = {See also \texttt{http://root.cern.ch/}.},
	Pages = {81-86},
	Publisher = {Proceedings AIHENP'96 Workshop, Lausanne, Sep. 1996},
	Title = {ROOT - An Object Oriented Data Analysis Framework},
	Volume = {389},
	Year = 1997}

@misc{CLHEP,
	Note = {\texttt{http://proj-clhep.web.cern.ch/proj-clhep/}}}

@article{FEAST1,
	Archiveprefix = {arXiv},
	Author = {Polizzi, Eric},
	Doi = {10.1103/PhysRevB.79.115112},
	Eprint = {0901.2665},
	Issue = {11},
	Journal = {Phys. Rev. B},
	Month = {Mar},
	Numpages = {6},
	Pages = {115112},
	Primaryclass = {cs.CE},
	Publisher = {American Physical Society},
	Title = {Density-matrix-based algorithm for solving eigenvalue problems},
	Volume = {79},
	Year = {2009}}

@phdthesis{ThijsThesis,
    author = "Cornelissen, Thijs Gerrit",
    title = "{Track Fitting in the ATLAS Experiment}",
    reportNumber = "CERN-THESIS-2006-072",
    school = "Amsterdam U.",
    year = "2006"
}

@Article{Lai:1999wy,
     author    = "Lai, H. L. and others",
 collaboration = "CTEQ",
     title     = "{Global QCD analysis of parton structure of the nucleon:
                  CTEQ5 parton distributions}",
     journal   = "Eur. Phys. J. C",
     volume    = "12",
     year      = "2000",
     pages     = "375",
     eprint    = "hep-ph/9903282",
     archivePrefix = "arXiv",
     doi       = "10.1007/s100529900196",
     SLACcitation  = "%%CITATION = HEP-PH/9903282;%%"
}

@Article{Pumplin:2002vw,
     author    = "Pumplin, J. and others",
     title     = "{New Generation of Parton Distributions with Uncertainties from Global QCD Analysis}",
     journal   = "JHEP",
     volume    = "07",
     year      = "2002",
     pages     = "012",
     doi       = "10.1088/1126-6708/2002/07/012",
     eprint    = "hep-ph/0201195",
     archivePrefix = "arXiv",
     SLACcitation  = "%%CITATION = HEP-PH/0201195;%%"
}

@article{Sjostrand:2014zea,
      author         = "Sj{\"o}strand, Torbj{\"o}rn and Ask, Stefan and Christiansen,
                        Jesper R. and Corke, Richard and Desai, Nishita and Ilten,
                        Philip and Mrenna, Stephen and Prestel, Stefan and
                        Rasmussen, Christine O. and Skands, Peter Z.",
      title          = "{An introduction to PYTHIA 8.2}",
      journal        = "Comput. Phys. Commun.",
      volume         = "191",
      year           = "2015",
      pages          = "159",
      doi            = "10.1016/j.cpc.2015.01.024",
      eprint         = "1410.3012",
      archivePrefix  = "arXiv",
      primaryClass   = "hep-ph",
      reportNumber   = "LU-TP-14-36, MCNET-14-22, CERN-PH-TH-2014-190,
                        FERMILAB-PUB-14-316-CD, DESY-14-178, SLAC-PUB-16122,
                        --FERMILAB-PUB-14-316-CD",
      SLACcitation   = "%%CITATION = ARXIV:1410.3012;%%"
}

@Article{Nason:2004rx,
      author         = "Nason, Paolo",
      title          = "{A new method for combining NLO QCD with shower Monte Carlo algorithms}",
      journal        = "JHEP",
      volume         = "11",
      pages          = "040",
      doi            = "10.1088/1126-6708/2004/11/040",
      year           = "2004",
      eprint         = "hep-ph/0409146",
      archivePrefix  = "arXiv",
}

@Article{Frixione:2007vw,
      author         = "Frixione, Stefano and Nason, Paolo and Oleari, Carlo",
      title          = "{Matching NLO QCD computations with parton shower
                        simulations: the POWHEG method}",
      journal        = "JHEP",
      volume         = "11",
      pages          = "070",
      doi            = "10.1088/1126-6708/2007/11/070",
      year           = "2007",
      eprint         = "0709.2092",
      archivePrefix  = "arXiv",
      primaryClass   = "hep-ph",
}

@Article{Alioli:2010xd,
      author         = "Alioli, Simone and Nason, Paolo and Oleari, Carlo and Re,
                        Emanuele",
      title          = "{A general framework for implementing NLO calculations in
                        shower Monte Carlo programs: the POWHEG BOX}",
      journal        = "JHEP",
      volume         = "06",
      pages          = "043",
      doi            = "10.1007/JHEP06(2010)043",
      year           = "2010",
      eprint         = "1002.2581",
      archivePrefix  = "arXiv",
      primaryClass   = "hep-ph",
}

@Article{PERF-2012-05,
    author         = "{ATLAS Collaboration}",
    title          = "{A neural network clustering algorithm for the ATLAS silicon pixel detector}",
    journal        = "JINST",
    volume         = "9",
    year           = "2014",
    pages          = "P09009",
    doi            = "10.1088/1748-0221/9/09/P09009",
    reportNumber   = "CERN-PH-EP-2014-120",
    eprint         = "1406.7690",
    archivePrefix  = "arXiv",
    primaryClass   = "hep-ex",
}

@Article{STDM-2012-23,
    author         = "{ATLAS Collaboration}",
    title          = "{Measurement of the \(Z/\gamma^*\) boson transverse momentum distribution in \(pp\) collisions at \(\sqrt{s} = 7\,\text{TeV}\) with the ATLAS detector}",
    journal        = "JHEP",
    volume         = "09",
    year           = "2014",
    pages          = "145",
    doi            = "10.1007/JHEP09(2014)145",
    reportNumber   = "CERN-PH-EP-2014-075",
    eprint         = "1406.3660",
    archivePrefix  = "arXiv",
    primaryClass   = "hep-ex",
}

@Article{IDET-2013-01,
    author         = "{ATLAS Collaboration}",
    title          = "{Operation and performance of the ATLAS semiconductor tracker}",
    journal        = "JINST",
    volume         = "9",
    year           = "2014",
    pages          = "P08009",
    doi            = "10.1088/1748-0221/9/08/P08009",
    reportNumber   = "CERN-PH-EP-2014-049",
    eprint         = "1404.7473",
    archivePrefix  = "arXiv",
    primaryClass   = "hep-ex",
}

@Article{IDET-2015-01,
    author         = "{ATLAS Collaboration}",
    title          = "{Performance of the ATLAS Transition Radiation Tracker in Run~1 of the LHC: tracker properties}",
    journal        = "JINST",
    volume         = "12",
    year           = "2017",
    pages          = "P05002",
    doi            = "10.1088/1748-0221/12/05/P05002",
    reportNumber   = "CERN-EP-2016-311",
    eprint         = "1702.06473",
    archivePrefix  = "arXiv",
    primaryClass   = "hep-ex",
}

@Article{PERF-2015-06,
    author         = "{ATLAS Collaboration}",
    title          = "{A measurement of material in the ATLAS tracker using secondary hadronic interactions in \(7\,\text{TeV}\) \(pp\) collisions}",
    journal        = "JINST",
    volume         = "11",
    year           = "2016",
    pages          = "P11020",
    doi            = "10.1088/1748-0221/11/11/P11020",
    reportNumber   = "CERN-EP-2016-137",
    eprint         = "1609.04305",
    archivePrefix  = "arXiv",
    primaryClass   = "hep-ex",
}

@Article{PERF-2015-07,
    author         = "{ATLAS Collaboration}",
    title          = "{Study of the material of the ATLAS inner detector for Run~2 of the LHC}",
    journal        = "JINST",
    volume         = "12",
    year           = "2017",
    pages          = "P12009",
    doi            = "10.1088/1748-0221/12/12/P12009",
    reportNumber   = "CERN-EP-2017-081",
    eprint         = "1707.02826",
    archivePrefix  = "arXiv",
    primaryClass   = "hep-ex",
}

@Article{PERF-2017-01,
    author         = "{ATLAS Collaboration}",
    title          = "{Electron reconstruction and identification in the ATLAS experiment using the 2015 and 2016 LHC proton--proton collision data at \(\sqrt{s} = 13\,\text{TeV}\)}",
    journal        = "Eur. Phys. J. C",
    volume         = "79",
    year           = "2019",
    pages          = "639",
    doi            = "10.1140/epjc/s10052-019-7140-6",
    reportNumber   = "CERN-EP-2018-273",
    eprint         = "1902.04655",
    archivePrefix  = "arXiv",
    primaryClass   = "hep-ex",
}

@Article{PIX-2018-001,
    author         = "Abbott, B. and others",
    title          = "{Production and integration of the ATLAS Insertable B-Layer}",
    journal        = "JINST",
    volume         = "13",
    year           = "2018",
    pages          = "T05008",
    doi            = "10.1088/1748-0221/13/05/T05008",
    eprint         = "1803.00844",
    archivePrefix  = "arXiv",
    primaryClass   = "physics.ins-det",
}

@Booklet{ATL-SOFT-PUB-2020-001,
    author         = "{ATLAS Collaboration}",
    title          = "{ATLAS Computing Acknowledgements}",
    howpublished   = "{ATL-SOFT-PUB-2020-001}",
    url            = "https://cds.cern.ch/record/2717821",
}

@Report{ATLAS-TDR-4,
    author         = "{ATLAS Collaboration}",
    title          = "{ATLAS Inner Detector: Technical Design Report, 1}",
    type           = "ATLAS-TDR-4",
    institution    = "CERN",
    year           = "1997",
    reportNumber   = "CERN-LHCC-97-016",
    url            = "https://cds.cern.ch/record/331063",
}

@Report{ATLAS-TDR-5,
    author         = "{ATLAS Collaboration}",
    title          = "{ATLAS Inner Detector: Technical Design Report, 2}",
    type           = "ATLAS-TDR-5",
    institution    = "CERN",
    year           = "1997",
    reportNumber   = "CERN-LHCC-97-017",
    url            = "https://cds.cern.ch/record/331064",
}

@Article{CMS-TRK-11-002,
    author         = "{CMS Collaboration}",
    title          = "{Alignment of the CMS tracker with LHC and cosmic ray data}",
    journal        = "JINST",
    volume         = "9",
    year           = "2014",
    pages          = "P06009",
    doi            = "10.1088/1748-0221/9/06/P06009",
    reportNumber   = "CERN-PH-EP-2014-028",
    eprint         = "1403.2286",
    archivePrefix  = "arXiv",
    primaryClass   = "hep-ex",
}

@Booklet{ATLAS-CONF-2012-141,
    author         = "{ATLAS Collaboration}",
    title          = "{Study of alignment-related systematic effects on the ATLAS Inner Detector track reconstruction}",
    howpublished   = "{ATLAS-CONF-2012-141}",
    url            = "https://cds.cern.ch/record/1483518",
    year           = "2012",
}
 
\clearpage % ATLAS Collaboration author list
% Reference date of IDTR-2019-05 is 2020-04-21
% Author list last updated on date 10-OCT-20
% Data extracted on 10-Oct-2020 for paper reference IDTR-2019-05
% at 12:58pm
 
\begin{flushleft}
\hypersetup{urlcolor=black}
{\Large The ATLAS Collaboration}

\bigskip

\AtlasOrcid[0000-0002-6665-4934]{G.~Aad}$^\textrm{\scriptsize 102}$,    
\AtlasOrcid[0000-0002-5888-2734]{B.~Abbott}$^\textrm{\scriptsize 128}$,    
\AtlasOrcid{D.C.~Abbott}$^\textrm{\scriptsize 103}$,    
\AtlasOrcid[0000-0002-2788-3822]{A.~Abed~Abud}$^\textrm{\scriptsize 36}$,    
\AtlasOrcid[0000-0002-1002-1652]{K.~Abeling}$^\textrm{\scriptsize 53}$,    
\AtlasOrcid[0000-0002-2987-4006]{D.K.~Abhayasinghe}$^\textrm{\scriptsize 94}$,    
\AtlasOrcid[0000-0002-8496-9294]{S.H.~Abidi}$^\textrm{\scriptsize 166}$,    
\AtlasOrcid[0000-0002-8279-9324]{O.S.~AbouZeid}$^\textrm{\scriptsize 40}$,    
\AtlasOrcid{N.L.~Abraham}$^\textrm{\scriptsize 155}$,    
\AtlasOrcid[0000-0001-5329-6640]{H.~Abramowicz}$^\textrm{\scriptsize 160}$
% ...
\bigskip
\\

% ...
$^{36}$CERN, Geneva; Switzerland.\\
% ...
$^{40}$Niels Bohr Institute, University of Copenhagen, Copenhagen; Denmark.\\
% ...
$^{53}$II. Physikalisches Institut, Georg-August-Universit\"{a}t G\"ottingen, G\"ottingen; Germany.\\
% ...
$^{94}$Department of Physics, Royal Holloway University of London, Egham; United Kingdom.\\
% ...
$^{102}$CPPM, Aix-Marseille Universit\'e, CNRS/IN2P3, Marseille; France.\\
$^{103}$Department of Physics, University of Massachusetts, Amherst MA; United States of America.\\
% ...
$^{128}$Homer L. Dodge Department of Physics and Astronomy, University of Oklahoma, Norman OK; United States of America.\\
% ...
$^{155}$Department of Physics and Astronomy, University of Sussex, Brighton; United Kingdom.\\
% ...
$^{160}$Raymond and Beverly Sackler School of Physics and Astronomy, Tel Aviv University, Tel Aviv; Israel.\\
% ...
$^{166}$Department of Physics, University of Toronto, Toronto ON; Canada.\\
% ...

\end{flushleft}

% Created with Glance <Atlas.Glance@cern.ch>

\FloatBarrier
\clearpage
\FloatBarrier
 
\end{document}